\documentclass[epsfig,10pt]{article} 
\usepackage[utf8]{inputenc}
\usepackage{tikz}
\usepackage{longtable}
\usepackage[top=1in, left=0.95in, bottom=1.1in, right=0.95in]{geometry}
\usepackage[colorlinks=true,
linkcolor=blue,
urlcolor=red,
citecolor=red]{hyperref}
\usepackage[toc]{appendix}
\usepackage{amssymb, amsmath,mathrsfs}
\usepackage{times} 
\usepackage{mathptmx} 

\usepackage{multicol}
\usepackage{multirow}
\usepackage{ytableau}
\usepackage[numbers,sort&compress]{natbib}
\numberwithin{equation}{section}

\newcommand{\lrd}{\overleftrightarrow{D}}
\newcommand{\bin}[3]{(\overline{#1} #3 #2)}
\newcommand{\lra}[1]{\langle #1 \rangle}
\newcommand{\caly}{\mathcal{Y}}
\newcommand{\calb}{\mathcal{B}}

\begin{document}
\begin{center}


{\Large \textbf  {Flavor and CP Symmetries in the Standard Model Effective Field Theory}}\\[10mm]

Hao Sun$^{a, b}$\footnote{sunhao@itp.ac.cn}, Jiang-Hao Yu$^{a, b, c}$\footnote{jhyu@itp.ac.cn}\\[10mm]

\noindent 
$^a${\em \small CAS Key Laboratory of Theoretical Physics, Institute of Theoretical Physics, Chinese Academy of Sciences,    \\ Beijing 100190, P. R. China}  \\
$^b${\em \small School of Physical Sciences, University of Chinese Academy of Sciences,   Beijing 100049, P.R. China}   \\
$^c${\em \small School of Fundamental Physics and Mathematical Sciences, Hangzhou Institute for Advanced Study, UCAS, Hangzhou 310024, China}
\\[10mm]

\date{\today}   
          
\end{center}

\begin{abstract}


The CP properties of effective operators are closely related to the spacetime and internal symmetries, such as the flavor symmetry, in the standard model effective field theory (SMEFT). In this work, we utilize the flavor symmetry to organize and reduce numbers of independent Wilson coefficients of the SMEFT operators. We classify the dimension-6 and dimension-8 baryon/lepton-number-conserving operators based on their CP properties. The $U(1)^4$ rephasing symmetry is applied to distinguish CP-violating and CP-odd operators which leads to reduction of the independent CP-violating phases. The $U(3)^5$ and $U(2)^5$ flavor symmetries let us classify the CP-violating phases as the flavor invariants, which can be enumerated by the Hilbert series and obtained explicitly by the Young tensor method. Then after introducing the minimal flavor violation (MFV) hypothesis, we present the flavor structures of the dimension-6 and dimension-8 SMEFT operators under the MFV hypothesis utilizing the spurion method.

\end{abstract}

\newpage

\tableofcontents

\newpage

\section{Introduction}

The standard model (SM) is a successful theory describing the strong, weak and electromagnetic interactions with the electroweak unification at the electroweak scale $\Lambda_{\text{EW}}\sim 1\text{ TeV}$. The SM is a gauge field theory with the gauge group $SU(3)_c\times SU(2)_{EW} \times U(1)_Y$, and the dynamic fields are the gauge bosons of the gauge group, the 3-flavor leptons and quarks, and one scalar field, the Higgs. In the past several decades, the search for new particles around $1\text{ TeV}$ has failed, which implies there is a considerable gap from the SM scale to the new physics scale. Under such a circumstance, the effective field theory (EFT) method is beneficial. The EFT with the gauge symmetry $SU(3)_c\times SU(2)_{EW} \times U(1)_Y$ and the SM fields is the standard model effective field theory (SMEFT). Beyond the renormalizable operators, the dimension 5 operators are first written by Weinberg~\cite{Weinberg:1979sa}, and the complete and independent dimension 6 operators are first obtained in Ref.~\cite{Buchmuller:1985jz,Grzadkowski:2010es}. Subsequently, the higher dimensional operators have been obtained in Ref.~\cite{Henning:2015alf,Lehman:2014jma,Liao:2016hru,Li:2020gnx,Murphy:2020rsh,Li:2020xlh,Liao:2020jmn,Harlander:2023psl}. 


Generally, there are huge sets of effective operators at higher mass dimensions because of the complicated flavor structures, so it is important to classify and organize these operators systematically. When flavor symmetry is absent, the operators in the SMEFT Lagrangian are all flavor multiples and an effective organization of these operators is to utilize the repeated fields and the associated symmetric group representations~\cite{Li:2020gnx,Li:2020xlh}.
Because the irreducible representations of the symmetric groups correspond to some Young diagrams, the operators of specific representation can be obtained by the corresponding Young symmetrizer applying on the flavor space. In practice, such a decomposition enables one to write down the independent components of the operator in the flavor space by the semi-standard Young tableaux.

The CP violation is important in new physics searches beyond the standard model, especially the matter-anti-matter asymmetry, etc. Since the CP-violation phase in the CKM-matrix~\cite{Cabibbo:1963yz,Kobayashi:1973fv} is tiny, the new CP-violation phases in the high-dimension operators of the SMEFT should exist to explain the baryon asymmetry of the universe. The CP-violation effects of the SM are highly suppressed, thus the experiments could be sensitive to the CP violations of the high-dimension operators. 
The parity (P) is a space-time transformation, while the charge conjugate (C) is related to the outer automorphism of the internal symmetries of the SMEFT, including the gauge symmetry and other global symmetries.
The CP-odd operators are of negative eigenvalues under the CP transformations,
\begin{equation}
    \text{CP-odd:}\quad \mathcal{O} \xrightarrow{CP} -\mathcal{O}\,,
\end{equation}
which could be the combinations of the operators involving the fermions.
The literature presents the CP-odd operators of the SMEFT at dimension 6~\cite{Alonso:2013hga} and the part of the CP-odd bosonic operators of the SMEFT at dimension 8~\cite{Remmen:2019cyz,Durieux:2024zrg}. Nevertheless, the CP-odd operators are not necessarily CP-violating, since the CP-violations depend on whether the complex phase of the Wilson coefficients can be absorbed or not by rephasing symmetry. Only when the CP-violating phases in the corresponding Wilson coefficients can not be eliminated by the field redefinitions, the CP-odd operators are CP-violating ones, indicating that they are invariant under the rephasing symmetry. The Hilbert series has presented the CP-odd operators and the CP-violating operators of the SMEFT up to dimension 14 with the rephasing symmetry added~\cite{Kondo:2022wcw}, which implies considerable differences between these two kinds of operators.

\begin{figure}[htb]
    \centering
    \includegraphics[scale=0.5]{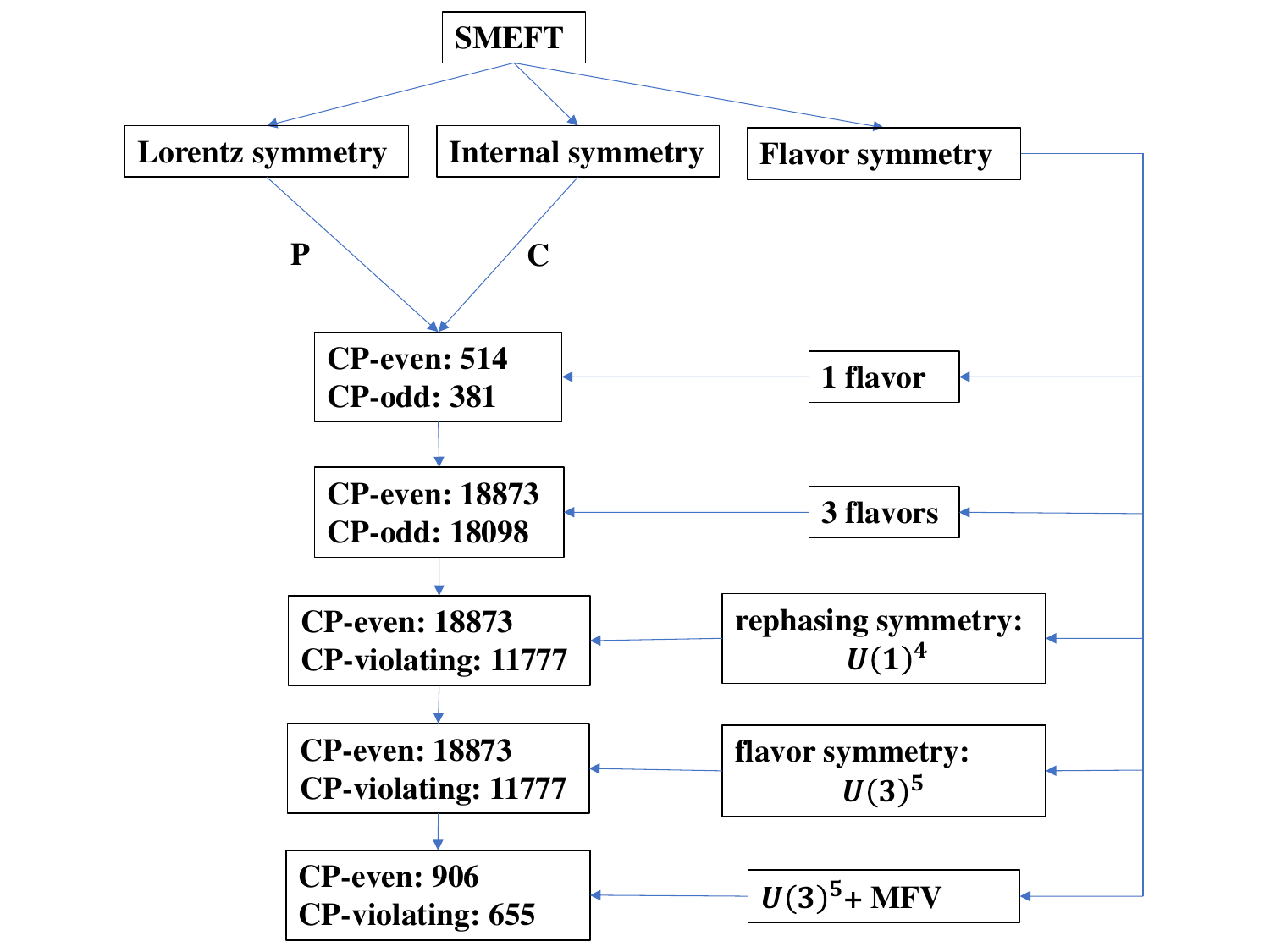}
    \caption{The logic flow in this paper, where the numbers are of the baryon/lepton-number-conserving dimension-8 operators. The reduction of the CP-violating operators due to the flavor symmetry and the MFV hypothesis is transparent.}
    \label{fig:logic_diagram}
\end{figure}

The assuming flavor symmetry of the SMEFT can reduce the independent parameters, and help to organize the effective operators more concisely.
At the same time, the flavor symmetry affects the CP violations of the SMEFT since it is an internal symmetry. On the one hand, the flavor symmetry reduces the independent operators, on the other hand, different flavor symmetries lead to different rephasing symmetries. Since the rephasing symmetry is a subgroup of the flavor symmetry, it is satisfied automatically and the CP-odd operators are always equivalent to the CP-violating ones. Because the Yukawa interactions in the SM Lagrangian are not flavor universal, it seems the choices of the flavor symmetries are arbitrary. If it is required the flavor symmetry can generate the SM Yukawa interactions and suppress the non-standard flavor-violating effects, the choices are quite limited. In this paper, we consider two flavor symmetries, the $U(3)^5$ symmetry and the $U(2)^5$ symmetry. 
The $U(3)^5$ is the largest flavor symmetry consistent with the SM gauge symmetry, and the single assumption of the $U(3)^5$ symmetry does not affect the CP-violating phases since the rephasing symmetry is the same with the case without the flavor symmetry, as shown in Fig.~\ref{fig:logic_diagram}.
Nevertheless, the flavor symmetry develops the Wilson coefficients to the spurions covariant under that, and keeps the Lagrangian flavor invariant,
\begin{equation}
\label{eq:spurion_langrangian}
    \mathcal{L} \supset c f(\mathbf{T}) \cdot \mathcal{O}(\psi) \,,
\end{equation}
where $c$ are  complex coefficients, $\mathbf{T}$ are the spurion. The dot '$\cdot$' implies the spurion part $f(\mathbf{T})$ and the operator $\mathcal{O}$ composed by the dynamic fields form flavor invariants. Thus the flavor transformations of the Wilson coefficients, which have been developed to the spurions $f(\mathbf{T})$ now, and the operators $\mathcal{O}$ are inverse to each other. 
Consequently, the assuming flavor symmetry implies that the independent CP-phases of the SMEFT can also be extracted by the flavor invariants~\cite{Bonnefoy:2021tbt,Bonnefoy:2023bzx}, which claims that the flavor-violating primary invariants linear in the Wilson coefficients correspond the independent CP-violating phases. Similar arguments are also applied to the seesaw model~\cite{Wang:2021wdq,Yu:2021cco,Yu:2022ttm} and the SMEFT with sterile neutrinos~\cite{Grojean:2024qdm}. 
The flavor invariants under the $U(3)^5$ symmetry have presented consistent results with the ones obtained from the rephasing symmetry at dimension 6~\cite{Bonnefoy:2021tbt,Kondo:2022wcw}. 
These flavor invariants form a subset of the primary invariants. The primary invariants are the ones algebraically independent, which means neither themselves nor their products can be expressed by other invariants. 
The primary invariants correspond to the physical parameters and can be counted by the Hilbert series.
In particular, the primary invariants should be distinguished from the basic invariants, which are linearly independent but not algebraically independent. The numbers of the basic invariant usually exceed the numbers of the physical parameters. In practice, we usually construct the basic invariants first and then reduce them to the primary ones.
In reality, the construction of the basic invariants is difficult, which exploits the numeric method~\cite{Bonnefoy:2021tbt,Wang:2021wdq,Yu:2021cco,Yu:2022ttm,Grojean:2024qdm}. 
In addition to other methods~\cite{Darvishi:2023ckq,Darvishi:2024cwe}, as discussed in Sec.~\ref{sec:flavor_symmetry}, we present a new method utilizing the Young tableaux to construct the basic invariants systematically and analytically.


The flavor symmetry is broken in the SMEFT, and the experiments imply that the flavor violations beyond the leading order (LO) are suppressed. At the LO, the flavor-changing neutral current (FCNC) interactions of the quarks are at the loop level, thus the FCNC interactions are significantly suppressed by the loop integral, the off-diagonal elements of the CKM-matrix, and the GIM mechanics~\cite{Glashow:1970gm} within the SM, thus should be sensitive to new physics effects. Nevertheless, the experiments testing the FCNC processes present the consistent results with the SM prediction. For example, the rate of the leptonic decay $K^0_L\rightarrow \mu^+\mu^-$, which corresponds to the process $d\overline{s}\rightarrow \mu^+\mu^-$ at the quark level, is $(6.84\pm 0.11)\times 10^{-9}$ and is in agreement with the SM prediction~\cite{ParticleDataGroup:2024cfk}.
Thus, although the higher-dimension operators present other FCNC processes originating from new physics, their modifications to the SM results should be very small, which constrains the cutoff scale a bound of $O(10^5\text{ TeV})$~\cite{Isidori:2010kg}. Nevertheless, the cutoff scale can be pulled down by the minimal flavor violating (MFV) hypothesis~\cite{Chivukula:1987py,DAmbrosio:2002vsn,Bonnefoy:2020yee,Aoude:2020dwv,Bruggisser:2021duo,Bruggisser:2022rhb,Bartocci:2023nvp,Bartocci:2024fgj}. The MFV hypothesis contains two ingredients, the flavor symmetry and a set of symmetry-breaking terms. 
In the SM, the symmetry-breaking terms are the Yukawa terms, thus the MFV hypothesis states that all the flavor violation effects in the SMEFT are from the Yukawa terms, which means all the Wilson coefficients are some functions of the Yukawa matrices. The $U(3)^5$ symmetry together with the MFV hypothesis presents strong constraints of the high-dimension flavor violating contributions, and reduces the independent parameters greatly, while the minimal breaking of the $U(2)^5$ symmetry retains more structures and contains richer phenomena~\cite{Barbieri:2011ci,Barbieri:2012uh,Blankenburg:2012nx,Greljo:2015mma,Barbieri:2015yvd,Buttazzo:2017ixm}.

In this paper, we analyze the flavor structures and the CP properties of the SMEFT operators conserving the baryon/lepton number of dimensions 6 and 8 to organize and classify the large number of operators. We present the SMEFT effective operators of specific CP properties explicitly up to dimension 8, including both the bosonic operators and the fermionic operators. In particular, we consider the 3 flavors of the fermions.
We discuss the relations between the flavor symmetry and the CP properties, and the flavor-invariant description of the CP-violating phases. In particular, we discuss the Young tableaux method to construct the primary invariants.
We adopt the $U(3)^5$ symmetry and the MFV hypothesis so that the independent parameters of the effective Lagrangian are reduced. 
We present the flavor structures of the SMEFT operators of dimensions 6 and 8 under the MFV hypothesis. A similar analysis of the dimension-6 operators has been presented in Ref.~\cite{Faroughy:2020ina}, while the dimension-8 operators have new flavor structures in the classes $\psi^4\phi^2$ and $\psi^4\phi D$. 
For these two classes with new structures, we also present their flavor structures under the $U(2)^5$ symmetry, which retains more operators.
To obtain the most general flavor structures of the operators, we adopt the spurion method, which constructs the flavor-invariant operators via some spurions and has been used in many other theories~\cite{Sun:2022snw,Sun:2022ssa,Song:2025snz}. After the spurions get their vacuum expectation values (VEVs), the MFV operators return to the SMEFT ones. 
On the other hand, we enumerate the CP-odd operators under the MFV hypothesis, which is equivalent to the CP-violating operators. At the same time, we adopt a specific weak basis, which means the VEVs of the spurions are fixed, the flavor symmetry is broken explicitly and every operator splits into several SMEFT operators, of which a CP-violating operator also splits into some CP-violating operators sharing the same CP-violating phase. 
In particular, both the numbers of CP-violating operators before and after the fix of the VEVs are less than the ones without any flavor symmetry. 
In practice, we perform our analysis on the down-basis, and we do not consider the right-handed neutrinos and the baryon/lepton-number violations throughout the discussion.
A logic flow of the paper is presented in Fig.~\ref{fig:logic_diagram}, where the numbers are of the baryon/lepton-number-conserving dimension-8 operators. Together with Fig.~\ref{fig:logic_diagram}, we compare the different assumptions in the flavor sector of the SMEFT as follows,
\begin{itemize}
    \item When there is no flavor symmetry, the rephasing symmetry $U(1)^4$ emerges at the LO Lagrangian as an accidental symmetry, which acts only on the dynamic fields, and the Wilson coefficients are regarded as general complex numbers.
    \item When some flavor symmetry is assumed, the Wilson coefficients are developed to the covariant spurions, and the Lagrangian is formally flavor invariant. The spurions extend the building blocks of the SMEFT, and the Lagrangian takes the form of Eq.~\eqref{eq:spurion_langrangian}. In general, every distinct operator presents an independent spurion.
    \item The MFV hypothesis states the spurions are limited and are the Yukawa matrices, which means the spurion part in Eq.~\eqref{eq:spurion_langrangian} are composed by the 3 Yukawa matrices,
    \begin{equation}
        f(\mathbf{T}) \equiv f(\mathbf{Y}_u\,,\mathbf{Y}_e\,,\mathbf{Y}_e)\,,
    \end{equation}
    which generally takes an expansion form, and we have used the bold notation to imply that the Yukawa matrices have been regarded as spurions.
\end{itemize}

The paper is organized as follows. In Sec.~\ref{sec:flavor_symmetry}, we review the flavor organizations of the effective operators without flavor symmetry. 
In Sec.~\ref{sec:flavor}, we review the rephasing symmetry and the CP property of the SMEFT without flavor symmetry. 
We present the completed operators at dimensions 6 and dimension 8 classified by their CP properties.
In Sec.~\ref{sec:flavor_symmetry}, we discuss the flavor symmetry and its effects on the CP-violating phases of the SMEFT. In particular, we consider the $U(3)^5$ symmetry and the $U(2)^5$ symmetry and discuss the Young tableaux method applying on the flavor-invariant descriptions of the CP-violating phases.
In Sec.~\ref{sec:MFV}, we introduce the MFV hypothesis and present the independent flavor structures of the dimension-6 and dimension-8 operators in detail of the SMEFT with the flavor symmetry $U(3)^5$ and $U(2)^5$ respectively. We enumerate the CP-violating operators under the MFV hypothesis and find they are reduced compared to the ones without flavor symmetry. Finally, we conclude this paper in Sec.~\ref{sec:con}. We present the fermionic operator basis of dimension 6 and dimension 8 in App.~\ref{app:dim6} and App.~\ref{app:dim8}, respectively.

\section{Flavor Structures of SMEFT}

\label{eq:flavor_structure}

The SMEFT is an effective theory with the gauge group $SU(3)_c\times SU(2)_{EW} \times U(1)_Y$, of which the associated gauge fields are $G$, $W$, and $B$, respectively, where the first one is the gluon field, while the last two will combine to form the weak gauge bosons $Z\,, W^\pm$ and the photon $A$ after the breaking of the $SU(2)_{EW}$. 
In addition to the gauge bosons, there are 3-flavor left/right-handed fermions, containing the leptons and quarks, and a scalar field, the Higgs field. 
The representations under the gauge group of all these fields are presented in Tab.~\ref{tab:1}. In particular, the right-handed neutrinos are not included in the SMEFT, since it is of trivial representation of the gauge group. 
In this section, we discuss the flavor structures of the SMEFT when flavor symmetry is absent, and illustrate an organization of the large number of operators by the repeated fields and the symmetric group.

The LO Lagrangian of the SMEFT is the standard model Lagrangian, and can be separated into different sectors,
\begin{equation}
\label{eq:smeft_lo}
    \mathcal{L}_{\text{SM}} = \mathcal{L}_{\text{gauge bosons}}+\mathcal{L}_{\text{fermions}} + \mathcal{L}_{\text{Higgs}}+\mathcal{L}_{\text{Yukawa}}\,.
\end{equation}
The first two sectors contain the dynamic terms of the gauge bosons and the fermions,
\begin{align}
    \mathcal{L}_{\text{gauge bosons}} &= -\frac{1}{4}{B^{\mu\nu}}{B_{\mu\nu}}-\frac{1}{4}{W^I}^{\mu\nu}{W^I_{\mu\nu}}-\frac{1}{4}{G^A}^{\mu\nu}{G^A_{\mu\nu}}\,, \\
    \mathcal{L}_{\text{fermions}} &= \bar{q}i\gamma^\mu D_\mu q + \bar{u}_Ri\gamma^\mu D_\mu u_R + \bar{d}_Ri\gamma^\mu D_\mu d_R + \bar{l}i\gamma^\mu D_\mu l + \bar{e}_Ri\gamma^\mu D_\mu e_R\,.
\end{align}
The left-handed fermions are $SU(2)_{EW}$ doublets,
\begin{equation}
    q=\left(\begin{array}{c}
u_L \\d_L
    \end{array}\right)\,,\quad l=\left(\begin{array}{c}
\nu_L \\ e_L 
    \end{array}\right)\,,
\end{equation}
while the right-handed fermions are $SU(2)_{EW}$ singlets. In particular, the summations in the flavor space are left implicitly, which means the fermions in the $\mathcal{L}_{\text{fermion}}$ are flavor triplets
\begin{equation}
\label{eq:flavor_multiples}
    \psi = \left(\begin{array}{c}
\psi_1\\\psi_2\\\psi_3 
    \end{array}\right)\,,\quad \psi=q,u_R,d_R,l,e_R\,,
\end{equation}
thus each term in the $\mathcal{L}_{\text{fermion}}$ is a short notation of the form
\begin{equation}
\label{eq:notation}
    \bar{\psi}i\gamma^\mu D_\mu \psi = \left(\begin{array}{ccc}
\bar{\psi}_1 & \bar{\psi}_2 & \bar{\psi}_3 
    \end{array}\right) i\gamma^\mu D_\mu  \left(\begin{array}{c}
\psi_1\\\psi_2\\\psi_3 
    \end{array}\right)= \sum_{p=1}^3\bar{\psi}^p i\gamma^\mu D_\mu\psi_p\,,\quad \psi=q\,,u_R\,,d_R\,,l\,,e_R\,,
\end{equation}
where $p$ is the flavor index of the fermions.

To generate the mass terms, one more scalar field, the Higgs field $H(x)$, is needed. Its dynamic term and self-interactions are described in the $\mathcal{L}_{\text{Higgs}}$
\begin{equation}
    \mathcal{L}_{\text{Higgs}} = D_\mu H^\dagger D^\mu H + V(H)\,,
\end{equation}
and its interactions with the fermions take the form of Yukawa interactions
\begin{equation}
\label{eq:L_higgs}
    \mathcal{L}_{\text{Yukawa}} =  \bar{q}\tilde{H}Y_u u_R + \bar{q}H Y_d d_R + \bar{l}H Y_e e_R + \text{h.c.} \,,
\end{equation}
where $\tilde{H}=i\sigma^2 H^*$ is its dual field.
The Yukawa couplings $Y_u\,, Y_d$ and $Y_e$ are general $3\times 3$ complex matrices in the flavor space, and the terms above are implicit for the flavor summations, which can be written explicitly as
\begin{equation}
    \mathcal{L}_{\text{Yukawa}} = \sum_{p,r=1}^3\bar{q}^p \tilde{H}\left(Y_u\right){}_p^r u_R{}_r + \sum_{p,r=1}^3\bar{q}^p H\left(Y_d\right){}_p^r d_R{}_r + \sum_{p,r=1}^3\bar{l}^p H\left(Y_e\right){}_p^r e_R{}_r + \text{h.c.} \,.
\end{equation}

\begin{table}
\renewcommand{\arraystretch}{1.8}
    \centering
    \begin{tabular}{c|c|c|c|c}
\hline
\multicolumn{2}{c|}{Field} & $SU(3)_C$ & $SU(2)_{EW}$ & $U(1)_Y$ \\
\hline
\multirow{3}{*}{Gauge boson} & $G$ & $\mathbf{8}$ & $\mathbf{1}$ & $0$ \\
& $W$ & $\mathbf{1}$ & $\mathbf{3}$ & $0$ \\
& $B$ & $\mathbf{1}$ & $\mathbf{1}$ & $0$ \\
\hline
\multirow{5}{*}{Fermion} & $l$ & $\mathbf{1}$ & $\mathbf{2}$ & $-\frac{1}{2}$ \\
& $e_R$ & $\mathbf{1}$ & $\mathbf{1}$ & $1$ \\
& $q$ & $\mathbf{3}$ & $\mathbf{2}$ & $\frac{1}{6}$ \\
& $u_R$ & ${\mathbf{3}}$ & $\mathbf{1}$ & $\frac{2}{3}$ \\
& $d_R$ & ${\mathbf{3}}$ & $\mathbf{1}$ & $-\frac{1}{3}$ \\
\hline
Scalar & $H$ & $\mathbf{1}$ & $\mathbf{2}$ & $\frac{1}{2}$ \\
\hline
    \end{tabular}
    \caption{The fields and their representations under the gauge group of the SMEFT.}
    \label{tab:1}
\end{table}

Moving on to the high dimension, there are a lot of effective operators of the SMEFT beyond the LO, which mainly originates from that the fermions are flavor multiples as shown in Eq.~\eqref{eq:flavor_multiples}.
Because the operators are not necessarily the flavor invariants, when the flavor symmetry is absent, they are generally flavor multiples, 
\begin{equation}
    \mathcal{L} = \sum_d \frac{c}{\Lambda^{d-4}} \mathcal{O}^d_{pr\dots s}\,,
\end{equation}
where $p\,,r\,,\dots\,,s$ are the flavor indices ranging from $1$ to $3$, thus such a term in the Lagrangian presents $3\times 3 \times \dots \times 3$ operators. For example, each term of the Yukawa interactions at the LO in Eq.~\eqref{eq:L_higgs} actually presents $3\times 3=9$ operators, and the 4-fermion operator such as 
\begin{equation}
    \mathcal{O}_{le}=(\bar l^p \gamma_\mu l_r)(\bar e^s \gamma^\mu e_t)\,,
\end{equation}
at dimension 6 presents $3^4 = 81$ operators in the flavor space.

Beyond the SM, there are repeated fields, and it reduces the number of the operators in the flavor space due to the permutation symmetry in principle. If the repeated fields are bosons, they should be symmetric under the permutation, since their flavor number are 1. However, if the repeated fields are fermions, their permutation symmetries are complex because of their nontrivial flavor numbers. We assume there are $N$ repeated fields in an operator, and their flavor number is $n_f$ generally. Because all the fields are of some specific representations of the Lorentz group and the gauge group as shown in Tab.~\ref{tab:1}, the permutation symmetry of the $N$ repeated fields are determined by the permutation symmetries of the Lorentz group and the gauge group simultaneously,
\begin{equation}
    \text{Sym}_\mathcal{O} = \text{Sym}_{\text{Lorentz}} \times \text{Sym}_{\text{gauge}} \times \text{Sym}_{\text{statistics}}\,,
\end{equation}
where the last factor $\text{Sym}_{\text{statistics}}$ is related to the spin-statistics of the repeated fields,
\begin{align}
    \text{Sym}_{\text{statistics}} & = 1\,,\quad \text{The repeated field is boson,} \\
    \text{Sym}_{\text{statistics}} & = -1\,,\quad \text{The repeated field is fermion.} 
\end{align}

The permutations of some objects are described by the symmetric groups, thus the effective operators can be organized by the irreducible representations under the symmetric group of the repeated fields. If there are $N$ repeated fields with flavor number $n_f$, the operators, as flavor multiples, can be decomposed as the irreducible representations of the symmetric group $S_N$. If the repeated field is a boson, the flavor number is 1, and only the symmetric representation is non-vanishing. On the contrary, if the repeated field is a fermion, other symmetric group representations could emerge. In general, if the number of the repeated field $N$ exceeds the flavor number $n_f$, $N>n_f$,  the non-vanishing representations of the operators are truncated by the condition that the number of the asymmetric fields can not exceed the flavor number $n_f$.

For example, the dimension-6 operator
\begin{equation}
    \mathcal{O}_{quqd}^{(1)} = (\bar q^{jp} u_r) \epsilon_{jk} (\bar q^{ks} d_t)\,,
\end{equation}
has a pair of repeated quarks, and it can be decomposed into the irreducible representations of the symmetric group $S_2$, which correspond to the Young diagrams that
\begin{equation}
\ytableausetup
 {boxsize=1.25em}
\ytableausetup
 {aligntableaux=top}
    [2] = \ydiagram{2}\,,\quad [1^2] = \ydiagram{1,1}\,.
\end{equation}
If the flavor number of the quark is 1, $n_f=1$, only the symmetric $[2]$ representation is non-vanishing, while if the flavor number is 3, $n_f=3$, both representations are non-vanishing. Thus we can decouple and organize the operators as 
\begin{equation}
    \mathcal{O}_{quqd}^{(1)} = \mathcal{Y}(\ytableaushort{ps})(\bar q^{jp} u_r) \epsilon_{jk} (\bar q^{ks} d_t) + \mathcal{Y}(\ytableaushort{p,s})(\bar q^{jp} u_r) \epsilon_{jk} (\bar q^{ks} d_t)\,,
\end{equation}
where $\mathcal{Y}$ is the Young symmetrizer of the corresponding representation.

The organization of the effective operators by the repeated fields and their irreducible representations of the associated symmetric group has been used in the Young tensor constructions of the dimension-8 operators~\cite{Li:2020gnx} and the dimension-9 operators~\cite{Li:2020xlh}, which present consistent results with the other method such as the Hilbert series method~\cite{Henning:2015alf}. Besides, the flavor components of the operators are transparent since they can be obtained by the semi-standard Young tableaux directly.

\section{CP Property of SMEFT}
\label{sec:flavor}

In this section, we will discuss the rephasing symmetry of the SMEFT and the associated CP property without flavor symmetry.


\subsection{Charge Conjugate and Parity}

The charge conjugate (C) and the parity (P) are two important discrete symmetries of the elemental particles. Although the effective operators should be invariant under the Lorentz group and the gauge groups, they are not necessary to be invariant under these discrete symmetries, which means CP violation is allowed. 

Parity is a space-time transformation that sends $(t,\vec{x})$ to $(t,-\vec{x})$. If the field is not Lorentz singlet, it would transform under the parity. Thus the parity transformation of a general field $\Phi(t,\vec{x})$ is
\begin{equation}
    P\Phi(t,\vec{x})P^{-1} = \eta_P\mathcal{P}\Phi(t,-\vec{x})\,,
\end{equation}
where $\eta_P$ is its intrinsic party, satisfying $\eta_P^2=1$, and $\mathcal{P}$ is the parity transformation associated to $\Phi$. If $\Phi$ is scalar, $\mathcal{P}$ is trivial, otherwise it is not. For the Dirac spinor field, $\mathcal{P}=\gamma^0$, while for vector field, $\mathcal{P}=\text{diag}(1,-1,-1,-1)$, which is just the space-time parity transformation.

The charge conjugate interchanges the particle and antiparticle, thus it correlates some field and its complex conjugate,
\begin{equation}
    C\Phi(t,\vec{x})C^{-1} = \eta_C\mathcal{C}\Phi^*(t,\vec{x})\,,
\end{equation}
where, similarly $\eta_C$ is the intrinsic charge, while $\mathcal{C}$ is the charge-conjugate transformation corresponding to $\Phi$. Again, $\mathcal{C}$ is trivial for the scalar field while for others, there is
\begin{align}
    \text{Spinor field:}\quad & C\psi(t,\vec{x})C^{-1} = \eta_C i\gamma^0\gamma^2\overline{\psi}(t,\vec{x})^T\,, \\
    \text{Vector field:}\quad & CA_\mu(t,\vec{x})C^{-1} = \eta_C A^*_\mu(t,\vec{x})\,.
\end{align}
As discussed above, the parity and charge conjugate are isomorphic to $Z_2$ groups
\begin{equation}
    \{1,\quad P\}\,,\{1,\quad C\} \cong Z_2\,,
\end{equation}
which define two outer automorphisms of the Lorentz group $SO(3,1)$ and the internal group $I$ respectively~\cite{Grimus:1995zi,Buchbinder:2000cq}.  
For every elements a group, an outer automorphism maps it to another elements.  
Thus the parity and charge conjugate enlarge the groups by the semi-direct product
\begin{equation}
    SO(3,1)\rightarrow O(3,1) = SO(3,1)\rtimes \{1,\quad P\}\,,\quad I\rightarrow \tilde{I} = I \rtimes \{1,\quad C\}\,.
\end{equation}
In particular, if a field is of a specific representation $\mathbf{r}$ of the internal symmetry $I$, the charge conjugate transforms it to the dual representation $\mathbf{r}^*$.
Thus it is natural that the flavor symmetry affects the CP properties of the SMEFT as discussed in Sec.~\ref{sec:flavor_symmetry}.

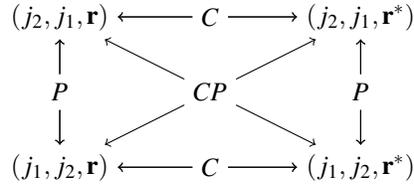
\begin{figure}[ht]
    \centering
    \begin{tikzpicture}
\node (1) at(0,0) {$(j_{1},j_{2},\bf r)$};
\node (2) at(4,0) {$(j_{1},j_{2},\bf r^{*})$};
\node (3) at(0,2) {$(j_{2},j_{1},\bf r)$};
\node (4) at(4,2) {$(j_{2},j_{1},\bf r^{*})$};
\node (5) at(2,0) {$C$};
\node (6) at(2,2) {$C$};
\node (7) at(0,1) {$P$};
\node (8) at(4,1) {$P$};
\node (9) at(2,1) {$CP$};
\draw[<-] (1) --(5);
\draw[->] (5) --(2);
\draw[<-] (1) --(5);
\draw[->] (5) --(2);
\draw[<-] (1) --(7);
\draw[->] (7) --(3);
\draw[<-] (1) --(7);
\draw[->] (7) --(3);
\draw[<-] (2) --(8);
\draw[->] (8) --(4);
\draw[<-] (2) --(8);
\draw[->] (8) --(4);
\draw[<-] (3) --(6);
\draw[->] (6) --(4);
\draw[<-] (3) --(6);
\draw[->] (6) --(4);
\draw[->] (9) --(1);
\draw[->] (9) --(2);
\draw[->] (9) --(3);
\draw[->] (9) --(4);
\end{tikzpicture}
    \caption{The actions of $C$ and $P$ on representations of $G = SO(3,1) \times I$. Here $\bf r$ is the irreducible representation of $I$ and $\bf r^{*}$ is its complex conjugation representation. $(j_{1},j_{2})$ label the representations of the Lorentz group $SO(3,1)$.}
    \label{CPaction}
\end{figure}

In summary, the CP transformation is the combination of them, which takes the form
\begin{align}
    \text{Scalar field:}\quad & CP\phi(t,\vec{x})CP^{-1} = \eta_{CP}\phi^*(t,-\vec{x})\,, \\
    \text{Spinor field:}\quad & CP\psi(t,\vec{x})CP^{-1} = \eta_{CP} -i\gamma^0\gamma^2\overline{\psi}(t,-\vec{x})^T\,, \\
    \text{Vector field:}\quad & CPA_\mu(t,\vec{x})CP^{-1} = \eta_{CP} \mathcal{P}_\mu^\nu A^*_\nu(t,-\vec{x})\,,
\end{align}
where $\eta_{CP}^2=1$. For simplicity we take $\eta_{CP}=1$. 
We note the symmetry group of the SMEFT is $G = SO(3,1) \times I$, where $I$ is the internal symmetry, including the gauge symmetry and some global symmetries, and a field $\Phi$ that
\begin{equation}
    \Phi(t,\vec{x}) \in (j_1,j_2,\mathbf{r})\,,
\end{equation}
where $\mathbf{r}$ is an irreducible representation of the internal group $I$. The C and P transformations of the field $\Phi$ are illustrated in Fig.~\ref{CPaction}

The composing fields determine the CP property of an operator. If it is invariant under the CP transformation, it is CP-even, and if it becomes its opposite, it is CP-odd,
\begin{align}
    \text{CP-even:}\quad &\mathcal{O}\xrightarrow{CP} \mathcal{O} \,,\\
    \text{CP-odd:}\quad &\mathcal{O}\xrightarrow{CP} -\mathcal{O} \,.
\end{align}


The CP properties of the pure meson operators are simple, since all the operators are real or imaginary, and there is no flavor structures. 
On the contrary, the CP properties of the fermionic operators are complicated because most of them are complex and there are flavor structures. As shown in Eq.~\eqref{eq:flavor_multiples}, all the fermions in the SMEFT are flavor triplets, thus a general bilinear $\psi \Gamma \chi$ is a flavor tensor with 9 components,
\begin{equation}
\label{eq:flavor_matrix}
    \overline{\psi} \Gamma \chi = \left(\begin{array}{ccc}
\overline{\psi}_1 \Gamma \chi_1 & \overline{\psi}_1 \Gamma \chi_2 & \overline{\psi}_1 \Gamma \chi_3 \\
\overline{\psi}_2 \Gamma \chi_1 & \overline{\psi}_2 \Gamma \chi_2 & \overline{\psi}_2 \Gamma \chi_3 \\
\overline{\psi}_3 \Gamma \chi_1 & \overline{\psi}_3 \Gamma \chi_2 & \overline{\psi}_3 \Gamma \chi_3 
    \end{array}\right)\,,
\end{equation}
where $\Gamma$ is any Dirac matrix. After the CP transformation, the bilinear becomes
\begin{equation}
    \overline{\psi} \Gamma \chi \xrightarrow{CP} \overline{\chi}\Gamma^C \psi = \left(\begin{array}{ccc}
\overline{\chi}_1 \Gamma^C \psi_1 & \overline{\chi}_1 \Gamma^C \psi_2 & \overline{\chi}_1 \Gamma^C \psi_3 \\
\overline{\chi}_2 \Gamma^C \psi_1 & \overline{\chi}_2 \Gamma^C \psi_2 & \overline{\chi}_2 \Gamma^C \psi_3 \\
\overline{\chi}_3 \Gamma^C \psi_1 & \overline{\chi}_3 \Gamma^C \psi_2 & \overline{\chi}_3 \Gamma^C \psi_3 
    \end{array}\right)\,,
\end{equation}
where $\Gamma^C = \mathcal{C}^{-1} \Gamma^T \mathcal{C} $, and in detail
\begin{align}
    1^C &= 1 \notag \,,\label{eq:dirac_1}\\
    \gamma^5{}^C &= \gamma^5 \,,\\
    \gamma^\mu{}^C &= -\gamma^\mu \,,\\
    (\gamma^5\gamma^\mu)^C &= \gamma^5\gamma^\mu\,,\\
    \sigma^{\mu\nu}{}^C &= - \sigma^{\mu\nu}\,.\label{eq:dirac_2}
\end{align}
Thus the CP transformation of a bilinear is
\begin{equation}
    \overline{\psi} \Gamma \chi \xrightarrow{CP} \overline{\chi}\Gamma^C \psi = \eta_\Gamma \overline{\chi}\Gamma\psi\,,
\end{equation}
where $\eta_\Gamma$ is a factor from $\Gamma$ according to Eq.~\eqref{eq:dirac_1} to Eq.~\eqref{eq:dirac_2}, so that a fermion bilinear of specific CP property are the combinations that
\begin{align}
    \text{CP-even:}\quad & \overline{\psi} \Gamma \chi + \eta_\Gamma \overline{\chi}\Gamma\psi \,, \\
    \text{CP-odd:}\quad & \overline{\psi} \Gamma \chi - \eta_\Gamma \overline{\chi}\Gamma\psi \,.
\end{align}
which contribute to 9 CP-even bilinears and 9 CP-odd bilinears, according to Eq.~\eqref{eq:flavor_matrix}. In particular, if the two fermions $\psi$ and $\chi$ are the same, the CP transformation is closed in the bilinear $\overline{\psi}\Gamma \psi$, there are only 3 CP-odd bilinears from the off-diagonal elements in Eq.~\eqref{eq:flavor_matrix}, and 6 CP-even bilinears. 

Moving on to the 4-fermion operators, which contain two fermion bilinears $(\overline{\psi}\Gamma_1\chi)(\overline{\lambda}\Gamma_2 \xi)$. We can get the CP transformation that
\begin{equation}
    (\overline{\psi}\Gamma_1\chi)(\overline{\lambda}\Gamma_2 \xi) \xrightarrow{CP} \eta_{\Gamma_1}\eta_{\Gamma_2} (\overline{\chi}\Gamma_1\psi)(\overline{\xi}\Gamma_2\lambda)\,,
\end{equation}
so that the combinations of specific CP properties are
\begin{align}
    \text{CP-even:}\quad &  (\overline{\psi}\Gamma_1\chi)(\overline{\lambda}\Gamma_2 \xi) + \eta_{\Gamma_1}\eta_{\Gamma_2} (\overline{\chi}\Gamma_1\psi)(\overline{\xi}\Gamma_2\lambda) \,, \\
    \text{CP-odd:}\quad & (\overline{\psi}\Gamma_1\chi)(\overline{\lambda}\Gamma_2 \xi) - \eta_{\Gamma_1}\eta_{\Gamma_2} (\overline{\chi}\Gamma_1\psi)(\overline{\xi}\Gamma_2\lambda) \,.
\end{align}
Suppose the two fermions in a single bilinear are different. In that case, the combinations present 81 CP-even and CP-odd operators respectively, while if the fermions in the two bilinears are the same, $\psi=\chi$ and $\lambda=\xi$, the CP-transformation is closed, the CP-even operators are reduced to 45, and the CP-odd operators are reduced to 36. In particular, if all the four fermions are the same, $\psi=\chi=\lambda=\xi$, the numbers are further reduced to 27 and 18 for the CP-even and CP-odd operators respectively.

In summary, we note an operator with fermions as $\mathcal{O}$, its CP-transformation takes the form
\begin{equation}
    \mathcal{O} \xrightarrow{CP} \mathcal{O}^C = \eta_\Gamma\eta_B \mathcal{O}^\dagger\,,
\end{equation}
where $\eta_\Gamma$ is the factor from the Dirac matrices, and $\eta_B$ is the factor from the bosonic fields. The combinations of specific CP properties are 
\begin{align}
    \text{CP-even:}\quad & \mathcal{O} + \eta_\Gamma\eta_B \mathcal{O}^\dagger\,, \\
    \text{CP-odd:}\quad & \mathcal{O} - \eta_\Gamma\eta_B \mathcal{O}^\dagger\,.
\end{align}
In particular, the CP-odd and CP-violating operators are in general not equivalent~\cite{Kondo:2022wcw}. The CP-violating operators are the CP-odd ones that are invariant under the rephasing symmetry, which will be discussed next.

\subsection{CP Property of Effective Operators}

\begin{table}[htb]
\renewcommand{\arraystretch}{1.4}
    \centering
    \begin{tabular}{|c|cc|c|c|}
\hline
class & \multicolumn{2}{|c|}{operator} & C & P \\
\hline
\multirow{4}{*}{$X^3$} & $\mathcal{O}_G$ & $f^{ABC}G^A_\nu{}^\mu G^B_\rho{}^\nu G^C_\mu{}^\rho$ & + & + \\
& $\mathcal{O}_{\tilde{G}}$ & $f^{ABC}\tilde{G}^A_\nu{}^\mu G^B_\rho{}^\nu G^C_\mu{}^\rho$ & + & - \\
& $\mathcal{O}_{W}$ & $\epsilon^{IJK}W^I_\nu{}^\mu W^J_\rho{}^\nu W^K_\mu{}^\rho $ & + & + \\
& $\mathcal{O}_{\tilde{W}}$ & $\epsilon^{IJK}\tilde{W}^I_\nu{}^\mu W^J_\rho{}^\nu W^K_\mu{}^\rho $ & + & - \\
\hline
\multirow{8}{*}{$X^2H^2$} & $\mathcal{O}_{HG}$ & $(H^\dagger H)G^A_{\mu\nu}G^A{}^{\mu\nu}$ & + & + \\
& $\mathcal{O}_{H\tilde{G}}$ & $(H^\dagger H)\tilde{G}^A_{\mu\nu}G^A{}^{\mu\nu}$ & + & - \\
& $\mathcal{O}_{HW}$ & $(H^\dagger H)W^I_{\mu\nu}W^I{}^{\mu\nu}$ & + & + \\
& $\mathcal{O}_{H\tilde{W}}$ & $(H^\dagger H)\tilde{W}^I_{\mu\nu}W^I{}^{\mu\nu}$ & + & - \\
& $\mathcal{O}_{HB}$ & $(H^\dagger H)B_{\mu\nu}B^{\mu\nu}$ & + & + \\
& $\mathcal{O}_{H\tilde{B}}$ & $(H^\dagger H)\tilde{B}_{\mu\nu}B^{\mu\nu}$ & + & - \\
& $\mathcal{O}_{HWB}$ & $(H^\dagger \tau^I H)W^I_{\mu\nu}B^{\mu\nu}$ & + & - \\
& $\mathcal{O}_{H\tilde{W}B}$ & $(H^\dagger \tau^I H)\tilde{W}^I_{\mu\nu}B^{\mu\nu}$ & + & - \\
\hline
\multirow{2}{*}{$H^4D^2$} & $\mathcal{O}_{H\square}$ & $(H^\dagger H)\square(H^\dagger H)$ & + & + \\
& $\mathcal{O}_{HD}$ & $(H^\dagger D^\mu H)^*(H^\dagger D_\mu H)$ & + & + \\
\hline
\multirow{1}{*}{$H^6$} & $\mathcal{O}_{H}$ & $(H^\dagger H)^3$ & + & + \\
\hline
    \end{tabular}
    \caption{The CP properties of the dimension-6 bosonic operators of the SMEFT.}
    \label{tab:cp_bosonic_d6}
\end{table}

In this subsection, we classify the effective operators of dimensions 6 and 8.
Part of the bosonic CP-conserving and CP-violating operators at the dimension 8 have been discussed previously, e.g.~\cite{Remmen:2019cyz,Durieux:2024zrg}, and in this subsection, we present the complete classification including both the bosonic and fermionic operators.

\subsubsection*{Dimension 6}

Considering the bosonic operators, there are four classes,
\begin{equation}
    X^3\,, \quad X^2H^2\,,\quad H^4D^2\,,\quad H^6\,,
\end{equation}
where all the operators are real, and CP-odd operators emerge in the classes with the gauges bosons due to the antisymmetric tensor $\epsilon_{\mu\nu\rho\lambda}$. We list the operators according to their CP properties following the convention in Ref.~\cite{Grzadkowski:2010es} in Tab.~\ref{tab:cp_bosonic_d6}.

We present the CP properties of the dimension-6 fermionic operators in Tab.~\ref{tab:CP_d6}, and present the dimension-6 fermionic operators in App.~\ref{app:dim6} for convenience. We present the combinations of specific CP properties and the numbers of the CP-even and CP-odd operators from them due to the flavor structures. These results are consistent with the Hilbert series result~\cite{Kondo:2022wcw} and the previous discussion~\cite{Alonso:2013hga}.

\begin{center}
    \centering
    \renewcommand{\arraystretch}{1.4}
    \begin{longtable}{|c|c|c|c|c|}
    \caption{The dimension 6 fermionic operators of specific CP properties. The second and the third columns are the CP-even and the CP-odd combinations of the operators, and the last two columns are the numbers considering the flavor components of the combinations. The explicit operators can be found in App.~\ref{app:dim6}. }
    \label{tab:CP_d6}\\
\hline
\multirow{2}{*}{class} & \multicolumn{2}{c|}{operators combination} & \multicolumn{2}{c|}{operators number} \\
\cline{2-5}
& CP-even & CP-odd & CP-even & CP-odd \\
\hline
\multirow{3}{*}{$\psi^2H^3 + \text{h.c.}$} & $\mathcal{O}_{eH} + \text{h.c.}$ & $\mathcal{O}_{eH} - \text{h.c.}$ & 9 & 9 \\
& $\mathcal{O}_{uH} + \text{h.c.}$ & $\mathcal{O}_{uH} - \text{h.c.}$ & 9 & 9 \\
& $\mathcal{O}_{dH} + \text{h.c.}$ & $\mathcal{O}_{dH} - \text{h.c.}$ & 9 & 9 \\
& & & 27 & 27 \\
\hline
\multirow{8}{*}{$\psi^2XH + \text{h.c.}$} & $\mathcal{O}_{eW} + \text{h.c.}$ & $\mathcal{O}_{eW} - \text{h.c.}$ & 9 & 9 \\
 & $\mathcal{O}_{eB} + \text{h.c.}$ & $\mathcal{O}_{eB} - \text{h.c.}$ & 9 & 9 \\
& $\mathcal{O}_{uG} + \text{h.c.}$ & $\mathcal{O}_{uG} - \text{h.c.}$ & 9 & 9 \\
& $\mathcal{O}_{uW} + \text{h.c.}$ & $\mathcal{O}_{uW} - \text{h.c.}$ & 9 & 9 \\
& $\mathcal{O}_{uB} + \text{h.c.}$ & $\mathcal{O}_{uB} - \text{h.c.}$ & 9 & 9 \\
& $\mathcal{O}_{dG} + \text{h.c.}$ & $\mathcal{O}_{dG} - \text{h.c.}$ & 9 & 9 \\
& $\mathcal{O}_{dW} + \text{h.c.}$ & $\mathcal{O}_{dW} - \text{h.c.}$ & 9 & 9 \\
& $\mathcal{O}_{dB} + \text{h.c.}$ & $\mathcal{O}_{dB} - \text{h.c.}$ & 9 & 9 \\
& & & 72 & 72 \\
\hline
\multirow{8}{*}{$\psi^2H^2D$} & $\mathcal{O}_{Hl}^{(1)}+\text{h.c.}$ & $\mathcal{O}_{Hl}^{(1)}-\text{h.c.}$ & 6 & 3 \\
& $\mathcal{O}_{Hl}^{(3)}+\text{h.c.}$ & $\mathcal{O}_{Hl}^{(1)}-\text{h.c.}$ & 6 & 3 \\
& $\mathcal{O}_{He}+\text{h.c.}$ & $\mathcal{O}_{He}-\text{h.c.}$ & 6 & 3 \\
& $\mathcal{O}_{Hq}^{(1)}+\text{h.c.}$ & $\mathcal{O}_{Hq}^{(1)}-\text{h.c.}$ & 6 & 3 \\
& $\mathcal{O}_{Hq}^{(3)}+\text{h.c.}$ & $\mathcal{O}_{Hq}^{(3)}-\text{h.c.}$ & 6 & 3 \\
& $\mathcal{O}_{Hu}+\text{h.c.}$ & $\mathcal{O}_{Hu}-\text{h.c.}$ & 6 & 3 \\
& $\mathcal{O}_{Hd}+\text{h.c.}$ & $\mathcal{O}_{Hd}-\text{h.c.}$ & 6 & 3 \\
& $\mathcal{O}_{Hud}+\text{h.c.}$ & $\mathcal{O}_{Hud}-\text{h.c.}$ & 9 & 9 \\
& & & 51 & 30 \\
\hline
\multirow{5}{*}{$(\overline{L}L)(\overline{L}L)$} & $\mathcal{O}_{ll} + \text{h.c.} $ & $\mathcal{O}_{ll} - \text{h.c.} $ & 27 & 18 \\
& $\mathcal{O}_{qq}^{(1)} + \text{h.c.} $ & $\mathcal{O}_{qq}^{(1)} - \text{h.c.} $ & 27 & 18 \\
& $\mathcal{O}_{qq}^{(3)} + \text{h.c.} $ & $\mathcal{O}_{qq}^{(3)} - \text{h.c.} $ & 27 & 18 \\
& $\mathcal{O}_{lq}^{(1)} + \text{h.c.} $ & $\mathcal{O}_{qq}^{(1)} - \text{h.c.} $ & 45 & 36 \\
& $\mathcal{O}_{lq}^{(3)} + \text{h.c.} $ & $\mathcal{O}_{qq}^{(3)} - \text{h.c.} $ & 45 & 36 \\
& & & 171 & 126 \\
\hline
\multirow{7}{*}{$(\overline{R}R)(\overline{R}R)$} & $\mathcal{O}_{ee} + \text{h.c.} $ & $\mathcal{O}_{ee} - \text{h.c.} $ & 21 & 15 \\
& $\mathcal{O}_{uu} + \text{h.c.} $ & $\mathcal{O}_{uu} - \text{h.c.} $ & 27 & 18 \\
& $\mathcal{O}_{dd} + \text{h.c.} $ & $\mathcal{O}_{dd} - \text{h.c.} $ & 27 & 18 \\
& $\mathcal{O}_{eu} + \text{h.c.} $ & $\mathcal{O}_{eu} - \text{h.c.} $ & 45 & 36 \\
& $\mathcal{O}_{ed} + \text{h.c.} $ & $\mathcal{O}_{ed} - \text{h.c.} $ & 45 & 36 \\
& $\mathcal{O}_{ud}^{(1)} + \text{h.c.} $ & $\mathcal{O}_{ud}^{(1)} - \text{h.c.} $ & 45 & 36 \\
& $\mathcal{O}_{ud}^{(3)} + \text{h.c.} $ & $\mathcal{O}_{ud}^{(3)} - \text{h.c.} $ & 45 & 36 \\
& & & 255 & 195 \\
\hline
\multirow{7}{*}{$(\overline{L}L)(\overline{R}R)$} & $\mathcal{O}_{le}+\text{h.c.}$ & $\mathcal{O}_{le}-\text{h.c.}$ & 45 & 36 \\
& $\mathcal{O}_{lu}+\text{h.c.}$ & $\mathcal{O}_{lu}-\text{h.c.}$ & 45 & 36 \\
& $\mathcal{O}_{ld}+\text{h.c.}$ & $\mathcal{O}_{ld}-\text{h.c.}$ & 45 & 36 \\
& $\mathcal{O}_{qe}+\text{h.c.}$ & $\mathcal{O}_{qe}-\text{h.c.}$ & 45 & 36 \\
& $\mathcal{O}_{qu}^{(1)}+\text{h.c.}$ & $\mathcal{O}_{qu}^{(1)}-\text{h.c.}$ & 45 & 36 \\
& $\mathcal{O}_{qu}^{(8)}+\text{h.c.}$ & $\mathcal{O}_{qu}^{(8)}-\text{h.c.}$ & 45 & 36 \\
& $\mathcal{O}_{qd}^{(1)}+\text{h.c.}$ & $\mathcal{O}_{qd}^{(1)}-\text{h.c.}$ & 45 & 36 \\
& $\mathcal{O}_{qd}^{(8)}+\text{h.c.}$ & $\mathcal{O}_{qd}^{(8)}-\text{h.c.}$ & 45 & 36 \\
& & & 360 & 288 \\
\hline
\multirow{1}{*}{$(\overline{L}R)(\overline{R}L) + \text{h.c.}$} & $\mathcal{O}_{ledq} + \text{h.c.}$ & $\mathcal{O}_{ledq} - \text{h.c.}$ & 81 & 81 \\
& & & 81 & 81 \\
\hline
\multirow{4}{*}{$(\overline{L}R)(\overline{R}L) + \text{h.c.}$} & $\mathcal{O}_{quqd}^{(1)} + \text{h.c.}$ & $\mathcal{O}_{quqd}^{(1)} - \text{h.c.}$ & 81 & 81 \\
& $\mathcal{O}_{quqd}^{(8)} + \text{h.c.}$ & $\mathcal{O}_{quqd}^{(8)} - \text{h.c.}$ & 81 & 81 \\
& $\mathcal{O}_{lequ}^{(1)} + \text{h.c.}$ & $\mathcal{O}_{lequ}^{(1)} - \text{h.c.}$ & 81 & 81 \\
& $\mathcal{O}_{lequ}^{(3)} + \text{h.c.}$ & $\mathcal{O}_{lequ}^{(3)} - \text{h.c.}$ & 81 & 81 \\
& & & 324 & 324 \\
\hline
    \end{longtable}
\end{center}

\subsubsection*{Dimension 8} 

For the bosonic operators, only the ones involving the gauge bosons could be CP-odd at dimension 8. These operators have been counted by the Hilbert series~\cite{Kondo:2022wcw} and constructed by the Young tensor method~\cite{Li:2020gnx}. Next, we list these operators according to their classes following the notations in Ref.~\cite{Li:2020gnx} in Tab.~\ref{tab:CP_bosonic_d8}. The operators involving the anomalous quartic gauge couplings of these results are consistent with Ref.~\cite{Durieux:2024zrg} 


\begin{center}
\renewcommand{\arraystretch}{1.8}
    \begin{longtable}{|c|cc|c|c|}
    \caption{The CP properties of the dimension-6 bosonic operators of the SMEFT.}
    \label{tab:CP_bosonic_d8}\\
\hline
class & \multicolumn{2}{|c|}{operator} & C & P \\
\hline
\multirow{43}{*}{$X^4$} & $\mathcal{O}_{G^4}^{(1)}$ & $(G^AG^B)(G^AG^B)$ & + & + \\
& $\mathcal{O}_{G^4}^{(2)}$ & $d^{ACE}d^{BDE}(G^AG^B)(G^CG^D)$ & + & + \\
& $\mathcal{O}_{G^4}^{(3)}$ & $f^{ACE}f^{BDE}(G^AG^B)(G^CG^D)$ & + & + \\
& $\mathcal{O}_{G^4}^{(4)}$ & $(G^AG^B)(G^A\tilde{G}^B)$ & + & - \\
& $\mathcal{O}_{G^4}^{(5)}$ & $d^{ACE}d^{BDE}(G^AG^B)(G^C\tilde{G}^D)$ & + & - \\
& $\mathcal{O}_{G^4}^{(6)}$ & $f^{ACE}f^{BDE}(G^AG^B)(G^C\tilde{G}^D)$ & + & - \\
& $\mathcal{O}_{G^4}^{(7)}$ & $(G^A\tilde{G}^B)(G^A\tilde{G}^B)$ & + & + \\
& $\mathcal{O}_{G^4}^{(8)}$ & $d^{ACE}d^{BDE}(G^A\tilde{G}^B)(G^C\tilde{G}^D)$ & + & + \\
& $\mathcal{O}_{G^4}^{(9)}$ & $f^{ACE}f^{BDE}(G^A\tilde{G}^B)(G^C\tilde{G}^D)$ & + & + \\
& $\mathcal{O}_{W^4}^{(1)}$ & $(W^IW^I)(W^JW^J)$ & + & + \\
& $\mathcal{O}_{W^4}^{(2)}$ & $(W^IW^I)(W^J\tilde{W}^J)$ & + & - \\
& $\mathcal{O}_{W^4}^{(3)}$ & $(W^I\tilde{W}^I)(W^J\tilde{W}^J)$ & + & + \\
& $\mathcal{O}_{W^4}^{(4)}$ & $(W^IW^J)(W^IW^J)$ & + & + \\
& $\mathcal{O}_{W^4}^{(5)}$ & $(W^IW^J)(W^I\tilde{W}^J)$ & + & - \\
& $\mathcal{O}_{W^4}^{(6)}$ & $(W^I\tilde{W}^J)(W^I\tilde{W}^J)$ & + & + \\
& $\mathcal{O}_{B^4}^{(1)}$ & $(B^2)(B^2)$ & + & + \\
& $\mathcal{O}_{B^4}^{(2)}$ & $(B^2)(B\tilde{B})$ & + & - \\
& $\mathcal{O}_{B^4}^{(3)}$ & $(B\tilde{B})(B\tilde{B})$ & + & + \\
& $\mathcal{O}_{G^2W^2}^{(1)}$ & $G^2W^2$ & + & + \\
& $\mathcal{O}_{G^2W^2}^{(2)}$ & $G^2(W^I\tilde{W}^I)$ & + & - \\
& $\mathcal{O}_{G^2W^2}^{(3)}$ & $(G^A\tilde{G}^A)(W^2)$ & + & - \\
& $\mathcal{O}_{G^2W^2}^{(4)}$ & $(G^A\tilde{G}^A)(W^I\tilde{W}^I)$ & + & + \\
& $\mathcal{O}_{G^2W^2}^{(5)}$ & $(G^AW^I)(G^A W^I)$ & + & + \\
& $\mathcal{O}_{G^2W^2}^{(6)}$ & $(G^AW^I)(G^A \tilde{W}^I)$ & + & - \\
& $\mathcal{O}_{G^2W^2}^{(7)}$ & $(G^A\tilde{W}^I)(G^A \tilde{W}^I)$ & + & + \\
& $\mathcal{O}_{G^2B^2}^{(1)}$ & $G^2B^2$ & + & + \\
& $\mathcal{O}_{G^2B^2}^{(2)}$ & $G^2(B\tilde{B})$ & + & - \\
& $\mathcal{O}_{G^2B^2}^{(3)}$ & $(G^A\tilde{G}^A)(B^2)$ & + & - \\
& $\mathcal{O}_{G^2B^2}^{(4)}$ & $(G^A\tilde{G}^A)(B\tilde{B})$ & + & + \\
& $\mathcal{O}_{G^2B^2}^{(5)}$ & $(G^AB)(G^A B)$ & + & + \\
& $\mathcal{O}_{G^2B^2}^{(6)}$ & $(G^AB)(G^A \tilde{B})$ & + & - \\
& $\mathcal{O}_{G^2B^2}^{(7)}$ & $(G^A\tilde{B})(G^A \tilde{B})$ & + & + \\
& $\mathcal{O}_{B^2W^2}^{(1)}$ & $B^2W^2$ & + & + \\
& $\mathcal{O}_{B^2W^2}^{(2)}$ & $B^2(W^I\tilde{W}^I)$ & + & - \\
& $\mathcal{O}_{B^2W^2}^{(3)}$ & $(B\tilde{B})(W^2)$ & + & - \\
& $\mathcal{O}_{B^2W^2}^{(4)}$ & $(B\tilde{B})(W^I\tilde{W}^I)$ & + & + \\
& $\mathcal{O}_{B^2W^2}^{(5)}$ & $(BW^I)(B W^I)$ & + & + \\
& $\mathcal{O}_{B^2W^2}^{(6)}$ & $(BW^I)(B \tilde{W}^I)$ & + & - \\
& $\mathcal{O}_{B^2W^2}^{(7)}$ & $(B\tilde{W}^I)(B \tilde{W}^I)$ & + & + \\
& $\mathcal{O}_{BG^3}^{(1)}$ & $d^{ABC}(BG^A)(G^BG^C)$ & + & + \\
& $\mathcal{O}_{BG^3}^{(2)}$ & $d^{ABC}(BG^A)(G^B\tilde{G}^C)$ & + & - \\
& $\mathcal{O}_{BG^3}^{(3)}$ & $d^{ABC}(B\tilde{G}^A)(G^BG^C)$ & + & - \\
& $\mathcal{O}_{BG^3}^{(4)}$ & $d^{ABC}(B\tilde{G}^A)(G^B\tilde{G}^C)$ & + & + \\
\hline
\multirow{6}{*}{$X^3\phi^2$} & $\mathcal{O}_{G^3H^2}^{(1)}$ & $f^{ABC}G^A_{\mu\nu}G^B_\lambda{}^\mu G^C{}^{\nu\lambda} H^\dagger H$ & + & + \\
& $\mathcal{O}_{G^3H^2}^{(2)}$ & $f^{ABC}G^A_{\mu\nu}G^B_\lambda{}^\mu \tilde{G}^C{}^{\nu\lambda} H^\dagger H$ & + & - \\
& $\mathcal{O}_{W^3H^2}^{(1)}$ & $\epsilon^{IJK}W^I_{\mu\nu}W^J_\lambda{}^\mu W^K{}^{\nu\lambda} H^\dagger H$ & + & + \\
& $\mathcal{O}_{W^3H^2}^{(2)}$ & $\epsilon^{IJK}W^I_{\mu\nu}W^J_\lambda{}^\mu \tilde{W}^K{}^{\nu\lambda} H^\dagger H$ & + & - \\
& $\mathcal{O}_{BW^2H^2}^{(1)}$ & $\epsilon^{IJK}B_{\mu\nu}W^I_\lambda{}^\mu W^J{}^{\nu\lambda} H^\dagger \tau^K H$ & + & + \\
& $\mathcal{O}_{BW^2H^2}^{(2)}$ & $\epsilon^{IJK}B_{\mu\nu}W^I_\lambda{}^\mu \tilde{W}^J{}^{\nu\lambda} H^\dagger \tau^K H$ & + & - \\
\hline
\multirow{10}{*}{$X^2\phi^4$} & $\mathcal{O}_{G^2H^4}^{(1)}$ & $G^2(H^\dagger H)^2$ & + & + \\
& $\mathcal{O}_{G^2H^4}^{(2)}$ & $G^A_{\mu\nu}\tilde{G}^A{}^{\mu\nu}(H^\dagger H)^2$ & + & - \\
& $\mathcal{O}_{W^2H^4}^{(1)}$ & $W^I_{\mu\nu}W^J{}^{\mu\nu}(H^\dagger \tau^I H)(H^\dagger \tau^J H)$ & + & + \\
& $\mathcal{O}_{W^2H^4}^{(2)}$ & $W^2(H^\dagger H)^2$ & + & + \\
& $\mathcal{O}_{W^2H^4}^{(3)}$ & $W^I_{\mu\nu}\tilde{W}^J{}^{\mu\nu}(H^\dagger \tau^I H)(H^\dagger \tau^J H)$ & + & - \\
& $\mathcal{O}_{W^2H^4}^{(4)}$ & $(W^I\tilde{W}^I)(H^\dagger H)^2$ & + & - \\
& $\mathcal{O}_{B^2H^4}^{(1)}$ & $B^2(H^\dagger H)^2$ & + & + \\
& $\mathcal{O}_{B^2H^4}^{(2)}$ & $B\tilde{B}(H^\dagger H)^2$ & + & - \\
& $\mathcal{O}_{BWH^4}^{(1)}$ & $B_{\mu\nu}W^I{}^{\mu\nu}(H^\dagger \tau^I H)(H^\dagger H)$ & + & + \\
& $\mathcal{O}_{BWH^4}^{(2)}$ & $B_{\mu\nu}\tilde{W}^I{}^{\mu\nu}(H^\dagger \tau^I H)(H^\dagger H)$ & + & - \\
\hline
\multirow{18}{*}{$X^2\phi^2D^2$} & $\mathcal{O}_{G^2H^2D^2}^{(1)}$ & $G^2(D^\mu H^\dagger D_\mu H)$ & + & + \\
& $\mathcal{O}_{G^2H^2D^2}^{(2)}$ & $(G^A\tilde{G}^A)(D^\mu H^\dagger D_\mu H)$ & + & - \\
& $\mathcal{O}_{G^2H^2D^2}^{(3)}$ & $G^A_\lambda{}^\mu G^A{}^{\nu\lambda} (D_\mu H^\dagger D_\nu H)$ & + & + \\
& $\mathcal{O}_{W^2H^2D^2}^{(1)}$ & $W^2(D^\mu H^\dagger D_\mu H)$ & + & + \\
& $\mathcal{O}_{W^2H^2D^2}^{(2)}$ & $i\epsilon^{IJK}W^I_\lambda{}^\mu W^J{}^{\nu\lambda} (D_\mu H^\dagger \tau^K D_\nu H)$ & - & + \\
& $\mathcal{O}_{W^2H^2D^2}^{(3)}$ & $(W^I\tilde{W}^I)(D^\mu H^\dagger D_\mu H)$ & + & - \\
& $\mathcal{O}_{W^2H^2D^2}^{(4)}$ & $i\epsilon^{IJK}W^I_\lambda{}^{[\mu}\tilde{W}^J{}^{\nu]\lambda}(D_\mu H^\dagger \tau^K D_\nu H)$ & - & - \\
& $\mathcal{O}_{W^2H^2D^2}^{(5)}$ & $W^I_\lambda{}^\mu W^I{}^{\nu\lambda}(D_\mu H^\dagger D_\nu H)$ & + & + \\
& $\mathcal{O}_{W^2H^2D^2}^{(6)}$ & $i\epsilon^{IJK}W^I_\lambda{}^{(\mu}\tilde{W}^J{}^{\nu)\lambda}(D_\mu H^\dagger \tau^K D_\nu H)$ & - & - \\
& $\mathcal{O}_{B^2H^2D^2}^{(1)}$ & $B^2 (D^\mu H^\dagger D_\mu H)$ & + & + \\
& $\mathcal{O}_{B^2H^2D^2}^{(2)}$ & $B\tilde{B} (D^\mu H^\dagger D_\mu H)$ & + & - \\
& $\mathcal{O}_{B^2H^2D^2}^{(3)}$ & $B^\mu{}_\lambda B^{\nu\lambda} (D_\mu H^\dagger D_\nu H)$ & + & + \\
& $\mathcal{O}_{BWH^2D^2}^{(1)}$ & $(BW^I)(D^\mu H^\dagger \tau^I D_\mu H)$ & + & + \\
& $\mathcal{O}_{BWH^2D^2}^{(2)}$ & $(B\tilde{W}^I)(D^\mu H^\dagger \tau^I D_\mu H)$ & + & - \\
& $\mathcal{O}_{BWH^2D^2}^{(3)}$ & $iB^{[\mu}{}_\lambda W^I{}^{\nu]\lambda}(D_\nu H^\dagger \tau^I D_\mu H)$ & + & + \\
& $\mathcal{O}_{BWH^2D^2}^{(4)}$ & $iB^{[\mu}{}_\lambda \tilde{W}^I{}^{\nu]\lambda}(D_\nu H^\dagger \tau^I D_\mu H)$ & + & - \\
& $\mathcal{O}_{BWH^2D^2}^{(5)}$ & $iB^{(\mu}{}_\lambda W^I{}^{\nu)\lambda}(D_\nu H^\dagger \tau^I D_\mu H)$ & + & + \\
& $\mathcal{O}_{BWH^2D^2}^{(6)}$ & $iB^{(\mu}{}_\lambda \tilde{W}^I{}^{\nu)\lambda}(D_\nu H^\dagger \tau^I D_\mu H)$ & + & - \\
\hline
\multirow{6}{*}{$X\phi^4D^2$} & $\mathcal{O}_{WH^4D^2}^{(1)}$ & $W^I{}^{\mu\nu}(H^\dagger H)(D_\mu H^\dagger \tau^I D_\nu H)$ & + & + \\
& $\mathcal{O}_{WH^4D^2}^{(2)}$ & $\tilde{W}^I{}^{\mu\nu}(H^\dagger H)(D_\mu H^\dagger \tau^I D_\nu H)$ & + & - \\
& $\mathcal{O}_{WH^4D^2}^{(3)}$ & $W^I{}^{\mu\nu}(H^\dagger \tau^I H)(D_\mu H^\dagger D_\nu H)$ & + & + \\
& $\mathcal{O}_{WH^4D^2}^{(4)}$ & $\tilde{W}^I{}^{\mu\nu}(H^\dagger \tau^I H)(D_\mu H^\dagger D_\nu H)$ & + & - \\
& $\mathcal{O}_{BH^4D^2}^{(1)}$ & $B^{\mu\nu}(H^\dagger H)(D_\mu H^\dagger \tau^I D_\nu H)$ & + & + \\
& $\mathcal{O}_{BH^4D^2}^{(2)}$ & $\tilde{B}^{\mu\nu}(H^\dagger H)(D_\mu H^\dagger \tau^I D_\nu H)$ & + & - \\
\hline
\multirow{1}{*}{$\phi^8$} & $\mathcal{O}_{H^8}$ & $(H^\dagger H)^3$ & + & + \\
\hline
\multirow{2}{*}{$\phi^6D^2$} & $\mathcal{O}_{H^6D^2}^{(1)}$ & $(H^\dagger H)\square(H^\dagger H)$ & + & + \\
& $\mathcal{O}_{H^6D^2}^{(2)}$ & $(H^\dagger H)|H^\dagger D_\mu H|^2$ & + & + \\
\hline
\multirow{3}{*}{$\phi^4D^4$} & $\mathcal{O}_{H^4D^4}^{(1)}$ & $(H^\dagger H)\square^2(H^\dagger H)$ & + & + \\
& $\mathcal{O}_{H^4D^4}^{(2)}$ & $H^\dagger D_\mu D_\nu H|^2$ & + & + \\
& $\mathcal{O}_{H^4D^4}^{(3)}$ & $(H^\dagger D_\mu H)^* \square(H^\dagger D^\mu H)$ & + & + \\
\hline
    \end{longtable}
\end{center}

\begin{table}[htb]
\renewcommand{\arraystretch}{1.8}
\begin{center}
\begin{minipage}[c]{0.4\textwidth}
    \begin{tabular}{|c|c|c|}
\hline
\multicolumn{3}{|c|}{dimension 6} \\
\hline
\multirow{2}{*}{Class} & \multicolumn{2}{c|}{operator number} \\ 
\cline{2-3}
& CP-even & CP-odd \\
\hline
$X^3$ & 2 & 2 \\
$H^6$ & 1 & 0 \\
$H^4D^2$ & 2 & 0 \\
$X^2H^2$ & 4 & 4 \\
\hline
\multirow{2}{*}{Total} & 9 & 6 \\
\cline{2-3}
& \multicolumn{2}{|c|}{15} \\
\hline
    \end{tabular}
    \end{minipage}\quad
    \begin{minipage}[c]{0.4\textwidth}
    \begin{tabular}{|c|c|c|}
\hline
\multicolumn{3}{|c|}{dimension 8} \\
\hline
\multirow{2}{*}{Class} & \multicolumn{2}{c|}{operator number} \\ 
\cline{2-3}
& CP-even & CP-odd \\
\hline
$X^4$ & 26 & 17 \\
$X^3H^2$ & 3 & 3 \\
$X^2H^4$ & 5 & 5 \\
$X^2H^2D^2$ &11 & 7 \\
$XH^4D^2$ & 3 & 3 \\
$H^6D^2$ & 2 & 0 \\
$H^4D^4$ & 3 & 0 \\
$H^8$ & 1 & 0 \\
\hline
\multirow{2}{*}{Total} & 54 & 35 \\
\cline{2-3}
& \multicolumn{2}{|c|}{89} \\
\hline
    \end{tabular}
    \end{minipage}
\end{center}
\caption{The CP properties of the bosonic operators at dimension 6 and dimension 8. The CP-odd operators and the CP-violating operators are equivalent and do not change with the flavor symmetry. }
\label{tab:odd_bosonic}
\end{table}

In summary, we present numbers of the CP-even and the CP-odd bosonic operators at dimension 6 and dimension 8 in Tab.~\ref{tab:odd_bosonic}. These results do not change when we discuss the flavor symmetries subsequently.

The dimension-8 fermionic operators of specific CP properties can also be obtained according to the discussion about the dimension 6 cases. 
We present the explicit operators in App.~\ref{app:dim8} following the notations in Ref.~\cite{Li:2020gnx}, and present their combinations of specific CP properties in Tab.~\ref{tab:CP_d8}.

\begin{center}
    \centering
    \renewcommand{\arraystretch}{1.4}
\begin{longtable}{|c|c|c|c|c|}
    \caption{The dimension 8 fermionic operators of specific CP properties. The second and the third columns are the CP-even and the CP-odd combinations of the operators, and the last two columns are the numbers considering the flavor components of the combinations. The explicit operators can be found in App.~\ref{app:dim8}. }
    \label{tab:CP_d8}\\
\hline
\multirow{2}{*}{class} & \multicolumn{2}{c|}{operators combination} & \multicolumn{2}{c|}{operators number} \\
\cline{2-5}
& CP-even & CP-odd & CP-even & CP-odd \\
\hline
\multirow{3}{*}{$\psi^2\phi^5$} & $\mathcal{O}_{quH^5} + \text{h.c.}$ & $\mathcal{O}_{quH^5} - \text{h.c.}$ & 9 & 9 \\
& $\mathcal{O}_{qdH^5} + \text{h.c.}$ & $\mathcal{O}_{qdH^5} - \text{h.c.}$ & 9 & 9 \\
& $\mathcal{O}_{leH^5} + \text{h.c.}$ & $\mathcal{O}_{leH^5} - \text{h.c.}$ & 9 & 9 \\
& & & 27 & 27 \\
\hline
\multirow{8}{*}{$\psi^2\phi^4D$} & $\mathcal{O}_{q^2H^4D}^{(1,2,4)} + \text{h.c.}$ & $\mathcal{O}_{q^2H^4D}^{(1,2,4)} - \text{h.c.}$ & 18 & 9 \\
\hline
    \end{longtable}
\end{center}

\subsection{Rephasing Symmetry and CP Violation}

The CP-odd operators in Tab.~\ref{tab:CP_d6} and Tab.~\ref{tab:CP_d8} are not necessarily CP-violating because of the rephasing symmetry. In this subsection, we introduce the rephasing symmetry and discuss its relation to the CP properties.

The Yukawa matrices are associated with the masses of the fermions after the spontaneous break of the gauge group $SU(2)_{EW}$, which the VEV of the Higgs field realizes, 
\begin{equation}
    H(x) = \frac{1}{\sqrt{2}}\left(\begin{array}{c}
         0\\v+h(x)
    \end{array}\right)\,,
\end{equation}
where $v$ is the VEV, and $h(x)$ is the fluctuation over the VEV.
The VEV of the Higgs field makes the gauge group $SU(3)_{c}\times SU(2)_{EW}\times U(1)_Y$ break to the subgroup $SU(3)_c\times U(1)_{EM}$, where $U(1)_{EM}$ represents the electric charge conservation. At the same time, the Yukawa terms in Eq.~\eqref{eq:L_higgs} become
\begin{equation}
    \mathcal{L}_{\text{Yukawa}}=(\bar{u}_L M_u u_R) + (\bar{d}_L M_d d_R) + (\bar{e}_L M_e e_R) + \text{h.c.} + \dots\,,  
\end{equation}
where the mass matrices 
\begin{equation}
    M_\psi = \frac{1}{\sqrt{2}}v Y_\psi\,,\quad \psi=u\,,d\,,e\,,
\end{equation}
are $3\times 3$ matrices in the flavor space, and the ellipse contains the interactions of the fermions and the fluctuation field $h(x)$. 
In particular, the neutrinos are massless when the right-handed neutrinos do not exist, $M_\nu = 0$. As mass matrices, the eigenvalues of $M_u\,,M_d$ and $M_e$ are real and positive, which means that there exists a set of unitary matrices satisfying
\begin{align}
    {U^u_L}^\dagger M_u U^u_R = \text{diag}(m_u\,,m_c\,,m_t) = D_u\,, \label{eq:diagonal_u}\\
    {U^d_L}^\dagger M_d U^d_R = \text{diag}(m_d\,,m_s\,,m_b) = D_d\,, \label{eq:diagonal_d}\\
    {U^e_L}^\dagger M_e U^e_R = \text{diag}(m_e\,,m_\nu\,,m_\tau) = D_e \label{eq:diagonal_e}\,,
\end{align}
and the corresponding transformations of the fermions
\begin{align}
    & u^0_L = {U^u_L}^\dagger u_L\,,\quad d^0_L = {U^d_L}^\dagger d_L\,,\quad u^0_R = {U^u_R}^\dagger u_R\,,\notag \\
    & d^0_R = {U^d_R}^\dagger d_R\,,\quad e^0_L = {U^e_L}^\dagger e_L\,,\quad e^0_R = {U^e_R}^\dagger e_R\,,
\end{align}
are the mass eigenstates, which form the mass basis. In terms of the mass basis, the Yukawa sector looks like
\begin{equation}
    \mathcal{L}_{\text{Yukawa}}=(\bar{u}^0_L D_u u^0_R) + (\bar{d}^0_L D_d d^0_R) + (\bar{e}^0_L D_e e^0_R) + \text{h.c.} + \dots\,.  
\end{equation}

At the same time, the charged interactions of the fermions after the spontaneous symmetry breaking are
\begin{equation}
    \mathcal{L}_{\text{CC}} = \frac{g}{\sqrt{2}}\left(\overline{u}_L\gamma^\mu d_L + \overline{\nu}_L\gamma^\mu e_L\right)W^+_\mu + \text{h.c.}\,,
\end{equation}
which can be expressed by the mass basis that
\begin{equation}
\label{eq:charged_current}
    \mathcal{L}_{\text{CC}} = \frac{g}{\sqrt{2}}\left(\overline{u}^0_L V_{CKM}\gamma^\mu d_L^0 + \overline{\nu}_L^0\gamma^\mu e_L^0\right)W^+_\mu + \text{h.c.}\,,
\end{equation}
where $V_{CKM}=U_L^u{}^\dagger U_L^d $ is a unitary matrix in the flavor space called the CKM-matrix~\cite{Cabibbo:1963yz,Kobayashi:1973fv}. It means that the 3 flavors of the quarks are mixed in the mass basis. While for the leptons, the 3 flavors do not mix.

As a $3\times 3$ unitary matrix, there are generally 9 real parameters. However, the field redefinitions can remove some unphysical ones. In principle, the redefinitions by the phases of the 6 quarks such as
\begin{equation}
\label{eq:redefinition}
    q_p \rightarrow e^{i\alpha_p} q_p\,,\quad p=1\,,2\,,3\,,\text{ and }q=u\,,d\,,
\end{equation}
can eliminate 5 phases, thus there are 4 physical parameters in the 
CKM-matrix and its standard parameterization~\cite{Chau:1984fp} is
\begin{equation}
    V_{CKM} = \left(\begin{array}{ccc}
c_{12}c_{13} & s_{12}c_{13} & s_{13}e^{-i\delta} \\
-s_{12}c_{23}-c_{12}s_{23}s_{13}e^{i\delta} & c_{12}c_{23}-s_{12}s_{23}s_{13}e^{i\delta} & s_{23}c_{13} \\
s_{12}s_{23}-c_{12}c_{23}s_{13}e^{i\delta} & -c_{12}s_{23}-s_{12}c_{23}s_{13}e^{i\delta} & c_{23}c_{13}
    \end{array}\right)\,,
\end{equation}
where $c_{ij}=\cos\theta_{ij}\,,s_{ij}=\sin\theta_{ij}$. The 4 parameters are the 3 angles $\theta_{12}\,,\theta_{13}\,,\theta_{23}$ and one phase $\delta$. The phase $\delta$ is closely related to the CP-violations of the SMEFT and will be discussed next

Return to the field redefinitions in Eq.~\eqref{eq:redefinition}, the remaining single redefinition is the universal rephasing of the 6 quarks,
\begin{equation}
    q_p \rightarrow e^{i\alpha} q_p\,,\quad p=1\,,2\,,3\,,\text{ and }q=u\,,d\,,
\end{equation}
which is a global symmetry of the SM Lagrangian, called the baryon number symmetry $U(1)_B$. Because the leptons do not mix at the SM Lagrangian, there are 3 more global symmetries for each flavor of the leptons respectively, noted as $U(1)_e$, $U(1)_\mu$, and $U(1)_\tau$. We factorize the flavor-universal one as the lepton number symmetry $U(1)_L$, and formulate the 4 $U(1)$ symmetries as
\begin{equation}
    U(1)^4=\left\{\begin{array}{l}
U(1)_{e-\mu}\times U(1)_{\mu-\tau} \\
U(1)_B \times U(1)_L
\end{array}\right.  \,,
\label{eq:rephasing}
\end{equation}
where $U(1)_B$ and $U(1)_L$ are flavor universal. $U(1)^4$ is called the rephasing symmetry, which is an accidental symmetry of the SM Lagrangian and could be violated by the high-dimension operators of the SMEFT. 


As mentioned before, not all the CP-odd operators are CP-violating, and only the ones that are invariant under the rephasing symmetry in Eq.~\eqref{eq:rephasing} are CP-violating since no field redefinition can be used to eliminate the CP-violating phase in the associated Wilson coefficients.

Since the bosons are trivial under the rephasing symmetry, all the CP-odd bosonic operators are always CP-violating. Considering the LO Lagrangian of the SMEFT, the CP-odd operators emerge only in the charged interactions of the quarks in Eq.~\eqref{eq:charged_current} due to the CKM-matrix. Because of the rephasing symmetry, these CP-odd operators are also CP-violating, which share the same CP-violating phase, $\delta$. Beyond the LO, the high-dimension operators could break the rephasing symmetry further, and only the ones that are CP-odd and rephasing symmetry conserving are CP-violating. 
For example, the dimension-6 operator
\begin{equation}
    \mathcal{O}_{eH}{}^p_r = (H^\dag H)(\bar l^p e_r H)\,,
\end{equation}
presents 3 CP-odd operators 
\begin{equation}
    \mathcal{O}_{eH}{}^p_r - \text{h.c.}\,,\quad p<r\,.
\end{equation}
However, these 3 operators are not CP-violating, since they are not invariant under the rephasing symmetry in Eq.~\eqref{eq:rephasing}. On the contrary, the 3 CP-odd operators in 
\begin{equation}
    \mathcal{O}_{uH}{}^p_r = (H^\dag H)(\bar q^p u_r H)\,,
\end{equation}
are all CP-violating.

Utilizing the Hilbert series, the numbers of the CP-odd and CP-violating operators are obtained up to dimension-14~\cite{Kondo:2022wcw}. In particular, there are 1422 and 22016 CP-odd operators at dimension 6 and dimension 8 respectively, while there are only 705 and 11777 CP-violating operators at dimension 6 and dimension 8 respectively. The detailed comparison of the dimension-8 operators can be found in Ref.~\cite{Kondo:2022wcw}.

\section{Flavor Symmetry of SMEFT}
\label{sec:flavor_symmetry}

In the previous section, we have discussed the CP property of the SMEFT without any flavor symmetry. The rephasing symmetry is important for the enumeration of the CP-violating operators, and there are a lot of CP-violating operators beyond the LO Lagrangian without any flavor symmetry. In this section, we will discuss the situation with some flavor symmetry, especially the non-abelian symmetry $U(3)^5$ and $U(2)^2$. The flavor invariants and the associated Young tableaux constructions are also discussed.

\subsection{$U(3)^5$ Symmetry}

Considering the LO Lagrangian of the SMEFT in Eq.~\eqref{eq:smeft_lo}, regardless of the Yukawa sector $\mathcal{L}_{\text{Yukawa}}$, the remaining three sectors\footnote{Actually only the $\mathcal{L}_{\text{fermion}}$ possesses non-trivial flavor structures.} possess a global symmetry 
\begin{equation}
    U(3)^5=U(3)_q \times U(3)_u \times U(3)_d \times U(3)_l \times U(3)_e\,,
\end{equation}
in the flavor space. It can be further decomposed as
\begin{equation}
    U(3)^5 = SU(3)^5\times U(1)^5\,,
\end{equation}
where
\begin{align}
    SU(3)^5 &= SU(3)_q \times SU(3)_u \times SU(3)_d \times SU(3)_l \times SU(3)_e\,, \\
    U(1)^5 &= U(1)_q \times U(1)_u \times U(1)_d \times U(1)_l \times U(1)_e\,.
\end{align}
The five fermions' representations under the $SU(3)^5$ are 
\begin{align}
    q & \in (\mathbf{3}\,,\mathbf{1}\,,\mathbf{1}\,,\mathbf{1}\,,\mathbf{1})\,, \notag \\
    u_R & \in (\mathbf{1}\,,\mathbf{3}\,,\mathbf{1}\,,\mathbf{1}\,,\mathbf{1})\,, \notag \\
    d_R & \in (\mathbf{1}\,,\mathbf{1}\,,\mathbf{3}\,,\mathbf{1}\,,\mathbf{1})\,, \notag \\
    l & \in (\mathbf{1}\,,\mathbf{1}\,,\mathbf{1}\,,\mathbf{3}\,,\mathbf{1})\,, \notag \\
    e_R & \in (\mathbf{1}\,,\mathbf{1}\,,\mathbf{1}\,,\mathbf{1}\,,\mathbf{3}) \,.
\end{align}
The $U(3)^5$ symmetry is the largest flavor symmetry of the SM fermions compatible with the gauge symmetry of the SM Lagrangian~\cite{Chivukula:1987py,Gerard1983FermionMS}. Sometimes, it is convenient to assume a subgroup of the $U(3)^5$ as the flavor symmetry, for example, the $U(2)^5$ or some finite groups~\cite{PhysRevD.19.3369,PhysRevD.21.3417,Frampton:1994rk}. We will talk about the $U(2)^5$ symmetry later.

The flavor symmetry $U(3)^5$ implies that the SM Lagrangian except the Yukawa interactions $\mathcal{L}_{\text{Yukawa}}$ is invariant under the unitary transformation
\begin{align}
    & \psi \rightarrow V_q \psi\,,\quad \psi=u,d\,,\notag \\
    & u_R \rightarrow V_u u_R\,,\quad d_R \rightarrow V_d d_R\,,\notag \\
    & \psi \rightarrow V_l \psi\,,\quad \psi=e,\nu\,,\notag \\
    & e_R \rightarrow V_e e_R \,,
\end{align}
where $(V_q\,,V_u\,,V_d\,,V_l\,,V_e)\in U(3)^5$. Such transformations are called the flavor rotations, and all the fermion triplets obtained by such transformations are the so-called weak basis. In particular, the mass basis and the weak basis are not equivalent.

Focusing on the weak basis, the $U(3)^5$ transformation induces the redefinitions of the Yukawa interactions,
\begin{align}
    \bar{u}_L M_u u_R & \rightarrow \bar{u'}_L (V_q^\dagger M_u V_u) u'_R \,,\\
    \bar{d}_L M_d d_R & \rightarrow \bar{d'}_L (V_q^\dagger M_d V_d) d'_R \,,\\
    \bar{e}_L M_e e_R & \rightarrow \bar{e'}_L (V_l^\dagger M_e V_e) e'_R \,.
\end{align}
Although the flavor rotation makes the weak basis arbitrary, there are some convenient choices. For example, if we choose $V_q = U_L^u\,,V_u=U_R^u$\,, the mass matrix $M_u$ becomes diagonal, as shown in Eq.~\eqref{eq:diagonal_u}, then we choose $V_d=U_R^d$, the down-quark mass matrix becomes
\begin{equation}
    {U_L^u}^\dagger M_d U_R^d = {U_L^u}^\dagger U_L^d {U_L^d}^\dagger M_d U_R^d = V_{CKM} D_d \,,
\end{equation}
according to Eq.~\eqref{eq:diagonal_d}, where $V_{CKM}=U_L^u{}^\dagger U_L^d $ is the CKM-matrix~\cite{Cabibbo:1963yz,Kobayashi:1973fv}. Its existence means that when $M_u$ is diagonal by a certain flavor rotation, $M_d$ is generally not, and leads to the quark flavor mixing, which emerges in the charged interactions in Eq.~\eqref{eq:charged_current} under the mass basis. As for the leptons, since the right-handed neutrinos do not exist, the transformation $V_l$ is arbitrary. We choose $V_l=U^e_L\,, V_e=U^e_R$, then the lepton mass matrix can always be diagonal
\begin{equation}
    {U_L^e}^\dagger M_e U^e_R = D_e \,,
\end{equation}
thus the lepton flavors do not mix in the LO Lagrangian under the weak basis as well.
In summary, $\mathcal{L}_{\text {Yukawa}}$ becomes
\begin{equation}
    \mathcal{L}_{\text{Yukawa}} = \bar{u}^0_L D_u u^0_R + \bar{d'}_L V_{CKM} D_d d^0_R  + \bar{e}^0_L D_e e^0_R + \text{h.c.}
\end{equation}
under this specific weak basis. The up-quark mass matrix is diagonal while the down-quark mass matrix is not, thus this weak basis is called the up-basis. In particular, only the left-handed down-quark $d'_L$ is not of its mass eigenstates in the up-basis. On the other hand, if we require the down-quark mass matrix diagonal, we obtain the down-basis, under which the $\mathcal{L}_{\text{Yukawa}}$ becomes
\begin{equation}
    \mathcal{L}_{\text{Yukawa}} = \bar{u'}_L V^\dagger_{CKM} D_u u^0_R + \bar{d}^0_L D_d d^0_R  + \bar{e}^0_L D_e e^0_R + \text{h.c.}\,,
\end{equation}
where only the left-handed up-quark $u'_L$ is not the mass eigenstates.
In this paper, when some specific weak basis is needed, we use the down-basis consistently. For convenience, we present the Yukawa interactions of the down-basis before the electro-weak symmetry breaking,
\begin{equation}
    \mathcal{L}_{\text{Yukawa}} = \bar{q}V_{CKM}^\dagger Y_u\tilde{H}u_R + \bar{q} Y_d H d_R + \bar{l} Y_e H e_R + \text{h.c.}\,,
\end{equation}
where $Y_\psi = \frac{\sqrt{2}}{v} D_\psi$ and we have dropped the superscripts of the fermions. 

The Yukawa interactions make the flavor symmetry $U(3)^5$ break explicitly. Because the flavor rotations induce the field redefinitions, the specific choice of a weak basis means the exploitation of the field redefinition. For example, in the down-basis, the $U(3)_l \times U(3)_e$ is exploited to diagonalize the lepton Yukawa matrix $Y_e$ with the 3 $U(1)$ symmetries remaining. The $U(3)_q\times U(3)_u \times U(3)_e$ is exploited to diagonalize the down-quark Yuakawa matrix $Y_d$ and generate the CKM-matrix of the standard parameterization, with the universal baryon number conservation $U(1)_B$ remaining. Thus the true symmetry is the rephasing symmetry,
\begin{equation}
    U(3)^5 = \left\{\begin{array}{l}
SU(3)^5 \\
U(1)^5
    \end{array}\right. 
\rightarrow \left.\begin{array}{r}
U(1)_{e-\mu}\times U(1)_{\mu-\tau} \\
U(1)_B \times U(1)_L
\end{array}\right\} = U(1)^4 \,.
\end{equation}
With the $U(3)^5$ symmetry assumption, the rephasing symmetry $U(1)^4$ emerges as the remaining symmetry of the flavor symmetry after the choice of a specific weak basis. 


\subsection{$U(2)^5$ Symmetry}


Cases are different for a subgroup of the $U(3)^5$, for example, the $U(2)^5$ symmetry.
Noticing the mass gaps between the first two particles and the last particle in a flavor triplet, we define the first 2 particles to form a doublet, while the last one is a singlet, 
\begin{equation}
    \left(\begin{array}{c}
\psi_1 \\ \psi_2 \\ \psi_3
    \end{array}\right) \rightarrow \left(\begin{array}{c}
\psi_1 \\ \psi_2
    \end{array}\right) + \psi_3\,,\quad \psi=q,u_R,d_R,l,e_R\,.
\end{equation}
Under such an arrangement, the general $U(3)^5$ is released to the subgroup
\begin{equation}
    U(2)^5 \in U(3)^5 \,,\quad \text{with  }U(2)^5 = U(2)_q \times U(2)_u \times U(2)_d \times U(2)_l \times U(2)_e\,,
\end{equation}
and the fermions' representations are 
\begin{align}
    q &\in (\mathbf{2}\,,\mathbf{1}\,,\mathbf{1}\,,\mathbf{1}\,,\mathbf{1})\,,\\
    u_R &\in (\mathbf{1}\,,\mathbf{2}\,,\mathbf{1}\,,\mathbf{1}\,,\mathbf{1})\,,\\
    d_R &\in (\mathbf{1}\,,\mathbf{1}\,,\mathbf{2}\,,\mathbf{1}\,,\mathbf{1})\,,\\
    l &\in (\mathbf{1}\,,\mathbf{1}\,,\mathbf{1}\,,\mathbf{2}\,,\mathbf{1})\,,\\
    e_R &\in (\mathbf{1}\,,\mathbf{1}\,,\mathbf{1}\,,\mathbf{1}\,,\mathbf{2})\,, \\
    q_3\,,u_R{}_3\,,d_R{}_3\,,l_3\,,e_R{}_3 &\in (\mathbf{1}\,,\mathbf{1}\,,\mathbf{1}\,,\mathbf{1}\,,\mathbf{1})\,.
\end{align}

The reduction of the symmetry group implies more abundant phenomena.
According to the Yukawa interactions in Eq.~\eqref{eq:L_higgs}, we parameterize the Yukawa matrices as 
\begin{equation}
    Y_{u/d} = y_{t/b}\left(\begin{array}{cc}
\Delta_{u/d} & x_{t/b}V_q \\
0 & 1 
    \end{array}\right)\,,\quad Y_{e} = y_{\tau}\left(\begin{array}{cc}
\Delta_{e} & x_lV_l \\
0 & 1 
    \end{array}\right)\,,
\end{equation}
where $y_{\tau,t,b}$ and $x_{\tau,t,b}$ are free coefficients expected to be of order $O(1)$.
The $\Delta_{u/d/e}$ are $2\times 2$ matrices, while $V_q\,,V_l$ are dimension-2 vectors. Writing the two-component fermions as
\begin{equation}
    \mathcal{Q} = \left(\begin{array}{c}
q_1 \\ q_2
    \end{array}\right)\,,\quad 
    \mathcal{U} = \left(\begin{array}{c}
u_1 \\ u_2
    \end{array}\right)\,,\quad 
    \mathcal{D} = \left(\begin{array}{c}
d_1 \\ d_2
    \end{array}\right)\,,\quad 
    \mathcal{L} = \left(\begin{array}{c}
l_1 \\ l_2
    \end{array}\right)\,,\quad 
    \mathcal{E} = \left(\begin{array}{c}
e_1 \\ e_2
    \end{array}\right)\,,
\end{equation}
the Yukawa interactions become
\begin{align}
    \mathcal{L}_{\text{Yukawa sector}} &= y_t \overline{\mathcal{Q}}_L\Delta_u \mathcal{U}_R + y_t \overline{t}_L t_R + y_tx_t \overline{\mathcal{Q}}_LV_q t_R + \text{h.c.} \notag \\
    &+ y_b \overline{\mathcal{Q}}_L\Delta_d \mathcal{D}_R + y_b \overline{b}_L b_R + y_bx_b \overline{\mathcal{Q}}_LV_q b_R + \text{h.c.} \notag \\
    &+ y_\tau \overline{\mathcal{L}}_L\Delta_e \mathcal{E}_R + y_\tau \overline{\tau}_L \tau_R + y_\tau x_\tau \overline{\mathcal{L}}_LV_l \tau_R + \text{h.c.}\,.\label{eq:yukawa_u2}
\end{align}
Next, we discuss the exploitations of the flavor symmetry and the resultant rephasing symmetry. Starting with the lepton sector with the symmetry $U(2)_l\times U(2)_e$, the vector $V_l$ can be reduced to 
\begin{equation}
    V_l = e^{i\phi_l} \left(\begin{array}{l}
0\\ \epsilon_l
    \end{array}\right)\,,
\end{equation}
by the exploitation of the $U(2)_l$, with a remaining $U(1)$ symmetry. Then the fix of $U(2)_e$ reduce the $U(2)_l\times U(2)_e$ to a single $U(1)$, and the matrix $\Delta_e$ becomes a real matrix,
\begin{equation}
    \Delta_e = O^T_e \left(\begin{array}{cc}
\delta'_e & 0\\0 & \delta_e
   \end{array}\right)\,, \quad \text{where }
   O_e = \left(\begin{array}{cc}
\cos\theta_e & \sin \theta_e \\-\sin\theta_e & \cos\theta_e
   \end{array}\right)\,.
\end{equation}
Nevertheless the remaining $U(1)$ is eliminated by the mix of the lepton doublet $\mathcal{L}$ and the $\tau$ since it is required to be invariant under that. Thus there is no remaining symmetry for the leptons. A similar argument applies to the quark sector, the flavor symmetry is eliminated completely, and the parameterizations of the Yukawa matrices are
\begin{equation}
    V_q = e^{i\phi_q}\left(\begin{array}{c}
0 \\ \epsilon_q 
    \end{array}\right)\,,\quad \Delta_u = U_u^\dagger \left(\begin{array}{cc}
\delta'_u & 0\\0 & \delta_u
   \end{array}\right)\,,\quad \Delta_d = U_d^\dagger \left(\begin{array}{cc}
\delta'_d & 0\\0 & \delta_d
   \end{array}\right)\,,
\end{equation}
where
\begin{equation}
    U_q = \left(\begin{array}{cc}
\cos\theta_q & \sin\theta_q e^{i\alpha_q} \\
-\sin\theta_q e^{-i\alpha_q} & \cos\theta_q
    \end{array}\right)\,,\quad q=u\,,d \,.
\end{equation}
In conclusion, there is no rephasing symmetry of the $U(2)^5$ flavor symmetry.

The $U(2)^5$ symmetry is more phenomenological abundant than the $U(3)^5$ symmetry since it presents less constraints~\cite{Barbieri:2011ci,Barbieri:2012uh,Blankenburg:2012nx,Greljo:2015mma,Barbieri:2015yvd,Buttazzo:2017ixm}. Even in the LO Lagrangian, the new phenomenology appears, for example, the mixing of the leptons in Eq.~\eqref{eq:yukawa_u2}, which is absent in the $U(3)^5$ symmetry.

\subsection{CP Property with Flavor Symmetry}

Although the flavor symmetries, including the $U(3)^5$, $U(2)^5$, and others, are broken in the SMEFT, they could be exact in the UV theory. To complement this symmetry-breaking pattern, the spurion method is convenient. In this method, some spurions are needed to keep the flavor symmetry formally, and the SMEFT operators are obtained by the VEVs of the spurions. 

Assuming a specific flavor symmetry, the spurions $\mathbf{T}$ should be some scalar fields covariant under the flavor group, and the terms in the Lagrangian should be the invariant composed by the spurions and the other fields under the flavor group,
\begin{equation}
    \mathcal{L} \supset c f(\mathbf{T}) \cdot \mathcal{O}(\psi) \,,
\end{equation}
where $c$ is the Wilson coefficient, $f(\mathbf{T})$ is a flavor tensor composed by the spurions, and $\mathcal{O}(\psi)$ is an operator composed by the other dynamic fields, and the dot '$\cdot$' between them means the composition of the invariant of them.

Focus on the LO Lagrangian, because the Yukawa interactions are the only terms breaking the flavor symmetry, the Yukawa matrices should be prompted to the spurions. For the $U(3)^5$ and the $U(2)^5$ flavor symmetries, the spurions and their representations at the LO Lagrangian are
\begin{equation}
    U(3)^5:\quad \left\{\begin{array}{l}
\mathbf{Y}_u \in (\mathbf{3},\bar{\mathbf{3}},\mathbf{1},\mathbf{1},\mathbf{1})\\
\mathbf{Y}_d \in (\mathbf{3},\mathbf{1},\bar{\mathbf{3}},\mathbf{1},\mathbf{1})\\
\mathbf{Y}_e \in (\mathbf{1},\mathbf{1},\mathbf{1},\mathbf{3},\bar{\mathbf{3}})
    \end{array}\right.\,,\quad U(2)^5:\quad \left\{\begin{array}{l}
\Delta_u \in (\mathbf{2}\,,\overline{\mathbf{2}}\,,\mathbf{1}\,,\mathbf{1}\,,\mathbf{1})\\
\Delta_d \in (\mathbf{2}\,,\mathbf{1}\,,\overline{\mathbf{2}}\,,\mathbf{1}\,,\mathbf{1})\\
\Delta_e \in (\mathbf{1}\,,\mathbf{1}\,,\mathbf{1}\,,\mathbf{2}\,,\overline{\mathbf{2}})\\
V_q \in (\mathbf{2}\,,\mathbf{1}\,,\mathbf{1}\,,\mathbf{1}\,,\mathbf{1}) \\
V_l \in (\mathbf{1}\,,\mathbf{1}\,,\mathbf{1}\,,\mathbf{2}\,,\mathbf{1}) 
    \end{array}\right.\,.
\end{equation}
For the high-dimension operators, there are more independent spurions in principle. For example, the dimension-8 Lagrangian contains the terms such as
\begin{align}
    \mathcal{L} &\supset C_{u^2H^2D} \bin{u^p}{u_r}{\gamma^\mu}(H^\dagger \lrd_\mu H)(H^\dagger H) \notag \\
    &+ C_{quH^5}\bin{q^p}{u_r}{\tilde{H}}(H^\dagger H)^2 \notag \\
    &+ C^{(1)}_{q^4D^2}\bin{q^p}{q_r}{\gamma^\mu}D^2\bin{q^s}{q_t}{\gamma_\mu}\,,
\end{align}
where the coefficients can be developped to the spurions. Under the $U(3)^5$ symmetry, the representations of these spurions can be assigned as
\begin{align}
     C_{u^2H^2D} & \rightarrow  \mathbf{C}_{u^2H^2D} \in (\mathbf{1},\mathbf{3}\times \bar{\mathbf{3}},\mathbf{1},\mathbf{1},\mathbf{1}) \,, \\
     C_{quH^5} & \rightarrow \mathbf{C}_{quH^5} \in (\mathbf{3},\bar{\mathbf{3}},\mathbf{1},\mathbf{1},\mathbf{1})\,, \\
     C^{(1)}_{q^4D^2} &\rightarrow \mathbf{C}^{(1)}_{q^4D^2} \in (\mathbf{3}\times \bar{\mathbf{3}}\times \mathbf{3}\times \bar{\mathbf{3}},\mathbf{1},\mathbf{1},\mathbf{1},\mathbf{1})\,.
\end{align}

According to the previous discussion, the effective operators of the SMEFT can be classified into the CP-even and CP-odd sectors according to the CP transformations of the fields, and the CP-odd operators that are rephasing invariant are CP-violating. As shown before, the CP-odd and the CP-violating operators are not equivalent when the flavor symmetry is absent, for example, the Hilbert series gives that there are 1422 CP-odd operators at dimension 6, but there are only 705 of them are CP-violating~\cite{Kondo:2022wcw}.

The cases are different for a specific flavor symmetry. Because the flavor symmetry is internal, it affects the CP properties. Firstly, the flavor symmetry constrains the independent Wilson coefficients, thus the independent operators are reduced in principle.
Secondly, as a subgroup, the rephasing symmetry depends on flavor symmetry. 
Besides, the CP-odd and the CP-violating operators are equivalent since the flavor symmetry guarantees the rephasing symmetries. This statement is general, for any specific symmetry, the CP-violating operators and the CP-odd operators are equivalent since the rephasing symmetry is satisfied automatically as a subgroup of the flavor symmetry. 
Considering the $U(3)^5$ and the $U(2)^5$ symmetries discussed before. For the $U(3)^5$ symmetry, the rephasing symmetry $U(1)^4$ is a subgroup, so the flavor symmetry guarantees the rephasing symmetry, while for the $U(2)^5$ symmetry, the rephasing symmetry is trivial, and is satisfied automatically.

This means the single assumption of $U(3)^5$ flavor symmetry does not affect the CP-violations of the SMEFT, since it is the maximal flavor symmetry and its break pattern is fixed by the Yukawa matrices. Nevertheless, the cases could be different for the other flavor symmetries such as the $U(2)^5$~\cite{Barbieri:2011ci,Barbieri:2012uh,Blankenburg:2012nx,Greljo:2015mma,Barbieri:2015yvd,Buttazzo:2017ixm}.

The assumption of the $U(3)^5$ symmetry presents another way to express the CP-violating phases by the flavor invariants linear in the Wilson coefficients. As discussed before, although the aforementioned rephasing symmetry acts only on the fields and exits without any flavor symmetry, while the $U(3)^5$ symmetry is supposed on the Lagrangian including both the Wilson coefficients and the fields, the equivalence between them is realized by the the fact that under the flavor symmetry, the Wilson coefficients become the spurions transforming according to the associated operators so that the invariants linear in them correspond to the operators. In particular, the Hilbert series is used in both the method, we compare them in follows,
\begin{itemize}
    \item Adopting the rephasing symmetry, the Hilbert series is utilized~\cite{Kondo:2022wcw} to count the independent CP-odd operators satisfying the rephasing symmetry. The Hilbert series of the CP-odd operators can be obtained by the extension of the Lorentz group and the internal group with the automorphism defined by the parity and the charge conjugate, which takes the form
    \begin{equation}
        \mathcal{H}_{\text{CP-odd}} = \frac{1}{2}(\mathcal{H}_++\mathcal{H}_-)\,,
    \end{equation}
    where $\mathcal{H}_\pm$ are the Hilbert series on the different branches separated by the CP transformation~\cite{Henning:2017fpj,Graf:2020yxt,Sun:2022aag}.
    \item  The Hilbert series can also be used to count the flavor invariants, which are part of the primary invariants. For some building blocks $q$, the Hilbert series takes the form 
\begin{equation}
    \mathcal{H}(q) = \frac{N(q)}{D(q)}\,,
\end{equation}
where the denominator takes the form
\begin{equation}
    D(q) = \prod_{r=1}^p(1-q^{d_r})\,.
\end{equation}
The number of the factors, $p$, is equal to the number of the primary invariants.
Although the Hilbert series can be used to count their numbers, there is no systematic method to construct them explicitly. 
    
\end{itemize}
In the rest of this section, we will discuss the flavor invariants of the dimension-8 operators and introduce the Young tensor method to construct the invariants.

\subsection{Flavor Invariant Description}

The assumption of the flavor symmetry implies that the flavor-invariant description of the CP-violating phases would be beneficial. Because the CP-violating observables are invariant under the flavor symmetry, they could be expressed in a flavor-invariant way.
Actually, the single CP-violating phase $\delta$ at LO can be expressed by the Jarlskog invariant~\cite{Jarlskog:1985ht,Jarlskog:1985cw,Bernabeu:1986fc} as
\begin{equation}
    J_4 = \text{Im}\left(\text{Tr}[Y_u Y_u^\dagger\,,Y_dY_d^\dagger ]^3\right)\,.
\end{equation}
Beyond the LO, the high-dimension Wilson coefficients $C$ are included. The CP-violating primary invariants linear in the Wilson coefficients correspond to the CP-violating operators counted by the Hilbert series with the rephasing symmetry added.
We can argue this equivalence by 3 aspects,
\begin{description}
    \item[Linearity and CP violations] If we specify a weak basis, and suppose the parameters in Yukawa matrices of dimension 4 are $\theta$, the parameters of the high-dimension Wilson coefficients are $\lambda_i$, the linearity means the invariants $I$ take an expansion form
    \begin{equation}
    \label{eq:invariant}
        I(Y,C) \rightarrow I(\theta,\lambda) = \sum_i f(\theta) \lambda_i\,,
    \end{equation}
    and the CP-violations mean the parameters $\lambda_i$ are complex, which correspond to the CP-odd operators.
    \item[Invariance] The flavor invariance $U(3)^5$ guarantees the rephasing symmetry $U(1)^4$, which means that when a specific weak basis is adopted, every term in Eq.~\eqref{eq:invariant} are invariant under the rephasing symmetry.
    \item[Primary] A primary invariant can not be expressed by the secondary invariants that are factorized. A secondary invariant $I'$ takes the form 
    \begin{equation}
        I'(Y,C) \rightarrow I'(\theta,\lambda) = g(\theta)\sum_i f(\theta) \lambda_i\,,
    \end{equation}
    after the specification of a weak basis, which is nothing but a primary invariant multiplied by an overall factor $g(\theta)$, thus are equivalent to the primary one.
\end{description}
Thus the flavor-invariants composed by the Wilson coefficients and the CP-odd operators satisfying the rephasing symmetry are equivalent.

The application of the flavor-invariant description to the dimension-6 operators with the $U(3)^5$ flavor symmetry gets 699 CP-violating flavor invariants~\cite{Bonnefoy:2021tbt}, which is consistent with the Hilbert series result utilizing the flavor symmetry.\footnote{699 is the number of the CP-violating fermionic operators obtained by the 705 minus the 6 CP-violating mesonic operators.}
Considering the interference between the dimension-4 and the dimension-6 operators, even more CP-violating invariants emerge~\cite{Bonnefoy:2023bzx}.

The flavor-invariant description also applies to the dimension-8 operators. For example, the Wilson coefficient $C_{u^2H^2D}$ of a dimension-8 operator
\begin{align}
    \mathcal{O}_{u^2H^4D} = \bin{u^p}{u_r}{\gamma^\mu}(H^\dagger \lrd_\mu H)(H^\dagger H)\,,
\end{align}
is a spurion in the flavor space, which transforms as
\begin{equation}
    C_{u^2H^2D} \rightarrow V_u C_{u^2H^2D}V_u^\dagger \,,\quad V_u \in SU(3)_u\,.
\end{equation}
As discussed before, 3 components of these operators are CP-odd, and they satisfy the rephasing symmetry and thus are CP-violating. The 3 CP-violating phases in the coefficient $C_{u^2H^2D}$ can be expressed by the flavor invariant that
\begin{equation}
    L_{1100}(Y_uC_{u^2H^2D}Y_u^\dagger )\,,\quad L_{2200}(Y_uC_{u^2H^2D}Y_u^\dagger )\,,\quad L_{1122}(Y_uC_{u^2H^2D}Y_u^\dagger )\,,
\end{equation}
where
\begin{equation}
    L_{abcd}(C) = \text{Im}\text{Tr}(X_u^a X_d^b X_u^c X_d^d C)\,,\quad a,b,c,d=0,1,2, \text{ and } a\neq c\,,b\neq d\,,
\end{equation}
with $X_u = Y_u Y_u^\dagger\,,X_d = Y_d Y_d^\dagger$. 
The vanishing of the 3 invariants is the sufficient and necessary condition for the vanishing of the CP-violations. 
On the other hand, the coefficient of the complex operator 
\begin{equation}
    \mathcal{O}_{quH^5}=\bin{q^p}{u_r}{\tilde{H}}(H^\dagger H)^2\,,
\end{equation}
presents 9 CP-violating phases, which correspond to the flavor invariants that,
\begin{align}
    & L_{0000}(C_{quH^5}Y_u^\dagger) \,,\quad L_{1000}(C_{quH^5}Y_u^\dagger) \,,\quad L_{0100}(C_{quH^5}Y_u^\dagger) \,,\notag \\
    & L_{1100}(C_{quH^5}Y_u^\dagger) \,,\quad L_{0110}(C_{quH^5}Y_u^\dagger) \,,\quad L_{2200}(C_{quH^5}Y_u^\dagger) \,,\notag \\
    & L_{0220}(C_{quH^5}Y_u^\dagger) \,,\quad L_{1220}(C_{quH^5}Y_u^\dagger) \,,\quad L_{0122}(C_{quH^5}Y_u^\dagger) \,.
\end{align}
For the 4-fermion operators, similar invariants exist.
For example, we consider the real operator
\begin{equation}
    \mathcal{O}_{q^4D^2}^{(1)}=\bin{q^p}{q_r}{\gamma^\mu}D^2\bin{q^s}{q_t}{\gamma_\mu}\,.
\end{equation}
As mentioned before, its coefficient $C_{q^4D^2}$ presents 18 CP-violating phases, which also can be expressed by the invariants. We define
\begin{equation}
    \text{Tr}_A(M^{(1)},M^{(2)},C) = M^{(1)}{}_p^r M^{(1)}{}_s^t C_{rt}^{ps}\,,\quad  \text{Tr}_B(M^{(1)},M^{(2)},C) = M^{(1)}{}_p^t M^{(1)}{}_s^r C_{rt}^{ps}\,,
\end{equation}
and
\begin{align}
    A_{efgh}^{abcd}(C) &= \text{Im}\text{Tr}_A(X_u^aX_d^bX_u^cX_d^d,X_u^eX_d^fX_u^gX_d^h,C) \,,\\
    B_{efgh}^{abcd}(C) &= \text{Im}\text{Tr}_B(X_u^aX_d^bX_u^cX_d^d,X_u^eX_d^fX_u^gX_d^h,C) \,,
\end{align}
then the 18 flavor invariants are 
\begin{align}
    & A_{1100}^{0000}(C_{q^4D^2}) \,,\quad A_{1100}^{1000}(C_{q^4D^2}) \,,\quad A_{1100}^{0100}(C_{q^4D^2}) \,,\notag \\
    & A_{2200}^{0000}(C_{q^4D^2}) \,,\quad A_{1100}^{1100}(C_{q^4D^2}) \,,\quad A_{1100}^{0200}(C_{q^4D^2}) \,,\notag \\
    & A_{2200}^{0100}(C_{q^4D^2}) \,,\quad A_{1122}^{0000}(C_{q^4D^2}) \,,\quad A_{2200}^{1100}(C_{q^4D^2}) \,,\notag \\
    & A_{1122}^{1000}(C_{q^4D^2}) \,,\quad A_{1122}^{0100}(C_{q^4D^2}) \,,\quad A_{0122}^{1100}(C_{q^4D^2}) \,,\notag \\
    & A_{2200}^{1200}(C_{q^4D^2}) \,,\quad B_{1100}^{0000}(C_{q^4D^2}) \,,\quad B_{1100}^{0100}(C_{q^4D^2}) \,,\notag \\
    & B_{2100}^{0200}(C_{q^4D^2}) \,,\quad A_{1122}^{1200}(C_{q^4D^2}) \,,\quad B_{1200}^{1000}(C_{q^4D^2}) \,.
\end{align}
Actually, the primary invariance can be constructed analytically via the Young tableaux, which will be discussed in the next subsection.

\subsection{Flavor Invariants Construction}

The primary invariants are important since they are algebraically independent and correspond to the physical parameters, which are the ones that can not be removed by the field redefinitions.
The identification of the primary invariants is difficult. 
As discussed before, the Hilbert series method can enumerate the primary invariants~\cite{Jenkins:2009dy},
but it is difficult to construct them explicitly. The common method is to construct an overcomplete set of the invariants and reduce them by finding their relations called syzygies. Thus it is useful to take the plethystic logarithm (PL) of the Hilbert series $\mathcal{H}(q)$. The $PL(\mathcal{H}(q))$ is an expansion of the building blocks $q$ and encodes the important information that the leading positive terms give the numbers and the orders of the basic invariants, and the leading negative terms give the orders of their relations called syzygies. 

The basic invariants are generally linearly independent but not algebraically independent. After the exploitation of the syzygies, they are reduced to the primary invariants. 
For example, we consider the group tensors composed of the two distinct matrices $X_u$ and $X_d$ of $U(3)$ adjoint representation. The PL of the corresponding Hilbert series is
\begin{equation}
\label{eq:pl}
    PL(\mathcal{H}(X_u,X_d)) = (X_u + X_d) + (X_u^2+X_d^2+X_uX_d) + (X_u^3+X_u^2X_d+X_uX_d^2+X_d^3) + X_u^2X_d^2 + X_u^3 X_d^3 - X_u^6X_d^6\,,
\end{equation}
which means there are 11 basic invariants, while the one of order $X_u^3X_d^3$ is not primary due to a syzygy of order $X_u^6X_d^6$. Thus there are 10 primary invariants.
In practice, the syzygies are obtained by numeric method and the $PL(\mathcal{H}(q))$ is instructive.

Actually, even the basic invariants are difficult to find. In this subsection, we present the Young tensor method as an analytic and systematic method to find the basic invariants and expect to pave the way to the primary ones.

As discussed in Ref.~\cite{Song:2024fae}, the symmetrization relations between the tensors of the Yong tableaux method are equivalent to the Cayley-Hamilton relations. In principle, the primary invariants correspond to the group tensors that can not be factorized, for example, $f^{ABC}\,,d^{ABC}\,,d^{ABE}d^{CDE}$ can not be factorized, while $\delta^{AB}\delta^{CD}$ is factorized. These tensors can be obtained by the Young tableaux and after the symmetrization of the tensors via the Young symmetrizers, the relations among them can be used to find the primary invariants. 
This method is systematic and we outline the basic steps here,
\begin{itemize}
    \item Utilize the Young tableaux to obtain the independent tensors at each order. These tensors are interpreted from the Young tableaux, which can be constructed from the outer product such as 
    \begin{equation}
        \ydiagram{2,2,2}\dots\ydiagram{1,1,1} \subset \ydiagram{2,1} \otimes \ydiagram{2,1} \otimes \dots \,.
    \end{equation}
    \item Assign different repetitions and use the associated Young symmetrizers to obtain the relations among the symmetrized tensors. These relations generally mix the factorized and the unfactorized tensors. The independent unfactorized tensors correspond to the primary invariants.
\end{itemize}

Next, we consider the primary invariants composed of the building blocks $X_u\,,X_d$ in terms of their orders. The PL has been presented in Eq.~\eqref{eq:pl}.
\begin{itemize}
    \item Order-1: The Young tensor method presents no constraints, thus there is one primary invariant for each of $X_u$ and $X_d$,
    \begin{equation}
        \text{Tr}(X_u)\,,\quad \text{Tr}(X_d)\,.
    \end{equation}
    \item Order-2: There is only one unfactorized tensor $\delta^{AB}$, which is symmetric, thus there is one primary invariant for each of $X_u^2\,,X_d^2$ and $X_uX_d$,
    \begin{equation}
        \text{Tr}(X_u^2)\,,\quad \text{Tr}(X_d^2)\,,\quad \text{Tr}(X_uX_d)\,.
    \end{equation}
    \item Order-3: The 2 tensors at this order are the totally symmetric tensor $d^{ABC}$ and the totally anti-symmetric tensor $f^{ABC}$. Since there are only 2 distinct building blocks, there must be repetition. The symmetrization of the repeated building blocks implies that only $d^{ABC}$ should be reserved. Thus there is one primary invariant for each of ${X_u}^3\,,{X_d}^3\,,{X_u}^2X_d$ and ${X_d}^2X_u$,
    \begin{equation}
        \text{Tr}(X_u^3)\,,\quad \text{Tr}(X_d^3)\,,\quad \text{Tr}(X_u^2X_d)\,,\quad \text{Tr}(X_uX_d^2)\,.
    \end{equation}
    \item Order-4: The Young tableaux present 8 independent tensors
\begin{align}
    \calb_1 &= d^{ABE}d^{CDE}\,, \quad &\calb_2 &= d^{ABE}f^{CDE}\,, \notag \\
    \calb_3 &= f^{ABE}f^{CDE}\,, \quad &\calb_4 &= \delta^{AB}\delta^{CD}\,, \notag \\
    \calb_5 &= f^{ABE}d^{CDE}\,, \quad &\calb_6 &= \delta^{AC}\delta^{BD}\,, \notag \\
    \calb_7 &= d^{ACE}d^{BDE}\,, \quad &\calb_8 &= d^{ACE}f^{BDE}\,.
\end{align}
Suppose there is only one kind of building block, $X_u$ or $X_d$, the tensors should be symmetrized by the symmetrizer of $S_4$ group $\caly_{[4]}$, which takes the form that
\begin{equation}
    \caly_{[4]}=\left(
\begin{array}{cccccccc}
 \frac{1}{9} & 0 & -\frac{1}{9} & \frac{2}{27} & 0 & \frac{4}{27} & \frac{2}{9} & 0 \\
 0 & 0 & 0 & 0 & 0 & 0 & 0 & 0 \\
 0 & 0 & 0 & 0 & 0 & 0 & 0 & 0 \\
 \frac{1}{3} & 0 & -\frac{1}{3} & \frac{2}{9} & 0 & \frac{4}{9} & \frac{2}{3} & 0 \\
 0 & 0 & 0 & 0 & 0 & 0 & 0 & 0 \\
 \frac{1}{3} & 0 & -\frac{1}{3} & \frac{2}{9} & 0 & \frac{4}{9} & \frac{2}{3} & 0 \\
 \frac{1}{9} & 0 & -\frac{1}{9} & \frac{2}{27} & 0 & \frac{4}{27} & \frac{2}{9} & 0 \\
 0 & 0 & 0 & 0 & 0 & 0 & 0 & 0 \\
\end{array}
\right)\,.
\end{equation}
There are only 4 non-vanishing tensors after symmetrization, and it is obvious that they are linear to each other. Denoting the tensor after symmetrization, which are usually polynomials, by 
\begin{equation}
    \calb \xrightarrow{\text{Symmetrization}} \caly(\calb)\,,
\end{equation}
Their linear relation can be expressed as
\begin{equation}
    \caly(\calb_4) = \caly(\calb_6)=3\caly(\calb_1) = 3\caly(\calb_7)\,.
\end{equation}
or 
\begin{equation}
\label{eq:cons1}
    \caly(\delta^{AB}\delta^{CD}) = \caly(\delta^{AC}\delta^{BD}) = 3\caly(d^{ABE}d^{CDE}) = 3 \caly(d^{ACE}d^{BDE})\,.
\end{equation}
The original unfactorized $\calb_1\,,\calb_7$ are identical with the factorized ones $\calb_4\,,\calb_6$ after the symmetrization, and correspond to no primary tensor. Thus there exists no primary invariant in this case.

If we assign the first two building blocks $X_u$ and the last two building blocks $X_d$, the 8 tensors must be symmetrized by the Young symmetrizer that
\begin{equation}
    \caly_{[2]}\caly_{[2]} = \left(
\begin{array}{cccccccc}
 1 & 0 & 0 & 0 & 0 & 0 & 0 & 0 \\
 0 & 0 & 0 & 0 & 0 & 0 & 0 & 0 \\
 0 & 0 & 0 & 0 & 0 & 0 & 0 & 0 \\
 0 & 0 & 0 & 1 & 0 & 0 & 0 & 0 \\
 0 & 0 & 0 & 0 & 0 & 0 & 0 & 0 \\
 \frac{1}{2} & 0 & -\frac{1}{2} & -\frac{1}{6} & 0 & \frac{2}{3} & 1 & 0 \\
 -\frac{1}{3} & 0 & -\frac{1}{6} & \frac{1}{9} & 0 & \frac{2}{9} & \frac{1}{3} & 0 \\
 0 & 0 & 0 & 0 & 0 & 0 & 0 & 0 \\
\end{array}
\right)\,.
\end{equation}
It is a rank-3 matrix, and its application presents a relation among the symmetrized tensors that
\begin{equation}
    \caly(d^{ACE}d^{BDE}) = -\frac{1}{2}\caly(d^{ABE}d^{CDE})+\frac{1}{6}\caly(\delta^{AB}\delta^{CD})+\frac{1}{3}\caly(\delta^{AC}\delta^{BD})\,.
\end{equation}
This relation can be interpreted as the two unfactorized tensors $\caly(d^{ACE}d^{BDE})\,,\caly(d^{ABE}d^{CDE})$ are not independent and they are related by the other two factorized tensors $\caly(\delta^{AB}\delta^{CD})$ and $\caly(\delta^{AC}\delta^{BD})$. Thus there is only one primary invariant composed by two $X_u$ and $X_d$, which corresponds the the tensor $\caly(d^{ABE}d^{CDE})$, and can be interpreted to the trace
\begin{equation}
    \caly(d^{ABE}d^{CDE}) \rightarrow \text{Tr}(X_u^2X_d^2)\,.
\end{equation}
The cases where there are 3 repeated building blocks present the same symmetrizer 
\begin{equation}
    \mathcal{Y}_{[3]} = \mathcal{Y}_{[4]}\,,
\end{equation}
thus present no primary invariant.
\item Order-5: There are 32 tensors, and we present the relations among the symmetrized tensors, rather than listing all the tensors here.
\begin{align}
    X_u^5\,,X_d^5:\quad  &\caly(d^{ABC}\delta^{DE}) = 3\caly(d^{ACF}d^{BFG}d^{DEG}) = 3\caly(d^{ACF}d^{DFG}d^{BEG}) = \caly(d^{BDE}\delta^{AC}) \notag \\
        &= 3\caly(d^{ABF}d^{CFG}d^{DEG}) = 3\caly(d^{ABF}d^{DFG}d^{CEG})+3\caly(d^{ADF}d^{BFG}d^{CEG})\,, \\
    X_u^4X_d\,,X_d^4X_u:\quad & \caly(d^{ABC}\delta^{DE})=\caly(d^{ABC}\delta^{DE})\,,\notag \\
        &\caly(d^{ACF}d^{BFG}d^{DEG}) = \caly(d^{ACF}d^{BEG}d^{DFG}) = \frac{1}{3}\caly(d^{BDE}\delta^{AC}) = \caly(d^{ABF}d^{DEG}d^{CFG}) \notag \\
        &= \caly(d^{ABF}d^{CEG}d^{DFG}) = \caly(d^{ADF}d^{BFG}d^{CEG})\,, \\
    X_u^3X_d^2\,,X_u^2X_d^3:\quad & \caly(d^{ACF}d^{BFG}d^{DEG})=\caly(d^{ABF}d^{CFG}d^{DEG})=\frac{1}{3}\caly(d^{BDE}\delta^{AC})\,,\notag \\
        &\caly(d^{ACF}d^{BEG}d^{DFG})=\caly(d^{ABF}d^{DFG}d^{CEG})\,, \notag \\
        &\caly(d^{ACF}f^{BEG}f^{DFG})=\caly(d^{ABF}f^{CEG}f^{DFG})\,, \notag \\
        &\caly(d^{ABC}\delta^{DE})=3\caly(d^{ACF}d^{BEG}d^{DFG})-\caly(d^{ACF}f^{BEG}f^{DFG})\,, \notag \\
        &\caly(d^{ADF}d^{BFG}d^{CEG})=\caly(d^{ACF}d^{BFG}d^{DEG}) + \frac{1}{6}\caly(d^{ACF}f^{BEG}f^{DFG})\,.
\end{align}
There is no independent unfactorized tensor thus there is no primary invariant in this order.
\end{itemize}
Thus we obtain all the 10 primary invariants composed of the two building blocks $X_u$ and $X_d$. 

For the invariants linear in the high-dimension Wilson coefficients $C$, the arguments are similar. We fix a building block as the Wilson coefficient $C$, and other building blocks are $X_u$ and $X_d$ dependently to form invariants. Following the steps above, the primary invariants can be obtained according to the building block repetitions order by order in principle.
For example, we consider a simple case where a high-dimension coefficient $C$ is included, which transforms in the same way as the $X_u\,,X_d$. As discussed before, the Young tensor method presents no constraints up to order 3, and the resultant basic invariants are 
\begin{equation}
    \text{Tr}(C)\,,\quad \text{Tr}(X_uC)\,,\quad \text{Tr}(X_dC)\,,\quad \text{Tr}(X_u^2C)\,,\quad \text{Tr}(X_uX_dC)\,,\quad \text{Tr}(X_d^2 C)\,. 
\end{equation}
Considering the order 4, it has been shown there is no basic invariant when there are 3 repeated building blocks, thus we consider only the case where there are exactly two repeated building blocks, for example, we assign the first two building blocks are repeated, the Young symmetrizer is 
\begin{equation}
    \mathcal{Y}_{[2]} = \left(
\begin{array}{cccccccc}
 1 & 0 & 0 & 0 & 0 & 0 & 0 & 0 \\
 0 & 1 & 0 & 0 & 0 & 0 & 0 & 0 \\
 0 & 0 & 0 & 0 & 0 & 0 & 0 & 0 \\
 0 & 0 & 0 & 1 & 0 & 0 & 0 & 0 \\
 0 & 0 & 0 & 0 & 0 & 0 & 0 & 0 \\
 \frac{1}{2} & 0 & -\frac{1}{2} & -\frac{1}{6} & 0 & \frac{2}{3} & 1 & 0 \\
 -\frac{1}{3} & 0 & -\frac{1}{6} & \frac{1}{9} & 0 & \frac{2}{9} & \frac{1}{3} & 0 \\
 0 & -\frac{1}{2} & 0 & 0 & 0 & 0 & 0 & 0 \\
\end{array}
\right)\,,
\end{equation}
which is a rank-4 matrix, thus there are 4 independent tensors after the symmetrization. However, there are relations that
\begin{align}
    \mathcal{Y}(\delta^{AC}\delta^{BD}) &= -\frac{3}{2}\mathcal{Y}(d^{ACE}d^{BDE}) \,,\\
    \mathcal{Y}(d^{ACE}f^{BDE}) &= -\frac{1}{2}\mathcal{Y}(d^{ABE}f^{CDE})\,,
\end{align}
thus there are only two independent unfactorized tensors, $d^{ABE}d^{CDE}$ and $d^{ABE}f^{CDE}$, which correspond to 4 basic invariants
\begin{equation}
    \text{Tr}(X_uX_d^2 C)\,,\quad \text{Tr}(X_u^2X_d C)\,,\quad \text{Tr}(X_dX_u^2 C)\,,\quad \text{Tr}(X_d^2X_u C)\,. 
\end{equation}
For the higher orders, we can perform the algorithm repeatedly and find all the basic invariants.

\section{Minimal Flavor Violation Hypothesis}
\label{sec:MFV}

In the point of EFT, all the Wilson coefficients of the effective operators are independent, and up to now, the complete set of the SMEFT operators up to dimension 12 has been constructed~\cite{Weinberg:1979sa,Buchmuller:1985jz,Grzadkowski:2010es,Lehman:2014jma,Liao:2016hru,Li:2020gnx,Li:2020xlh,Harlander:2023psl}. 
However, the Wilson coefficients are constrained by the experiments.
To be consistent with the experiment results involving the FCNC processes and not push the new physics scale too high, the minimal flavor violation (MFV) hypothesis is introduced, which states that the Yukawa matrices are the only flavor-symmetry-breaking sources.

Specifically, we consider a dimension-$d$ effective operator $\mathcal{O}^d_{pr\dots s}$, where $pr\dots s$ are its flavor indices. The MFV hypothesis implies that such an operator appears in the effective Lagrangian in the form that
\begin{equation}
\label{eq:expansion}
    \mathcal{L} \supset \frac{c}{\Lambda^{d-4}} f(Y_u,Y_d,Y_e)^{pr\dots,s} \mathcal{O}^d_{pr\dots s}\,,
\end{equation}
where $c$ is the Wilson coefficient, and $f(Y_u,Y_d,Y_e)^{pr\dots,s}$ is a polynomial composed of the Yukawa matrices. For example, the dimension-4 Yukawa interactions satisfy
\begin{equation}
    f_\psi(Y_u,Y_d,Y_e) = Y_\psi\,,\quad \psi=u,d,e\,,
\end{equation}
and $c=1$.

Generally, the polynomial is a series
\begin{equation}
    f(Y_u,Y_d,Y_e) = f_0 + O(Y_\psi) + O(Y_\psi^2) + \dots\,, \quad \psi=u,d,e\,,
\end{equation}
thus a power-counting scheme is needed. Because the Yukawa matrices are proportional to the fermion masses, their hierarchy can help to define the expansion parameters. Adopting the down-basis, the three Yukawa matrices take the form
\begin{align}
    Y_u &\propto y_t V_{CKM}^\dagger \text{diag}(\frac{y_u}{y_t},\frac{y_c}{y_t},1) \sim y_t\,, \\
    Y_d & \propto y_b \text{diag}(\frac{y_d}{y_b},\frac{y_s}{y_b},1) \sim y_b\,, \\
    Y_e & \propto y_\tau \text{diag}(\frac{y_e}{y_\tau},\frac{y_\mu}{y_\tau},1) \sim y_\tau\,,
\end{align}
and 
\begin{equation}
    \frac{y_\tau}{y_t} \approx \frac{y_b}{y_t} \approx O(10^{-3})\,,
\end{equation}
where we have set $\frac{\sqrt{2}}{v} m = y$,
thus we can define the expansion parameters as
\begin{equation}
\label{eq:power_counting}
    \frac{Y_u}{y_t}\sim 1\,, \quad \frac{Y_d}{y_t},\frac{Y_e}{y_t} \sim 10^{-3}\,.
\end{equation}
Consequently, it looks like the $Y_u$ can be any power in the expansion, but it is false since the ring generated of it is finite-generated~\cite{Grinstein:2023njq}. In this paper, we usually expand to $O(Y_u)$ or $O(Y_u^2)$ instead of any power of it~\cite{Kagan:2009bn,Feldmann:2008ja}, since the terms with more powers of $Y_u$ can be absorbed by field redefinitions~\cite{Greljo:2022cah}. 

Under the MFV hypothesis, the effective operators can be classified in terms of the Yukawa matrices they couple, which has been done for the dimension-6 operators in Ref.~\cite{Faroughy:2020ina}, and in the rest of this section, we will summarize the result and focus on the same analysis for the dimension-8 operators. In particular, we present the MFV operators up to $10^{-3}$ according to Eq.~\eqref{eq:power_counting}. 

\subsection{Dimension-6 MFV Operators}

As discussed before, the most efficient way to construct MFV effective operators is to promote the Yuakwa matrices to the dynamic degrees of freedom,
\begin{equation}
    Y_\psi \rightarrow \mathbf{Y}_\psi(x)\,,\quad \psi=u,d,e\,,
\end{equation}
called spurions, which transform under the global $U(3)^5$ symmetry as
\begin{align}
    \mathbf{Y}_u &\in (\mathbf{3},\bar{\mathbf{3}},\mathbf{1},\mathbf{1},\mathbf{1})\,,\\
    \mathbf{Y}_d &\in (\mathbf{3},\mathbf{1},\bar{\mathbf{3}},\mathbf{1},\mathbf{1})\,,\\
    \mathbf{Y}_e &\in (\mathbf{1},\mathbf{1},\mathbf{1},\mathbf{3},\bar{\mathbf{3}})\,,
\end{align}
and to require the operators to be invariant under the $U(3)^5$. Such management assumes that the underlying UV theory shares the same flavor symmetry $U(3)^5$. After the fields $\mathbf{Y}_\psi(x)$ are frozen and take their VEVs, these operators become the SMEFT ones.


We adopt the Warsa basis~\cite{Grzadkowski:2010es} of the dimension-6 operators, the explicit operators are presented in App.~\ref{app:dim6}. Apart from the pure boson sectors, there are 4 fermionic sectors, $\psi^2H^3\,,\psi^2XH\,,\psi^2H^2D$ and $\psi^4$. Next, we will discuss the MFV operators of these classes separately, a similar discussion can be found in Ref.~\cite{Faroughy:2020ina}.

\begin{itemize}
    \item There are 9 flavorful operators in the class $\psi^2H^2D$, and only $\mathcal{O}_{Hud}$ is complex. For the real ones, the expansion of the coefficients is
    \begin{equation}
        f(\mathbf{Y})_p^r = \delta_p^r + f_2 \left(\mathbf{Y}_\psi^\dagger \mathbf{Y}_\psi\right){}_p^r\,.
    \end{equation}
    For $\mathcal{O}_{Hl}^{(1,3)}\,, \mathcal{O}_{He}$ and $\mathcal{O}_{Hd}$, the expansion is truncated at $\delta_p^r$, thus these MFV operators are diagonal, while for $\mathcal{O}_{Hq}^{(1,3)}$ and $\mathcal{O}_{Hu}$, the second term is preserved,
    \begin{align}
        \mathcal{O}_{Hl}^{(1,3)}\,, \mathcal{O}_{He}\,, \mathcal{O}_{Hd}: \quad & f(\mathbf{Y})_p^r = \delta_p^r \,, \\
        \mathcal{O}_{Hq}^{(1,3)}: \quad & f(\mathbf{Y})_p^r = \delta_p^r + f_2 \left(\mathbf{Y}_u \mathbf{Y}_u^\dagger\right){}_p^r \,, \\
        \mathcal{O}_{Hu}: \quad & f(\mathbf{Y})_p^r = \delta_p^r + f_2 \left(\mathbf{Y}_u^\dagger \mathbf{Y}_u\right){}_p^r \,.
    \end{align}
    As for the complex one $\mathcal{O}_{Hud}$, it vanishes at leading order and only the quadratic term survives,
    \begin{equation}
        \mathcal{O}_{Hud}:\quad f(\mathbf{Y})_p^r =  \left(\mathbf{Y}_u^\dagger \mathbf{Y}_d\right){}_p^r \,,
    \end{equation}
    where we follow the convention that the coefficient of the first term in the expansion is always absorbed in the overall number $c$ in Eq.~\eqref{eq:expansion}.
    \item The operators in the classes $\psi^2 H^3\,, \psi^2XH$ are similar to the dimension-4 Yukawa interactions, thus the linear term in the expansion is included $f(\mathbf{Y}) = \mathbf{Y}_\psi+\dots\,,\psi =u,d,e$. Besides, the coefficients of the operators containing down-quark $d(x)$ can be expanded to the next order, thus we have
    \begin{equation}
        f(\mathbf{Y}) = \mathbf{Y}_d + f_3(\mathbf{Y}_u \mathbf{Y}_u^\dagger)\mathbf{Y}_d + \dots\,,
    \end{equation}
for the operators $\mathcal{O}_{dH}\,,\mathcal{O}_{dG}\,,\mathcal{O}_{dW}$ and $\mathcal{O}_{dB}$.

    \item For the class $\psi^4$, we assume the general operators take the form $(\bar{\psi}_1{}^p \Gamma_1\psi_2{}_r )(\bar{\psi}_3{}^s\Gamma_2\psi_4{}_t)$, where $p,r,s,t$ are flavor indices. It should be noted that when there are repeated fields among the four fermions, not all the terms in the expansion are independent. If the four fermions are the same, $(\bar{\psi}{}^p \Gamma_1\psi{}_r )(\bar{\psi}{}^s\Gamma_2\psi{}_t)$, where $\psi=q,u,d,l,e$, the leading terms are generally $\delta_p^r\delta^t_s + f_0' \delta_p^t \delta_s^r$, but for the right-handed lepton $e$, the Fierz identity implies 
    \begin{equation}
        (\bar{\psi}^p\gamma_\mu \psi_r)(\bar{\psi}^s \gamma^\mu \psi_t) = (\bar{\psi}^s\gamma_\mu \psi_r)(\bar{\psi}^p \gamma^\mu \psi_t)\,,
    \end{equation}
    thus the second tensor vanishes. Only if $\psi=q,u$ the next order is preserved, $f_2 \delta_p^r\left(\mathbf{Y}_u\mathbf{Y}_u^\dagger\right){}_s^t + f'_2\delta_s^r\left(\mathbf{Y}_u\mathbf{Y}_u^\dagger\right){}_p^t$, or $f_2 \delta_p^r\left(\mathbf{Y}_u^\dagger\mathbf{Y}_u\right){}_s^t + f'_2\delta_s^r\left(\mathbf{Y}_u^\dagger\mathbf{Y}_u\right){}_p^t$. Thus we obtain 
    \begin{align}
        \mathcal{O}_{ll}\,, \mathcal{O}_{dd}:\quad & f(\mathbf{Y})_{ps}^{rt} = \delta_p^r\delta_s^t + f_0' \delta_p^t\delta_s^r\,, \label{eq:qqqq1}\\
        \mathcal{O}^{(1,3)}_{qq}:\quad & f(\mathbf{Y})_{ps}^{rt} = \delta_p^r\delta_s^t + f_0' \delta_p^t\delta_s^r + f_2\delta_p^r\left(\mathbf{Y}_u\mathbf{Y}_u^\dagger\right){}_s^t + f_2'\delta_s^r\left(\mathbf{Y}_u\mathbf{Y}_u^\dagger\right){}_p^t\,, \\
        \mathcal{O}_{ee}: \quad & f(\mathbf{Y})_{ps}^{rt} = \delta_p^r\delta_s^t \,, \\
        \mathcal{O}_{uu}: \quad & f(\mathbf{Y})_{ps}^{rt} = \delta_p^r\delta_s^t + f_0' \delta_p^t\delta_s^r+f_2\delta_p^r\left(\mathbf{Y}_u^\dagger\mathbf{Y}_u\right){}_s^t + f_2'\delta_s^r\left(\mathbf{Y}_u^\dagger\mathbf{Y}_u\right){}_p^t\,. \label{eq:qqqq2}
    \end{align}
    If there are only two different pairs of repeated fields, $(\bar{\psi}_1{}^p \Gamma_1\psi_1{}_r )(\bar{\psi}_2{}^s\Gamma_2\psi_2{}_t)$, the leading order in the expansion is $\delta_p^r \delta_s^t$, and if $\psi_1/\psi_2=q,u$, the second order survives. Thus we have
    \begin{align}
        \mathcal{O}_{lq}^{(1,3)}\,, \mathcal{O}_{qe}\,, \mathcal{O}_{qd}^{(1,8)}: \quad & f(\mathbf{Y})_{ps}^{rt} = \delta_p^r\delta_s^t + f_2 \delta_s^t\left(\mathbf{Y}_u\mathbf{Y}_u^\dagger\right){}_p^r\,, \label{eq:qqqq3} \\
        \mathcal{O}_{eu}\,, \mathcal{O}_{ud}^{(1,8)}\,, \mathcal{O}_{lu}: \quad & f(\mathbf{Y})_{ps}^{rt} = \delta_p^r\delta_s^t + f_2 \delta_s^t\left(\mathbf{Y}_u^\dagger\mathbf{Y}_u\right){}_p^r\,, \\
        \mathcal{O}_{ed}\,, \mathcal{O}_{le}\,, \mathcal{O}_{ld}: \quad & f(\mathbf{Y})_{ps}^{rt} = \delta_p^r\delta_s^t\,, \\
        \mathcal{O}_{qu}^{(1,8)}: \quad & f(\mathbf{Y})_{ps}^{rt} = \delta_p^r\delta_s^t +  f_2 \delta_s^t\left(\mathbf{Y}_u\mathbf{Y}_u^\dagger\right){}_p^r +f'_2 \delta_p^r\left(\mathbf{Y}_u^\dagger\mathbf{Y}_u\right){}_s^t +f''_2\left(\mathbf{Y}_u^\dagger\right){}_p^t \left(\mathbf{Y}_u\right){}_s^r \,.\label{eq:qqqq4}
    \end{align}
    If there are only two repeated fields, there are two operators $\mathcal{O}_{quqd}^{(1,8)}$, and their coefficients are
    \begin{equation}
        \mathcal{O}_{quqd}^{(1,8)}:\quad f(\mathbf{Y})_{ps}^{rt} = \left(\mathbf{Y}_u\right)_p^r\left(\mathbf{Y}_d\right)_s^t\,. 
    \end{equation}
    Finally, if there is no repeated field there are 3 operators, $\mathcal{O}_{lequ}^{(1,3)}\,, \mathcal{O}_{ledq}$, and the first two's coefficients are 
    \begin{equation}
    \label{eq:qqqq5}
        \mathcal{O}_{lequ}^{(1,3)}: \quad f(\mathbf{Y})_{ps}^{rt} = \left(\mathbf{Y}_e\right)_p^r\left(\mathbf{Y}_u\right)_s^t\,,
    \end{equation}
    where the last one's coefficient is zero up to $O(10^{-3})$.
\end{itemize}

In particular, the operators discussed above are all lepton-/baryon-number conserving. Actually, the operators violating the lepton/baryon number are eliminated under the MFV hypothesis, since all the Yukawa matrices carry no lepton-/baryon number. 

Under the MFV hypothesis, both the CP-even and CP-odd operators are reduced, because the Yukawa matrices are prompted to be spurions, and the independent parameters are reduced. We summarize the CP-even and CP-violating operators of dimension 6 in Tab.~\ref{tab:dim6}, where $n_\mathbf{Y}$ is the number of the operators when the Yukawa matrices are regarded as the spurions, and $n_Y$ is the number of the operators when the Yukawa matrices take their VEVs, which is taken as the down-basis in this paper. In particular, we present the CP-even and the CP-violating operators without any flavor symmetry in the column $n$.
As discussed before, the CP-odd operators are equivalent to the CP-violating ones, thus there are 26 CP-violating operators when the Yukawa matrices are regarded as spurions, which is the number of the independent CP-violating phases under the MFV hypothesis, and there are 327 CP-violating operators when the spurions take their VEVs, which is generally larger than the number of the independent CP-violating phases. In particular, the number of the CP-violating phases is reduced compared to 705 without any flavor symmetry.  

\begin{table}
\renewcommand{\arraystretch}{1.8}
\begin{center}
    \begin{tabular}{|c|c|c||c|c||c|c|}
\hline
\multirow{2}{*}{Class} & \multicolumn{2}{c||}{$n$} & \multicolumn{2}{c||}{$n_\mathbf{Y}$} &\multicolumn{2}{c|}{$n_Y$} \\ 
\cline{2-7}
& CP-even & CP-violating & CP-even & CP-violating & CP-even & CP-violating \\
\hline
bosonic sectors & 9 & 6 & 9 & 6 & 9 & 6 \\
\hline
$\psi^2H^3$ & 21 & 21 & 4 & 4 & 21 & 21 \\
\hline
$\psi^2XH$ & 60 & 60 & 11 & 11 & 60 & 60 \\
\hline
$\psi^2H^2D$ & 42 & 21 & 11 & 1 & 36 & 15 \\
\hline
$\psi^4$ & 774 & 597 & 50 & 4 & 447 & 231 \\
\hline
\multirow{2}{*}{Total} & 906 & 705 & 85 & 26 & 567 & 327 \\
\cline{2-7}
& \multicolumn{2}{c||}{1611} & \multicolumn{2}{c||}{111} & \multicolumn{2}{c|}{894}\\
\hline
    \end{tabular}
\end{center}
\caption{
The CP properties of the SMEFT dimension-6 effective operators, where $n$ corresponds to the operators without flavor symmetry, $n_\mathbf{Y}$ corresponds to the operators with the spurions $\mathbf{Y}_\psi$, and $n_Y$ corresponds to the cases when $\mathbf{Y}_\psi$'s become matrices $Y_\psi$'s. 
The CP-odd bosonic operators can also be found in Tab.~\ref{tab:odd_bosonic}.
}
\label{tab:dim6}
\end{table}

\subsection{Dimension-8 MFV Operators}

The same analysis can be applied to the dimension-8 operators of the SMEFT. Consequently, new flavor structures appear at dimension 8, whose MFV coefficients need more attention. Next, we will present the coefficient expansion of them by the different classes, and such an analysis provides a classification of the dimension-8 operators. In this paper, we adopt the explicit operators of Ref.~\cite{Li:2020gnx} but reformulate them more concisely so that their flavor structures are more transparent. The explicit operators are presented in App.~\ref{app:dim8}.

\paragraph{Current Operators} As shown previously the classes of the current operators can be divided into two parts, the first contains the Yukawa-like operators composed of scalar bilinears, including 
\begin{equation}
    \psi^2\phi^5\,, \psi^2\phi^3D^2\,, \psi^2X\phi^3\,,\psi^2X\phi D^2\,,\psi^2X^2\phi
\end{equation}
and the second is composed of vector bilinears, including
\begin{equation}
    \psi^2\phi^4D\,, \psi^2\phi^2D^3\,,\psi^2X\phi^2D\,,\psi^2X^2D\,.
\end{equation}
For the first part, all the coefficient expansions take the form
\begin{equation}
    f(\mathbf{Y})_p^r = \left(\mathbf{Y}_\psi\right){}_p^r\,,\quad \psi=u,d,e\,,
\end{equation}
but the for the ones containing the down-quark $d(x)$, it can be expanded to the next order,
\begin{equation}
    f(\mathbf{Y})_p^r = \left(\mathbf{Y}_d\right){}_p^r + f_3(\mathbf{Y}_u \mathbf{Y}_u^\dagger)\left(\mathbf{Y}_d\right){}_p^r\,.
\end{equation}
For the second part, the coefficients of the real operators are diagonal, 
\begin{equation}
    f(\mathbf{Y})_p^r = \delta_p^r\,,
\end{equation}
while for the ones containing the left quarks or the up-quark $u(x)$, the coefficients can be expanded to the next order, $f_2 (\mathbf{Y}_u \mathbf{Y}_u^\dagger)_p^r$ or $f_2 (\mathbf{Y}_u^\dagger \mathbf{Y}_u)_p^r$. For the complex operators containing the bilinear $(\bar{u}\Gamma d)$, the coefficients are
\begin{equation}
        f(\mathbf{Y})_p^r =  \left(\mathbf{Y}_u^\dagger \mathbf{Y}_d\right){}_p^r + \dots \,.
    \end{equation}

\paragraph{4-Fermion Operators} There are 4 classes
\begin{equation}
    \psi^4D^2\,, \psi^4\phi D\,, \psi^4\phi^2\,, \psi^4X\,.
\end{equation}
The classes $\psi^4D^2$ and $\psi^4X$ share the same flavor structures with the dimension-6 class $\psi^4$, thus their coefficients can be read from Eq.~\eqref{eq:qqqq1} to Eq.~\eqref{eq:qqqq5} according to the field content. While for the remaining two classes $\psi^4\phi^2$ and $\psi^4\phi D$, new flavor structures appear.

\begin{itemize}
    \item Class $\psi^4\phi^2$: Because of the two Higgs fields, new complex operators appear,
    \begin{equation}
    \label{eq:new_flavor_psi4phi2}
        \begin{array}{ll}
\mathcal{O}^{'(1)}_{q^2u^2H^2} =\bin{q^p}{u_r}{\tilde{H}}\bin{q^s}{u_t}{\tilde{H}}\,, & \mathcal{O}^{'(2)}_{q^2u^2H^2} =\bin{q^p}{u_r}{\tilde{H}\lambda^A}\bin{q^s}{u_t}{\tilde{H}\lambda^A} \\
\mathcal{O}^{'(1)}_{q^2d^2H^2} =\bin{q^p}{d_r}{\tilde{H}}\bin{q^s}{d_t}{\tilde{H}}\,, & \mathcal{O}^{'(2)}_{q^2d^2H^2} =\bin{q^p}{d_r}{\tilde{H}\lambda^A}\bin{q^s}{d_t}{\tilde{H}\lambda^A} \\
\mathcal{O}'_{l^2e^2H^2}=\bin{l^p}{e_r}{H}\bin{l^s}{e_t}{H} & \mathcal{O}_{q^2udH^2}^{'(1)}=\bin{q^p}{u_r}{\tilde{H}}\bin{d^s}{q_t}{H^\dagger} \\
\mathcal{O}_{q^2udH^2}^{'(2)}=\bin{q^p}{u_r}{\tilde{H}\lambda^A}\bin{d^s}{q_t}{H^\dagger\lambda^A} & \mathcal{O}'_{l^2udH^2}=\bin{l^p}{l_r}{H\gamma^\mu\tilde{H}^\dagger}\bin{u^s}{d_t}{\gamma_\mu} \\
\mathcal{O}_{qdleH^2}^{'(1)}=\bin{q^p}{d_r}{H}\bin{l^s}{e_t}{H} & \mathcal{O}_{qdleH^2}^{'(2)}=\bin{q^p}{d_r}{H\sigma^{\mu\nu}}\bin{l^s}{e_t}{H\sigma_{\mu\nu}} \\
\mathcal{O}'_{quleH^2} = (\overline{q^p}\tilde{H}u_r)(\overline{e^s}H^\dagger l_t)
        \end{array}\,,
    \end{equation}
where the quadratic terms dominate all the coefficients. Up to $O(10^{-3})$, only $\mathcal{O}_{q^2u^2H^2}^{'(1,2)}\,, \mathcal{O}_{q^2udH^2}^{'(1,2)}$, $\mathcal{O}'_{l^2udH^2}$, and $\mathcal{O}'_{quleH^2}$ are non zero,
\begin{align}
    \mathcal{O}_{q^2u^2H^2}^{'(1,2)}: \quad & f(\mathbf{Y})^{rt}_{ps}=\left(\mathbf{Y}_u\right){}_p^r \left(\mathbf{Y}_u\right){}_s^t + f'_2 \left(\mathbf{Y}_u\right){}_p^t \left(\mathbf{Y}_u\right){}_s^r \,, \\
    \mathcal{O}_{q^2udH^2}^{'(1,2)}: \quad & f(\mathbf{Y})^{rt}_{ps} = \left(\mathbf{Y}_u\right){}_p^r \left(\mathbf{Y}_d^\dagger\right){}_s^t + f'_2 \delta_p^t \left(\mathbf{Y}^\dagger_d \mathbf{Y}_u\right){}_s^r\,, \\
    \mathcal{O}'_{l^2udH^2}: \quad & f(\mathbf{Y})^{rt}_{ps} = \delta_p^r \left(\mathbf{Y}^\dagger_u \mathbf{Y}_d\right){}_s^t\,, \\
     \mathcal{O}'_{quleH^2}:\quad & f(\mathbf{Y})_{ps}^{rt} = \left(\mathbf{Y}_u\right)_p^r \left(\mathbf{Y}_e^\dagger\right)_s^t\,, \\
     \mathcal{O}_{q^2d^2H^2}^{'(1,2)}\,,\mathcal{O}_{l^2e^2H^2}\,,\mathcal{O}_{qdleH^2}^{'(1,2)}: \quad & f(\mathbf{Y})^{rt}_{ps}=0\,. \label{eq:dimension_8_u35}
\end{align}
    \item Class $\psi^4\phi D$: All the operators of this class possess different flavor structures from the dimension-6 ones because the single Higgs field carries non-zero hypercharge. Next, we will discuss their coefficients in terms of the repeated fields.

    If there are 3 repeated fields, the coefficient expansion depends on whether the operator contains the down-quark $d(x)$, if so, the expansion can take cubic terms, 
    \begin{align}
        \mathcal{O}_{q^3dHD}^{(1-6)}: \quad & f(\mathbf{Y})^{rt}_{ps} = \delta_p^r\left(\mathbf{Y}_d\right){}_s^t + f'_1\delta_s^r\left(\mathbf{Y}_d\right){}_p^t + f^{(1)}_3 \left(\mathbf{Y}_u\mathbf{Y}_u^\dagger\right){}_p^r\left(\mathbf{Y}_d\right){}_s^t + f^{(2)}_3 \left(\mathbf{Y}_u\mathbf{Y}_u^\dagger\right){}_s^r\left(\mathbf{Y}_d\right){}_p^t \notag \\
        & + f^{(3)}_3 \delta_p^r\left(\mathbf{Y}_u\mathbf{Y}_u^\dagger\mathbf{Y}_d\right){}_s^t + f^{(4)}_3 \delta_s^r\left(\mathbf{Y}_u\mathbf{Y}_u^\dagger\mathbf{Y}_d\right){}_p^t\,, \\
        \mathcal{O}_{qd^3HD}^{(1-6)}: \quad & f(\mathbf{Y})^{rt}_{ps} = \left(\mathbf{Y}_d\right){}_p^r \delta_s^t + f'_1\delta_s^r\left(\mathbf{Y}_d\right){}_p^t + f^{(1)}_3 \left(\mathbf{Y}_d\right){}_p^r\left(\mathbf{Y}_u\mathbf{Y}_u^\dagger\right){}_s^t + f^{(2)}_3 \left(\mathbf{Y}_u\mathbf{Y}_u^\dagger\right){}_s^r\left(\mathbf{Y}_d\right){}_p^t \notag \\
        & + f^{(3)}_3 \left(\mathbf{Y}_u\mathbf{Y}_u^\dagger\mathbf{Y}_d\right){}_p^r\delta_s^t + f^{(4)}_3 \delta_s^r\left(\mathbf{Y}_u\mathbf{Y}_u^\dagger\mathbf{Y}_d\right){}_p^t\,,
    \end{align}
    otherwise, only the linear terms are reserved,
    \begin{align}
        \mathcal{O}^{(1-6)}_{q^3uHD}: \quad & f(\mathbf{Y})^{rt}_{ps} = \delta_p^r\left(\mathbf{Y}_u\right){}_s^t + \delta_s^r \left(\mathbf{Y}_u\right){}_p^t \,, \\
        \mathcal{O}^{(1-6)}_{qu^3HD}: \quad & f(\mathbf{Y})^{rt}_{ps} = \left(\mathbf{Y}_u\right){}_p^r\delta_s^t + \delta_s^r \left(\mathbf{Y}_u\right){}_p^t \,, \\
        \mathcal{O}^{(1-3)}_{l^3eHD}: \quad & f(\mathbf{Y})^{rt}_{ps} = \delta_p^r\left(\mathbf{Y}_e\right){}_s^t + \delta_s^r \left(\mathbf{Y}_e\right){}_p^t \,, \\
        \mathcal{O}^{(1,2)}_{le^3HD}: \quad & f(\mathbf{Y})^{rt}_{ps} = \left(\mathbf{Y}_e\right){}_p^r\delta_s^t + \delta_s^r \left(\mathbf{Y}_e\right){}_p^t \,.
    \end{align}
    When there are two repeated fields if the repeated ones are left-handed quark $q(x)$ or right-handed up-quark $u(x)$, the expansions are
    \begin{align}
        \mathcal{O}_{qu^2dHD}^{(1-6)}: \quad & f(\mathbf{Y})^{rt}_{ps} = \left(\mathbf{Y}_d\right){}_p^r \delta_s^t + f_3 \left(\mathbf{Y}_d\right){}_p^r\left(\mathbf{Y}_u^\dagger \mathbf{Y}_u\right){}_s^t + f'_3 \left(\mathbf{Y}_u\right){}_p^t\left(\mathbf{Y}_u^\dagger \mathbf{Y}_d\right){}_s^r + f''_3 \left(\mathbf{Y}_u\mathbf{Y}_u^\dagger\mathbf{Y}_d\right){}_p^r\delta_s^t\,, \\
        \mathcal{O}_{u^2leHD}^{(1-3)}: \quad & f(\mathbf{Y})^{rt}_{ps} = \delta_p^r\left(\mathbf{Y}_e\right){}_s^t + f_3\left(\mathbf{Y}_u^\dagger \mathbf{Y}_u\right){}_p^r\left(\mathbf{Y}_e\right){}_s^t\,,\\
        \mathcal{O}_{q^2leHD}^{(1-6)}: \quad & f(\mathbf{Y})^{rt}_{ps} = \delta_p^r\left(\mathbf{Y}_e\right){}_s^t + f_3\left(\mathbf{Y}_u \mathbf{Y}_u^\dagger\right){}_p^r\left(\mathbf{Y}_e\right){}_s^t\,,\\
    \end{align}
    otherwise, the expansions are dominant by the linear terms
    \begin{align}
        \mathcal{O}_{qud^2HD}^{(1-6)}\,, \mathcal{O}_{qul^2HD}^{(1-6)}\,, \mathcal{O}_{que^2HD}^{(1-3)}: \quad & f(\mathbf{Y})^{rt}_{ps} = \left(\mathbf{Y}_u\right){}_p^r\delta_s^t\,,\\
        \mathcal{O}_{qdl^2HD}^{(1-6)}\,,\mathcal{O}_{qde^2HD}^{(1-3)}: \quad & f(\mathbf{Y})^{rt}_{ps} = \left(\mathbf{Y}_d\right){}_p^r\delta_s^t\,,\\
        \mathcal{O}_{d^2leHD}^{(1-3)}: \quad & f(\mathbf{Y})^{rt}_{ps} = \delta_p^r\left(\mathbf{Y}_e\right){}_s^t\,.
    \end{align}
\end{itemize}

Thus all the MFV operators of the (lepton number/baryon-number conserving) dimension-8 operators are presented. Apart from the same flavor structures with the dimension-6 case, new structures appear in the classes $\psi^4\phi D$ and $\psi^4\phi^2$. We summarize the result in Tab.~\ref{tab:dim8}. Similarly, the number of the CP-violating phases (655) is reduced compared to the number without flavor symmetry (11777).

\begin{table}
\renewcommand{\arraystretch}{1.8}
\begin{center}
    \begin{tabular}{|c|c|c||c|c||c|c|}
\hline
\multirow{2}{*}{Class} & \multicolumn{2}{c||}{$n$} & \multicolumn{2}{c||}{$n_\mathbf{Y}$} & \multicolumn{2}{c|}{$n_Y$} \\ 
\cline{2-7}
& CP-even & CP-violating & CP-even & CP-violating & CP-even & CP-violating \\
\hline
bosonic sectors & 54 & 35 & 54 & 35 & 54 & 35 \\
\hline
$\psi^2\phi^5$ & 27 & 21 & 4 & 4 & 21 & 21 \\
\hline
$\psi^2\phi^4D$ & 69 & 33 & 14 & 4 & 54 & 21 \\
\hline
$\psi^2\phi^3D^2$ & 162 & 126 & 24 & 24 & 126 & 126 \\
\hline
$\psi^2\phi^2D^3$ & 93 & 33 & 21 & 1 & 63 & 21 \\
\hline
$\psi^2X\phi^3$ & 99 & 81 & 15 & 15 & 81 & 81 \\
\hline
$\psi^2X\phi^2D$ & 414 & 330 & 86 & 86 & 258 & 258 \\
\hline
$\psi^2X\phi D^2$ & 216 & 180 & 33 & 33 & 180 & 180 \\
\hline
$\psi^2X^2\phi$ & 432 & 372 & 67 & 67 & 372 & 372 \\
\hline
$\psi^2X^2D$ & 291 & 183 & 62 & 28 & 186 & 117 \\
\hline
$\psi^4D^2$ & 2205 & 1086 & 98 & 6 & 840 & 462 \\
\hline
$\psi^4\phi^2$ & 2895 & 1566 & 94 & 18 & 1032 & 561 \\
\hline
$\psi^4\phi D$ & 5454 & 3387 & 175 & 175 & 2559 & 2559 \\
\hline
$\psi^4X$ & 6462 & 4344 & 159 & 159 & 2043 & 2043 \\
\hline
\multirow{2}{*}{Total} & 18873 & 11777 & 906 & 655 & 7869 & 6857 \\
\cline{2-7}
& \multicolumn{2}{c||}{30650} & \multicolumn{2}{c||}{1561} & \multicolumn{2}{c|}{14726}\\
\hline
    \end{tabular}
\end{center}
\caption{The CP properties of the SMEFT dimension-8 effective operators, where $n$ corresponds to the operators without flavor symmetry, $n_\mathbf{Y}$ corresponds to the operators with the spurions $\mathbf{Y}_\psi$, and $n_Y$ corresponds to the cases when $\mathbf{Y}_\psi$'s become matrices $Y_\psi$'s. 
The CP-odd bosonic operators can also be found in Tab.~\ref{tab:odd_bosonic}.}
\label{tab:dim8}
\end{table}

\subsection{$U(2)^5$ Symmetry}

Adopting the MFV hypothesis, the effective operators of the $U(2)^5$ symmetry can also be obtained by the spurion technique. Different from the $U(3)^5$ symmetry, each Yukawa matrix can contribute more than 1 spurion. We list the spurions and their representations as follows
\begin{align}
    \Delta_u &\in (\mathbf{2}\,,\overline{\mathbf{2}}\,,\mathbf{1}\,,\mathbf{1}\,,\mathbf{1})\,, \\
    \Delta_d &\in (\mathbf{2}\,,\mathbf{1}\,,\overline{\mathbf{2}}\,,\mathbf{1}\,,\mathbf{1})\,, \\
    \Delta_e &\in (\mathbf{1}\,,\mathbf{1}\,,\mathbf{1}\,,\mathbf{2}\,,\overline{\mathbf{2}})\,, \\
    V_q &\in (\mathbf{2}\,,\mathbf{1}\,,\mathbf{1}\,,\mathbf{1}\,,\mathbf{1})\,, \\
    V_l &\in (\mathbf{1}\,,\mathbf{1}\,,\mathbf{1}\,,\mathbf{2}\,,\mathbf{1})\,, 
\end{align}
where $\Delta$'s are called the tensor spurions and $V$'s are called the vector spurions. Thus 5 spurions are needed in the $U(2)^5$ flavor symmetry and their power-counting are different~\cite{Faroughy:2020ina},
\begin{align}
    \Delta_u\,,\Delta_d\,,\Delta_e \sim 10^{-2}\,,\quad V_q\,,V_l \sim 10^{-1}\,,\label{eq:power_counting_u25}
\end{align}

Compared to the $U(3)^5$ flavor symmetry, there are two differences in the effective operators under the $U(2)^5$ symmetry. Firstly, the existing operators in the $U(3)^5$ symmetry would split into the ones that are independent individually, secondly, the vanishing operators in the $U(3)^5$ symmetry may exist. 
Because most of the flavor structures of the dimension-6 and dimension-8 operators are the same, we only discuss the effective operators of the classes $\psi^4\phi D$ and $\psi^4\phi^2$ here, since other cases are the same as the dimension-6 cases and can be found in Ref.~\cite{Faroughy:2020ina}

\subsubsection*{$\psi^4\phi^2$ Class}
The operators with the new flavor structures have been listed in Eq.~\eqref{eq:new_flavor_psi4phi2}. Considering the operators $\mathcal{O}_{q^2u^2H^2}^{'(1,2)}$ as an example, every single operator splits into several ones in the $U(2)^5$ symmetry. Including the spurions and truncating at $10^{-4}$ according to the power-counting in Eq.~\eqref{eq:power_counting_u25}, we list all the operators in terms of their powers,
\begin{align}
\mathcal{O}_{q^2u^2H^2}^{'(1,2)}:\quad & \notag \\
1:\quad & a_1 \epsilon_{ps}\epsilon^{rt}(\overline{q}^p u_r)(\overline{q}^s u_t) + a_2 (\overline{q}^3 u_3)(\overline{q}^3 u_3) \,,\\
10^{-1}:\quad & b_1\left(V_q\right)_p (\overline{q}^p u_3)(\overline{q}^3 u_3) \,,\\
10^{-2}:\quad & c_1\left(V_q\right)_p\left(V_q\right)_s (\overline{q}^p u_3)(\overline{q}^s u_3) \\
& +d_1\left(\Delta_u\right)_p^r (\overline{q}^p u_r)(\overline{q}^3 u_3) + d_2\left(\Delta_u\right)_p^r (\overline{q}^p u_3)(\overline{q}^3 u_r) \,,\\
10^{-3}:\quad & e_1\left(\Delta_u\right)_p^r \left(V_q\right)_s(\overline{q}^p u_r)(\overline{q}^s u_3) + e_2\left(\Delta_u\right)_s^r \left(V_q\right)_p(\overline{q}^p u_r)(\overline{q}^s u_3)\,,
\end{align}
where $a_i\,,b_i\,,c_i\,,d_i\,,e_i$'s are the Wilson coefficients, and for simplicity, we have dropped the quantities other than the fermions in each bilinear. In particular, the operator
\begin{equation}
    a_1 \epsilon_{ps}\epsilon^{rt}(\overline{q}^p u_r)(\overline{q}^s u_t) \,,
\end{equation}
is of power 1, which is composed of antisymmetric tensor $\epsilon$ and is absent in the $U(3)^5$ symmetry.

The effective operators of $\mathcal{O}_{q^2d^2H^2}^{'(1,2)}\,,\mathcal{O}_{l^2e^2H^2}\,,\mathcal{O}_{qdleH^2}^{'(1,2)}$ under the $U(2)^5$ symmetry can be obtained similarly
\begin{align}
\mathcal{O}_{q^2d^2H^2}^{'(1,2)}:\quad & \notag \\
1:\quad & a_1 \epsilon_{ps}\epsilon^{rt}(\overline{q}^p d_r)(\overline{q}^s d_t) + a_2 (\overline{q}^3 d_3)(\overline{q}^3 d_3) \,,\\
10^{-1}:\quad & b_1\left(V_q\right)_p (\overline{q}^p d_3)(\overline{q}^3 d_3) \,,\\
10^{-2}:\quad & c_1\left(V_q\right)_p\left(V_q\right)_s (\overline{q}^p d_3)(\overline{q}^s d_3) \\
& +d_1\left(\Delta_d\right)_p^r (\overline{q}^p d_r)(\overline{q}^3 d_3) + d_2\left(\Delta_d\right)_p^r (\overline{q}^p d_3)(\overline{q}^3 d_r) \,,\\
10^{-3}:\quad & e_1\left(\Delta_d\right)_p^r \left(V_q\right)_s(\overline{q}^p d_r)(\overline{q}^s d_3) + e_2\left(\Delta_d\right)_s^r \left(V_q\right)_p(\overline{q}^p d_r)(\overline{q}^s d_3)\,,
\end{align}
\begin{align}
\mathcal{O}_{l^2e^2H^2}:\quad & \notag \\
1:\quad & a_1 \epsilon_{ps}\epsilon^{rt}(\overline{l}^p e_r)(\overline{l}^s e_t) + a_2 (\overline{l}^3 e_3)(\overline{l}^3 e_3) \,,\\
10^{-1}:\quad & b_1\left(V_l\right)_p (\overline{l}^p e_3)(\overline{l}^3 e_3) \,,\\
10^{-2}:\quad & c_1\left(V_l\right)_p\left(V_l\right)_s (\overline{l}^p e_3)(\overline{l}^s e_3) \\
& +d_1\left(\Delta_e\right)_p^r (\overline{l}^p e_r)(\overline{l}^3 e_3) + d_2\left(\Delta_e\right)_p^r (\overline{l}^p e_3)(\overline{l}^3 e_r) \,,\\
10^{-3}:\quad & e_1\left(\Delta_e\right)_p^r \left(V_l\right)_s(\overline{l}^p e_r)(\overline{l}^s e_3) + e_2\left(\Delta_e\right)_s^r \left(V_l\right)_p(\overline{l}^p e_r)(\overline{l}^s e_3)\,,
\end{align}
\begin{align}
\mathcal{O}_{qdleH^2}^{'(1,2)}:\quad & \notag \\
1:\quad & a_1 (\overline{q}^3 d_3)(\overline{l}^3 e_3) \,,\\
10^{-1}:\quad & b_1\left(V_q\right)_p (\overline{q}^p d_3)(\overline{l}^3 e_3) + b_2\left(V_l\right)_p (\overline{q}^3 d_3)(\overline{l}^p e_3) \,,\\
10^{-2}:\quad & c_1\left(V_q\right)_p\left(V_l\right)_s (\overline{q}^p d_3)(\overline{l}^s e_3) \\
& +d_1\left(\Delta_d\right)_p^r (\overline{q}^p d_r)(\overline{l}^3 e_3) + d_2\left(\Delta_e\right)_s^t (\overline{q}^3 d_3)(\overline{l}^s e_t) \,,\\
10^{-3}:\quad & e_1\left(\Delta_d\right)_p^r \left(V_l\right)_s(\overline{q}^p d_r)(\overline{l}^s e_3) + e_2\left(\Delta_e\right)_s^t \left(V_l\right)_p(\overline{q}^p d_3)(\overline{l}^s e_t)\,.
\end{align}
Thus the absent operators $\mathcal{O}_{q^2d^2H^2}^{'(1,2)}\,,\mathcal{O}_{l^2e^2H^2}\,,\mathcal{O}_{qdleH^2}^{'(1,2)}$ under the $U(3)^5$
symmetry are nonvanishing under the $U(2)^5$ symmetry.

For the operators $\mathcal{O}_{q^2udH^2}^{'(1,2)}$ there are 
\begin{align}
    \mathcal{O}_{q^2udH^2}^{'(1,2)}:\quad & \notag \\
    1:\quad & a_1 (\overline{q}^3 u_3)(\overline{d}^3 q_3) + a_2\delta_p^r (\overline{q}^p u_3)(\overline{d}^3 q_r) \,,\\
    10^{-1}:\quad & b_1 \left(V_q\right)_p (\overline{q}^p u_3)(\overline{d}^3 q_3) + b_2\left(V_q^\dagger\right)^r(\overline{q}^3 u_3)(\overline{d}^3 q_r) \,,\\
    10^{-2}:\quad & c_1 \left(V_q\right)_p\left(V_q^\dagger\right)^r(\overline{q}^p u_3)(\overline{d}^3 q_r) \,,\\
    & +d_1 \left(\Delta_u\right)_p^r (\overline{q}^p u_r)(\overline{d}^3 q_3) + d_2\left(\Delta_d\right)_s^t (\overline{q}^3 u_3)(\overline{d}^s q_t) \,,\\
    10^{-3}:\quad & e_1 \left(\Delta_u\right)_p^r \left(V_q^\dagger\right)^t(\overline{q}^p u_r)(\overline{d}^3 q_t) + e_2 \delta_p^t\left(V_q^\dagger\Delta_u\right)^r (\overline{q}^p u_r)(\overline{d}^3 q_t) \notag \\
    &+ e_3\left(\Delta_d^\dagger\right)_s^t \left(V_q\right)_p (\overline{q}^p u_3)(\overline{d}^s q_t) + e_4\delta_p^t \left(\Delta_d^\dagger V_q\right)_s (\overline{q}^p u_3)(\overline{d}^s q_t)\,,
\end{align}
while for the $\mathcal{O}'_{l^2udH^2}$, the operators are
\begin{align}
    \mathcal{O}'_{l^2udH^2}:\quad & \notag \\
    1:\quad & a_1 (\overline{l}^3l_3)(\overline{u}^3 d_3) + a_2 \delta_p^r(\overline{l}^pl_r)(\overline{u}^3 d_3) \,,\\
    10^{-1}:\quad & b_1 \left(V_l\right)_p (\overline{l}^pl_3)(\overline{u}^3 d_3) + b_2 \left(V_l^\dagger\right)^r (\overline{l}^3l_r)(\overline{u}^3 d_3) \,,\\
    10^{-2}:\quad & c_1\left(V_l\right)_p \left(V_l^\dagger\right)^r(\overline{l}^pl_r)(\overline{u}^3 d_3) \,,\\
    10^{-3}:\quad & d_1 \delta_p^r \left(\Delta_u^\dagger V_q\right)_s (\overline{l}^pl_r)(\overline{u}^s d_3) + d_2 \delta_p^r\left(V_q^\dagger\Delta_d\right)^t(\overline{l}^pl_r)(\overline{u}^3 d_t) \notag \\
    & + d_3 \left(\Delta_u^\dagger V_q\right)_s (\overline{l}^3l_3)(\overline{u}^s d_3) +  d_4\left(V_q^\dagger\Delta_d\right)^t(\overline{l}^3l_3)(\overline{u}^3 d_t)
\end{align}

\subsubsection*{$\psi^4\phi D$ Class}

Because of the nonvanishing hypercharge of the Higgs field, the flavor structures of all the operators of the $\psi^4\phi D$ class are different from the dimension-6 cases.

Firstly, we notice the operators $\mathcal{O}_{q^3dHD}^{(1-6)}\,,\mathcal{O}_{q^3dHD}^{(1-6)}\,,\mathcal{O}_{l^3eHD}^{(1-3)}$ are similar since there are 3 repeated left-handed fermions, so their flavor structures are similar, we present explicitly the effective operators of $\mathcal{O}_{q^3dHD}^{(1-6)}$ under the $U(2)^5$ symmetry here,

\begin{align}
    \mathcal{O}_{q^3dHD}^{(1-6)}:\quad & \notag \\
    1:\quad & a_1(\overline{q}^3 q_3)(\overline{q}^3 u_3) + a_2 \delta_p^r(\overline{q}^p q_r)(\overline{q}^3 u_3) + a_3 \delta_p^r (\overline{q}^3 q_r)(\overline{q}^p u_3) + a_4 \epsilon_{ps} (\overline{q}^p q_3)(\overline{q}^s u_3) \,,\\
    10^{-1}:\quad & b_1 \left(V_q\right)_p (\overline{q}^p q_3)(\overline{q}^3 u_3) + b_2 \left(V_q\right)_p (\overline{q}^3 q_3)(\overline{q}^p u_3) + b_3 \left(V_q^\dagger\right)^r (\overline{q}^3 q_r)(\overline{q}^3 u_3) \notag \\
    &+ b_4 \left(V_q\right)_p\delta^r_s (\overline{q}^p q_r)(\overline{q}^s u_3) + b_5 \left(V_q\right)_p\delta^r_s (\overline{q}^s q_r)(\overline{q}^p u_3) + b_6 \left(V_q^\dagger\right)^r \epsilon_{ps}(\overline{q}^p q_r)(\overline{q}^s u_3) \,,\\
    10^{-2}:\quad & c_1 \left(V_q\right)_p \left(V_q^\dagger\right)^r (\overline{q}^p q_r)(\overline{q}^3 u_3) + c_2 \left(V_q\right)_p \left(V_q^\dagger\right)^r (\overline{q}^3 q_r)(\overline{q}^p u_3) + c_3 \left(V_q\right)_p \left(V_q\right)_s (\overline{q}^p q_3)(\overline{q}^s u_3) \\
    & +d_1 \left(\Delta_u\right)_s^t (\overline{q}^3 q_3)(\overline{q}^s u_t) + d_2 \left(\Delta_u\right)_p^t (\overline{q}^p q_3)(\overline{q}^3 u_t) + d_3 \left(V_q\right)_p\left(V_q^\dagger\right)^r \left(V_q\right)_s (\overline{q}^p q_r)(\overline{q}^s u_3) \notag \\
    &+ d_4 \left(\Delta_u\right)_s^t\delta_p^r (\overline{q}^p q_r)(\overline{q}^s u_t) + d_5 \left(\Delta_u\right)_p^t\delta_s^r (\overline{q}^p q_r)(\overline{q}^s u_t) \,,\\
    10^{-3}:\quad & e_1 \delta_p^r \left(V_q^\dagger \Delta_u\right)^t (\overline{q}^p q_r)(\overline{q}^3 u_t) + e_2 \left(\Delta_u\right)_p^t \left(V_u^\dagger\right)^r (\overline{q}^p q_r)(\overline{q}^3 u_t) + e_3 \left(\Delta_u\right)_s^t \left(V_q\right)_p (\overline{q}^p q_s)(\overline{q}^s u_t) \notag \\
    &+ e_4\left(\Delta_u\right)_p^t \left(V_q\right)_s(\overline{q}^p q_s)(\overline{q}^s u_t) + e_5 \delta_s^r \left(V_q^\dagger \Delta_u\right)^t (\overline{q}^3 q_r)(\overline{q}^s u_t) + e_6 \left(\Delta_u\right)_s^t \left(V_q^\dagger\right)^r (\overline{q}^3 q_r)(\overline{q}^s u_t) \notag \\
    &+ e_7 \left(V_q^\dagger \Delta_u\right)^t (\overline{q}^3 q_3)(\overline{q}^3 u_t)\,.
\end{align}
The other 2 cases can be obtained by the spurion replacements.
Secondly, the operators $\mathcal{O}_{qd^3HD}^{(1-6)}\,,\mathcal{O}_{qu^3HD}^{(1-6)}\,,\mathcal{O}_{le^3HD}^{(1-3)}$ have the same flavor structures since there are 3 repeated right-handed fermions. We present the explicit operators of only one of them, $\mathcal{O}_{qu^3HD}^{(1-6)}$,
\begin{align}
    \mathcal{O}_{qu^3HD}^{(1-6)}:\quad & \notag \\
    1:\quad & a_1\delta_p^r (\overline{d}^pd_r)(\overline{q}^3 d_3) + a_2\delta_p^t (\overline{d}^pd_3)(\overline{q}^3 d_t) + a3(\overline{d}^3d_3)(\overline{q}^3 d_3) + a_4 \epsilon^{rt}(\overline{d}^3d_r)(\overline{q}^3 d_t) \,,\\
    10^{-1}: \quad & b_1 \delta_p^r \left(V_q\right)_s (\overline{d}^pd_r)(\overline{q}^s d_3) + b_2 \delta_p^t \left(V_q\right)_s (\overline{d}^pd_3)(\overline{q}^s d_t) + b_3 \epsilon^{rt} \left(V_q\right)_s (\overline{d}^3d_r)(\overline{q}^s d_t) \notag \\
    &+ b_4  \left(V_q\right)_s (\overline{d}^3d_3)(\overline{q}^s d_3) \,,\\
    10^{-2}:\quad & c_1 \delta_p^r \left(\Delta_d\right)_s^t (\overline{d}^p d_r)(\overline{q}^s d_t) + c_2 \delta_p^t \left(\Delta_d\right)_s^r (\overline{d}^p d_r)(\overline{q}^s d_t) + c_3 \left(\Delta_d\right)_s^t (\overline{d}^3 d_3)(\overline{q}^s d_t) \notag \\
    &+ c_4 \left(\Delta_d\right)_s^r (\overline{d}^3 d_r)(\overline{q}^s d_3) \,,\\
    10^{-3}:\quad & d_1 \delta_p^r \left(V_q^\dagger \Delta_d\right)^t (\overline{d}^p d_r)(\overline{q}^3 d_t) + d_2 \delta_p^t \left(V_q^\dagger \Delta_d\right)^r (\overline{d}^p d_r)(\overline{q}^3 d_t) + d_3 \left(\Delta_d^\dagger V_q\right)_p (\overline{d}^p d_3)(\overline{q}^3 d_3) \notag \\
    & + d_4 \left(\Delta_d V_q^\dagger\right)^r (\overline{d}^3 d_r)(\overline{q}^3 d_3) + d_5 \left(V_q^\dagger \Delta_d\right)^t (\overline{d}^3 d_3)(\overline{q}^3 d_t)\,.
\end{align}
Next, we identify the operators $\mathcal{O}_{qu^2dHD}^{(1-6)}\,,\mathcal{O}_{qud^2HD}^{(1-6)}\,,\mathcal{O}_{u^2leHD}^{(1-3)}\,,\mathcal{O}_{d^2leHD}^{(1-3)}\,,\mathcal{O}_{que^2HD}^{(1-3)}\,,\mathcal{O}_{e^2qdHD}^{(1-3)}$ are of the same flavor structures, thus we only present the operators of $\mathcal{O}_{qu^2dHD}^{(1-6)}$ explicitly here,
\begin{align}
    \mathcal{O}_{qu^2dHD}^{(1-6)}:\quad & \notag \\
    1:\quad & a_1 \delta_p^r (\overline{u}^p u_r)(\overline{q}^3 d_3) + a_2 (\overline{u}^3 u_3)(\overline{q}^3 d_3) \,,\\
    10^{-1}:\quad & b_1 \delta_p^r \left(V_q\right)_s (\overline{u}^p u_r)(\overline{q}^s d_3) + b_2\left(V_q\right)_s (\overline{u}^3 u_3)(\overline{q}^s d_3) \,,\\
    10^{-2}:\quad & c_1 \delta_p^r \left(\Delta_d\right)_s^t (\overline{u}^p u_r)(\overline{q}^s d_t) + c_2 \left(\Delta_d\right)_s^t (\overline{u}^3 u_3)(\overline{q}^s d_t) + c_3 \left(\Delta_u\right)_s^r (\overline{u}^3 u_r)(\overline{q}^s d_3) \,,\\
    10^{-3}:\quad & d_1 \delta_p^r \left(V_q^\dagger \Delta_d\right)^t (\overline{u}^p u_r)(\overline{q}^3 d_t) + d_2 \left(V_q^\dagger \Delta_d\right)^t (\overline{u}^3 u_3)(\overline{q}^3 d_t) + d_3 \left(\Delta_u^\dagger V_q\right)_p  (\overline{u}^p u_3)(\overline{q}^3 d_3) \notag \\
    &+ d_4 \left(V_q^\dagger \Delta_u\right)^r d_1 (\overline{u}^3 u_r)(\overline{q}^3 d_3)
\end{align}
Lastly the operators $\mathcal{O}_{q^2leHD}^{(1-6)}\,,\mathcal{O}_{l^2quHD}^{(1-6)}\,,\mathcal{O}_{qdl^2HD}^{(1-6)}$ are of the same flavor structures under the $U(2)^5$ flavor symmetry, thus we present the operators  of $\mathcal{O}_{q^2leHD}^{(1-6)}$ here,
\begin{align}
    \mathcal{O}_{q^2leHD}^{(1-6)}:\quad & \notag \\
    1:\quad & a_1 \delta_p^r (\overline{q}^p q_r)(\overline{l}^3 e_3) + a_2 (\overline{q}^3 q_3)(\overline{l}^3 e_3) \,,\\
    10^{-1}:\quad & b_1 \delta_p^r \left(V_l\right)_s (\overline{q}^p q_r)(\overline{l}^s e_3) + b_2 \left(V_l\right)_s (\overline{q}^3 q_3)(\overline{l}^3 e_3) + b_3 \left(V_l\right)_p (\overline{q}^p q_3)(\overline{l}^3 e_3) + b_4 \left(V_l^\dagger\right)^r (\overline{q}^3 q_r) (\overline{l}^3 e_3) \,,\\
    10^{-2}:\quad & c_1 \left(V_q\right)_p \left(V_q^\dagger\right)^r (\overline{q}^p q_r) (\overline{l}^3 e_3) + c_2 \left(V_q\right)_p \left(V_l\right)_s (\overline{q}^p q_3) (\overline{l}^s e_3) + c_3 \left(V_q^\dagger\right)^r \left(V_l\right)_s (\overline{q}^3 q_r) (\overline{l}^s e_3) \,,\\
    10^{-3}:\quad & d_1 \left(V_q\right)_p \left(V_q^\dagger\right)^r \left(V_l\right)_s(\overline{q}^p q_r) (\overline{l}^s e_3) + d_2 \delta_p^r \left(\Delta_e\right)_r^s (\overline{q}^p q_r) (\overline{l}^s e_t) + d_3 \left(\Delta_e\right)_r^s (\overline{q}^3 q_3) (\overline{l}^s e_t) \\
    & +e_1 \delta_p^r \left(V_l^\dagger \Delta_e\right)^t (\overline{q}^p q_r) (\overline{l}^3 e_t) + e_2 \left(V_l^\dagger \Delta_e\right)^t (\overline{q}^3 q_3) (\overline{l}^3 e_t) + e_3 \left(V_q\right)_p \left(\Delta_e\right)_s^t (\overline{q}^p q_3) (\overline{l}^s e_t)\notag \\
    &+ e_4 \left(V_q^\dagger\right)^r \left(\Delta_e\right)_s^t (\overline{q}^s q_r) (\overline{l}^s e_t)\,.
\end{align}

In summary, the $U(2)^5$ as a subgroup of $U(3)^5$, is a less-constrained flavor symmetry, thus the involving operators are more abundant than the $U(3)^5$ case, which means richer phenomenologies. Besides, there are two advantages of the $U(2)^5$ symmetry compared to the $U(3)^5$, firstly, the split of the fermion triplet makes the power-counting more precise, secondly, more spurions make the flavor-changed process occur also in the leptonic sector, which is banned in the $U(3)^5$ symmetry. No matter the $U(3)^5$ or $U(2)^5$ symmetry, we adopt the MFV hypothesis, which is consistent with the observations and protects the cutoff scale of the SMEFT from being pushed too high.

\section{Conclusion}
\label{sec:con}

The flavor symmetry can be used to organize and classify the large number of operators of the SMEFT.
When the flavor symmetries are absent, the operators are organized by their flavor structures and CP properties. The flavor structures can be decomposed into irreducible representations of the symmetric groups, which is closely related to the Lorentz and gauge structures of the operators. 
The CP violation of the SMEFT is important in the research of new physics. In general, the CP-odd operators are not equivalent to CP-violating operators, and only the ones invariant under the rephasing symmetry are CP-violating, since there is no field redefinition to eliminate the complex phases of the associated Wilson coefficients. 
We present the complete CP-even and CP-odd operators at dimension 6 and dimension 8, including both the bosonic operators and the fermionic operators. 

The flavor symmetries such as $U(3)^5$ and $U(2)^5$ can reduce the independent parameters in principle, and affect the CP violation in two aspects. Firstly, the flavor symmetries reduce the independent operators, secondly, the flavor symmetries change the rephasing symmetry. Nevertheless, since the rephasing symmetry is a subgroup of the flavor symmetry, it is guaranteed by the flavor symmetry, thus the CP-odd and the CP-violating operators are always equivalent with flavor symmetry. 
However, the single assumption of the $U(3)^5$ symmetry does not change the CP-violating phases. At the same time, the subgroup $U(2)^5$ could be different.
At the same time, adding flavor symmetries, the flavor-invariant method can also be used to express the CP-violating phases by the flavor invariants. As discussed in this paper, the tensors obtained by the Young tableaux are equivalent to the traces obtained by the Cayley-Hamilton relations. Thus the flavor invariants can be obtained by the Young tensor method analytically, which can be applied to the constructions of the high-dimension flavor invariants of the SMEFT.

To reduce the parameters and classify the operators furthermore, 
we consider the MFV hypothesis to be consistent with the experiment results. 
We analyze the flavor structures and the CP properties of the SMEFT operators at dimensions 6 and 8 with the flavor symmetries $U(3)^5$. We use the spurion method to present their flavor structures explicitly. In dimension 6, we find 26 independent CP-violating phases, which is smaller than the 705 without flavor symmetry. In dimension 8, there are new flavor structures in the classes $\psi^4\phi^2$ and $\psi^4\phi D$. We find 655 independent CP-violating phases, and it is also smaller than the 11777 without the flavor symmetry. For the two classes $\psi^4\phi^2$ and $\psi^4\phi D$ with the new flavor structures in dimension 8, we also present their flavor structures assuming the $U(2)^5$ symmetry and its minimal breaking, which include more operators than the $U(3)^5$ symmetry.

The flavor symmetry and the MFV hypothesis are used to reduce and classify the effective operators of the SMEFT, and their relation to the CP violations presented in this paper is general. 
The Young tensor method can be used to obtain the flavor invariants of the SMEFT to extract the CP-violating phases in a basis-invariant way.
The analysis here can be extended to other flavor symmetries, e.g.~\cite{Kobayashi:2021pav,Calo:2022jqv,Kobayashi:2023zzc,Loisa:2024xuk,Palavric:2024gvu}. Based on the analysis, we can classify and organize the large number of SMEFT operators systematically, and reduce the parameters in the related experiments on the LHC.

\section*{Acknowledgement}
We thank Luo-Jia Kang for the useful discussion.
This work is supported by the National
Science Foundation of China under Grants No. 12347105, No. 12375099, and No. 12047503, and
the National Key Research and Development Program of China Grant No. 2020YFC2201501, No.
2021YFA0718304.

\newpage

\begin{appendix}

\section{Dimension-6 Operators}
\label{app:dim6}

Here the dimension-6 SMEFT operators are adopted from the Ref.~\cite{Grzadkowski:2010es}. , and only the baryon-number and lepton-number conserving fermionic operators are presented. In particular, we have simplified the operators by writing the right-handed fermions as
\begin{equation}
    u_R\,,d_R\,,e_R \rightarrow u\,, d\,, e\,.
\end{equation}

\begin{center}

\begin{minipage}[t]{4cm}

    \renewcommand{\arraystretch}{1.5}
    \begin{tabular}[t]{c|c}
    \multicolumn{2}{c}{$1: \psi^2H^3 + \hbox{h.c.}$} \\
    \hline
    $\mathcal{O}_{eH}$           & $(H^\dag H)(\bar l^p e_r H)$ \\
    $\mathcal{O}_{uH}$          & $(H^\dag H)(\bar q^p u_r \widetilde H )$ \\
    $\mathcal{O}_{dH}$           & $(H^\dag H)(\bar q^p d_r H)$\\
    \hline
    \end{tabular}
\end{minipage}
\quad
\begin{minipage}[t]{5.2cm}
    \renewcommand{\arraystretch}{1.5}
    \begin{tabular}[t]{c|c}
    \multicolumn{2}{c}{$2:\psi^2 XH+\hbox{h.c.}$} \\
    \hline
    $\mathcal{O}_{eW}$      & $(\bar l^p \sigma^{\mu\nu} e_r) \tau^I H W_{\mu\nu}^I$ \\
    $\mathcal{O}_{eB}$        & $(\bar l^p \sigma^{\mu\nu} e_r) H B_{\mu\nu}$ \\
    $\mathcal{O}_{uG}$        & $(\bar q^p \sigma^{\mu\nu} \lambda^A u_r) \widetilde H \, G_{\mu\nu}^A$ \\
    $\mathcal{O}_{uW}$        & $(\bar q^p \sigma^{\mu\nu} u_r) \tau^I \widetilde H \, W_{\mu\nu}^I$ \\
    $\mathcal{O}_{uB}$        & $(\bar q^p \sigma^{\mu\nu} u_r) \widetilde H \, B_{\mu\nu}$ \\
    $\mathcal{O}_{dG}$        & $(\bar q^p \sigma^{\mu\nu} \lambda^A d_r) H\, G_{\mu\nu}^A$ \\
    $\mathcal{O}_{dW}$         & $(\bar q^p \sigma^{\mu\nu} d_r) \tau^I H\, W_{\mu\nu}^I$ \\
    $\mathcal{O}_{dB}$        & $(\bar q^p \sigma^{\mu\nu} d_r) H\, B_{\mu\nu}$ \\
\hline    
\end{tabular}
\end{minipage}
\quad
\begin{minipage}[t]{5.4cm}
    \renewcommand{\arraystretch}{1.5}
    \begin{tabular}[t]{c|c}
    \multicolumn{2}{c}{$3:\psi^2H^2 D$} \\
    \hline
    $\mathcal{O}_{H l}^{(1)}$      & $(H^\dag i\overleftrightarrow{D}_\mu H)(\bar l^p \gamma^\mu l_r)$\\
    $\mathcal{O}_{H l}^{(3)}$      & $(H^\dag i\overleftrightarrow{D}^I_\mu H)(\bar l^p \tau^I \gamma^\mu l_r)$\\
    $\mathcal{O}_{H e}$            & $(H^\dag i\overleftrightarrow{D}_\mu H)(\bar e^p \gamma^\mu e_r)$\\
    $\mathcal{O}_{H q}^{(1)}$      & $(H^\dag i\overleftrightarrow{D}_\mu H)(\bar q^p \gamma^\mu q_r)$\\
    $\mathcal{O}_{H q}^{(3)}$      & $(H^\dag i\overleftrightarrow{D}^I_\mu H)(\bar q^p \tau^I \gamma^\mu q_r)$\\
    $\mathcal{O}_{H u}$            & $(H^\dag i\overleftrightarrow{D}_\mu H)(\bar u^p \gamma^\mu u_r)$\\
    $\mathcal{O}_{H d}$            & $(H^\dag i\overleftrightarrow{D}_\mu H)(\bar d^p \gamma^\mu d_r)$\\
    $\mathcal{O}_{H u d}$          & $i(\widetilde H ^\dag D_\mu H)(\bar u^p \gamma^\mu d_r)$\\
    \hline
    \end{tabular}
\end{minipage}

\vspace{0.25cm}

\begin{minipage}[t]{4.75cm}
    \renewcommand{\arraystretch}{1.5}
    \begin{tabular}[t]{c|c}
    \multicolumn{2}{c}{$4: (\bar LL)(\bar LL)$} \\
    \hline
    $\mathcal{O}_{ll}$        & $(\bar l^p \gamma_\mu l_r)(\bar l^s \gamma^\mu l_t)$ \\
    $\mathcal{O}_{qq}^{(1)}$  & $(\bar q^p \gamma_\mu q_r)(\bar q^s \gamma^\mu q_t)$ \\
    $\mathcal{O}_{qq}^{(3)}$  & $(\bar q^p \gamma_\mu \tau^I q_r)(\bar q^s \gamma^\mu \tau^I q_t)$ \\
    $\mathcal{O}_{lq}^{(1)}$                & $(\bar q^p\gamma^\mu q_r)(\bar l^s \gamma_\mu l_t)$ \\
    $\mathcal{O}_{lq}^{(3)}$                & $(\bar q^p \gamma^\mu \tau^I q_r)(\bar l^s \gamma_\mu \tau^I l_t)$ \\
    \hline
    \end{tabular}
\end{minipage}
\quad
\begin{minipage}[t]{5.25cm}
    \renewcommand{\arraystretch}{1.5}
    \begin{tabular}[t]{c|c}
    \multicolumn{2}{c}{$4:(\bar RR)(\bar RR)$} \\
    \hline
    $\mathcal{O}_{ee}$               & $(\bar e^p \gamma_\mu e_r)(\bar e^s \gamma^\mu e_t)$ \\
    $\mathcal{O}_{uu}$        & $(\bar u^p \gamma_\mu u_r)(\bar u^s \gamma^\mu u_t)$ \\
    $\mathcal{O}_{dd}$        & $(\bar d^p \gamma_\mu d_r)(\bar d^s \gamma^\mu d_t)$ \\
    $\mathcal{O}_{eu}$                      & $(\bar u^p \gamma^\mu u_r)(\bar e^s \gamma_\mu e_t)$ \\
    $\mathcal{O}_{ed}$                      & $(\bar d^p\gamma^\mu d_r)(\bar e^s \gamma_\mu e_t)$ \\
    $\mathcal{O}_{ud}^{(1)}$                & $(\bar u^p \gamma_\mu u_r)(\bar d^s \gamma^\mu d_t)$ \\
    $\mathcal{O}_{ud}^{(8)}$                & $(\bar u^p \gamma_\mu \lambda^A u_r)(\bar d^s \gamma^\mu \lambda^A d_t)$ \\
    \hline
    \end{tabular}
\end{minipage}
\quad
\begin{minipage}[t]{4.75cm}
    \renewcommand{\arraystretch}{1.5}
    \begin{tabular}[t]{c|c}
    \multicolumn{2}{c}{$4:(\bar LL)(\bar RR)$} \\
    \hline
    $\mathcal{O}_{le}$               & $(\bar l^p \gamma_\mu l_r)(\bar e^s \gamma^\mu e_t)$ \\
    $\mathcal{O}_{lu}$               & $(\bar u^p \gamma^\mu u_r)(\bar l^s \gamma_\mu l_t)$ \\
    $\mathcal{O}_{ld}$               & $(\bar d^p \gamma^\mu d_r)(\bar l^s \gamma_\mu l_t)$ \\
    $\mathcal{O}_{qe}$               & $(\bar q^p \gamma_\mu q_r)(\bar e^s \gamma^\mu e_t)$ \\
    $\mathcal{O}_{qu}^{(1)}$         & $(\bar q^p \gamma_\mu q_r)(\bar u^s \gamma^\mu u_t)$ \\
    $\mathcal{O}_{qu}^{(8)}$         & $(\bar q^p \gamma_\mu \lambda^A q_r)(\bar u^s \gamma^\mu \lambda^A u_t)$ \\
    $\mathcal{O}_{qd}^{(1)}$ & $(\bar q^p \gamma_\mu q_r)(\bar d^s \gamma^\mu d_t)$ \\
    $\mathcal{O}_{qd}^{(8)}$ & $(\bar q^p \gamma_\mu \lambda^A q_r)(\bar d^s \gamma^\mu \lambda^A d_t)$\\
    \hline
    \end{tabular}
\end{minipage}

\vspace{0.25cm}

\begin{minipage}[t]{3.75cm}
    \renewcommand{\arraystretch}{1.5}
    \begin{tabular}[t]{c|c}
    \multicolumn{2}{c}{$4:(\bar LR)(\bar RL)+\hbox{h.c.}$} \\
    \hline
    $\mathcal{O}_{ledq}$ & $(\bar d^p q_{jr})(\bar l^{is} e_t)$ \\
    \hline
    \end{tabular}
\end{minipage}
\quad
\begin{minipage}[t]{5.5cm}
    \renewcommand{\arraystretch}{1.5}
    \begin{tabular}[t]{c|c}
    \multicolumn{2}{c}{$4:(\bar LR)(\bar L R)+\hbox{h.c.}$} \\
    \hline
    $\mathcal{O}_{quqd}^{(1)}$ & $(\bar q^{jp} u_r) \epsilon_{jk} (\bar q^{ks} d_t)$ \\
    $\mathcal{O}_{quqd}^{(8)}$ & $(\bar q^{jp} \lambda^A u_r) \epsilon_{jk} (\bar q^{ks} \lambda^A d_t)$ \\
    $\mathcal{O}_{lequ}^{(1)}$ & $(\bar q^{kp} u_r)\epsilon_{jk}(\bar l^{js} e_t)  $ \\
    $\mathcal{O}_{lequ}^{(3)}$ & $ (\bar q^{kp} \sigma^{\mu\nu} u_r)\epsilon_{jk} (\bar l^{js} \sigma_{\mu\nu} e_t)$ \\
    \hline
    \end{tabular}
\end{minipage}
    
\end{center}

\newpage

\newgeometry{left=1cm, right=1cm, top=3cm, bottom=3cm}

\section{Dimension-8 Operators}
\label{app:dim8}

The dimension-8 operators here are the reformulations of the ones obtained in Ref.~\cite{Li:2020gnx}. Similar to the dimension-6 case, only the lepton-number and baryon-number conserving fermionic operators are presented, and the same simplification is adopted,
\begin{equation}
    u_R\,,d_R\,,e_R \rightarrow u\,, d\,, e\,.
\end{equation}

\fontsize{10pt}{\baselineskip}\selectfont

\small
\renewcommand\arraystretch{1.4}

\begin{minipage}[c][4cm][t]{14em} 
\begin{tabular}{c|c}
\multicolumn{2}{c}{1: $\psi^2\phi^5$} \\
\hline
$\mathcal{O}_{quH^5}$ & $\bin{q^p}{u_r}{\tilde{H}}(H^\dagger H)^2$ \\
$\mathcal{O}_{qdH^5}$ & $\bin{q^p}{d_r}{H}(H^\dagger H)^2$ \\
$\mathcal{O}_{leH^5}$ & $\bin{l^p}{e_r}{H}(H^\dagger H)^2$ \\
\hline
\end{tabular}
\end{minipage}
   \quad
\begin{minipage}[c][4cm][t]{14em} 
\begin{tabular}{c|c||c|c}
\multicolumn{4}{c}{2: $\psi^2\phi^4D$} \\
\hline
$\mathcal{O}_{q^2H^4D}^{(1)}$ & $\bin{q^p}{q_r}{\gamma^\mu}(H^\dagger\lrd_\mu H)(H^\dagger H)$ & $\mathcal{O}_{l^2H^4D}^{(1)}$ & $\bin{l^p}{l_r}{\gamma^\mu}(H^\dagger\lrd_\mu H)(H^\dagger H)$ \\
$\mathcal{O}_{q^2H^4D}^{(2)}$ & $\bin{q^p}{q_r}{\tau^I\gamma^\mu}(H^\dagger \tau^I H)(H^\dagger\lrd_\mu H)$ & $\mathcal{O}_{l^2H^4D}^{(2)}$ & $\bin{l^p}{l_r}{\tau^I\gamma^\mu}(H^\dagger \tau^I H)(H^\dagger\lrd_\mu H)$\\
$\mathcal{O}_{q^2H^4D}^{(3)}$ & $\bin{q^p}{q_r}{\tau^I\gamma^\mu}(H^\dagger \tau^I H)D_\mu(H^\dagger H)$ & $\mathcal{O}_{l^2H^4D}^{(3)}$ & $\bin{l^p}{l_r}{\tau^I\gamma^\mu}(H^\dagger \tau^I H)D_\mu(H^\dagger H)$ \\
$\mathcal{O}_{q^2H^4D}^{(4)}$ & $\bin{q^p}{q_r}{\tau^I\gamma^\mu}(H^\dagger \tau^I \lrd_\mu H)(H^\dagger H)$ & $\mathcal{O}_{l^2H^4D}^{(4)}$ & $\bin{l^p}{l_r}{\tau^I\gamma^\mu}(H^\dagger \tau^I \lrd_\mu H)(H^\dagger H)$ \\
$\mathcal{O}_{u^2H^4D}$ & $\bin{u^p}{u_r}{\gamma^\mu}(H^\dagger \lrd_\mu H)(H^\dagger H)$ & $\mathcal{O}_{e^2H^4D}$ & $\bin{e^p}{e_r}{\gamma^\mu}(H^\dagger \lrd_\mu H)(H^\dagger H)$\\
$\mathcal{O}_{d^2H^4D}$ & $\bin{d^p}{d_r}{\gamma^\mu}(H^\dagger \lrd_\mu H)(H^\dagger H)$ & $\mathcal{O}_{udH^4D}$ & $\bin{u^p}{d_r}{\gamma^\mu}(\tilde{H}^\dagger \lrd_\mu H)(H^\dagger H)$ \\
        \hline
\end{tabular}
\end{minipage}

\begin{minipage}[c][4cm][t]{22em} 
\begin{tabular}{c|c||c|c||c|c}
\multicolumn{6}{c}{3: $\psi^2\phi^3D^2$} \\
\hline
$\mathcal{O}_{quH^3D^2}^{(1)}$ & $\bin{q^p}{u_r}{D^\mu\tilde{H}}(H^\dagger \lrd_\mu H)$ & $\mathcal{O}_{qdH^3D^2}^{(1)}$ & $\bin{q^p}{d_r}{D^\mu H}(H^\dagger \lrd_\mu H)$ & $\mathcal{O}_{leH^3D^2}^{(1)}$ & $\bin{l^p}{e_r}{D^\mu H}(H^\dagger \lrd_\mu H)$ \\
$\mathcal{O}_{quH^3D^2}^{(2)}$ & $\bin{q^p}{u_r}{D_\mu\tilde{H}\sigma^{\mu\nu}}(H^\dagger \lrd_\nu H)$ & $\mathcal{O}_{qdH^3D^2}^{(2)}$ & $\bin{q^p}{d_r}{D_\mu H\sigma^{\mu\nu}}(H^\dagger \lrd_\nu H)$ & $\mathcal{O}_{leH^3D^2}^{(2)}$ & $\bin{l^p}{e_r}{D_\mu H\sigma^{\mu\nu}}(H^\dagger \lrd_\nu H)$ \\
$\mathcal{O}_{quH^3D^2}^{(3)}$ & $\bin{q^p}{u_r}{\tilde{H}}D^2(H^\dagger H)$ & $\mathcal{O}_{qdH^3D^2}^{(3)}$ & $\bin{q^p}{d_r}{H}D^2(H^\dagger H)$ & $\mathcal{O}_{leH^3D^2}^{(3)}$ & $\bin{l^p}{e_r}{H}D^2(H^\dagger H)$ \\
$\mathcal{O}_{quH^3D^2}^{(4)}$ & $\bin{q^p}{u_r}{\tilde{H}\sigma^{\mu\nu}}(D_\mu H^\dagger D_\nu H)$ & $\mathcal{O}_{qdH^3D^2}^{(4)}$ & $\bin{q^p}{d_r}{H\sigma^{\mu\nu}}(D_\mu H^\dagger D_\nu H)$ & $\mathcal{O}_{leH^3D^2}^{(4)}$ & $\bin{l^p}{e_r}{H\sigma^{\mu\nu}}(D_\mu H^\dagger D_\nu H)$ \\
$\mathcal{O}_{quH^3D^2}^{(5)}$ & $\bin{q^p}{u_r}{D^\mu\tilde{H}}D_\mu(H^\dagger H)$ & $\mathcal{O}_{qdH^3D^2}^{(5)}$ & $\bin{q^p}{d_r}{D^\mu H}D_\mu(H^\dagger H)$ & $\mathcal{O}_{leH^3D^2}^{(5)}$ & $\bin{l^p}{e_r}{D^\mu H}D_\mu(H^\dagger H)$ \\
$\mathcal{O}_{quH^3D^2}^{(6)}$ & $\bin{q^p}{u_r}{D_\mu\tilde{H}\sigma^{\mu\nu}}D_\nu(H^\dagger H)$ & $\mathcal{O}_{qdH^3D^2}^{(6)}$ & $\bin{q^p}{d_r}{D_\mu H\sigma^{\mu\nu}}D_\nu(H^\dagger H)$ & $\mathcal{O}_{leH^3D^2}^{(6)}$ & $\bin{l^p}{e_r}{D_\mu H\sigma^{\mu\nu}}D_\nu(H^\dagger H)$ \\
\hline
\end{tabular}
\end{minipage}

\begin{minipage}[c][4cm][c]{22em} 
\begin{tabular}{c|c||c|c||c|c}
\multicolumn{6}{c}{4: $\psi^2\phi^2D^3$} \\
\hline
$\mathcal{O}_{q^2H^2D^3}^{(1)}$ & $\bin{q^p}{q_r}{\gamma^\mu}D^2(H^\dagger\lrd_\mu H)$ & $\mathcal{O}_{u^2H^2D^3}^{(1)}$ & $\bin{u^p}{u_r}{\gamma^\mu}D^2(H^\dagger\lrd_\mu H)$ & $\mathcal{O}_{l^2H^2D^3}^{(1)}$ & $\bin{l^p}{l_r}{\gamma^\mu}D^2(H^\dagger\lrd_\mu H)$ \\
$\mathcal{O}_{q^2H^2D^3}^{(2)}$ & $\bin{q^p}{q_r}{\gamma_\mu\lrd_\nu}(D^\mu H^\dagger D^\nu H)$ & $\mathcal{O}_{u^2H^2D^3}^{(2)}$ & $\bin{u^p}{u_r}{\gamma_\mu\lrd_\nu}(D^\mu H^\dagger D^\nu H)$ & $\mathcal{O}_{l^2H^2D^3}^{(2)}$ & $\bin{l^p}{l_r}{\gamma_\mu\lrd_\nu}(D^\mu H^\dagger D^\nu H)$\\
$\mathcal{O}_{q^2H^2D^3}^{(3)}$ & $\bin{q^p}{q_r}{\tau^I\gamma^\mu}D^2(H^\dagger\tau^I\lrd_\mu H)$ & $\mathcal{O}_{d^2H^2D^3}^{(1)}$ & $\bin{d^p}{d_r}{\gamma^\mu}D^2(H^\dagger\lrd_\mu H)$ & $\mathcal{O}_{l^2H^2D^3}^{(3)}$ & $\bin{l^p}{l_r}{\tau^I\gamma^\mu}D^2(H^\dagger\tau^I\lrd_\mu H)$\\
$\mathcal{O}_{q^2H^2D^3}^{(4)}$ & $\bin{q^p}{q_r}{\tau^I\gamma_\mu\lrd_\nu}(D^\mu H^\dagger\tau^I D^\nu H)$ & $\mathcal{O}_{d^2H^2D^3}^{(2)}$ & $\bin{d^p}{d_r}{\gamma_\mu\lrd_\nu}(D^\mu H^\dagger D^\nu H)$ & $\mathcal{O}_{l^2H^2D^3}^{(4)}$ & $\bin{l^p}{l_r}{\tau^I\gamma_\mu\lrd_\nu}(D^\mu H^\dagger\tau^I D^\nu H)$\\
$\mathcal{O}_{e^2H^2D^3}^{(1)}$ & $\bin{e^p}{e_r}{\gamma^\mu}D^2(H^\dagger\lrd_\mu H)$ & $\mathcal{O}_{e^2H^2D^3}^{(2)}$ & $\bin{e^p}{e_r}{\gamma_\mu\lrd_\nu}(D^\mu H^\dagger D^\nu H)$ & $\mathcal{O}_{udH^2D^3}$ & $\bin{u^p}{d_r}{\gamma^\mu \lrd^\nu}(D_\mu\tilde{H}^\dagger D_\nu H)$ \\
\hline
\end{tabular}
\end{minipage}
   
\begin{minipage}[c][5cm][c]{22em} 
\begin{tabular}{c|c||c|c}
\multicolumn{4}{c}{5: $\psi^2X\phi^3$} \\
\hline
$\mathcal{O}_{GquH^3}$ & $\bin{q^p}{u_r}{\tilde{H}\lambda^A\sigma^{\mu\nu}}(H^\dagger H)G^A_{\mu\nu}$ & $\mathcal{O}_{WqdH^3}^{(2)}$ & $\bin{q^p}{d_r}{\tau^I H\sigma^{\mu\nu}}(H^\dagger H) W^I_{\mu\nu}$ \\
$\mathcal{O}_{WquH^3}^{(1)}$ & $\bin{q^p}{u_r}{H\sigma^{\mu\nu}}(H^\dagger \tau^I \tilde{H}) W^I_{\mu\nu}$ & $\mathcal{O}_{BqdH^3}$ & $\bin{q^p}{d_r}{H \sigma^{\mu\nu}}(H^\dagger H)B_{\mu\nu}$ \\
$\mathcal{O}_{WquH^3}^{(2)}$ & $\bin{q^p}{u_r}{\tau^I\tilde{H}\sigma^{\mu\nu}}(H^\dagger H) W^I_{\mu\nu}$ & $\mathcal{O}_{WleH^3}^{(1)}$ & $\bin{l^p}{e_r}{H\sigma^{\mu\nu}}(H^\dagger \tau^I H) W^I_{\mu\nu}$ \\
$\mathcal{O}_{BquH^3}$ & $\bin{q^p}{u_r}{\tilde{H}\sigma^{\mu\nu}}(H^\dagger H)B_{\mu\nu}$ & $\mathcal{O}_{WleH^3}^{(2)}$ & $\bin{l^p}{e_r}{\tau^I H\sigma^{\mu\nu}}(H^\dagger H) W^I_{\mu\nu}$\\
$\mathcal{O}_{GqdH^3}$ & $\bin{q^p}{d_r}{H\lambda^A\sigma^{\mu\nu}}(H^\dagger H)G^A_{\mu\nu}$ & $\mathcal{O}_{BleH^3}$ & $\bin{l^p}{e_r}{H \sigma^{\mu\nu}}(H^\dagger H)B_{\mu\nu}$ \\
$\mathcal{O}_{WqdH^3}^{(1)}$ & $\bin{q^p}{d_r}{H\sigma^{\mu\nu}}(H^\dagger \tau^I H) W^I_{\mu\nu}$ & \\
\hline
\end{tabular}
\end{minipage}

\newpage

\begin{minipage}[c][6cm][t]{22em} 
\begin{tabular}{c|c||c|c||c|c}
\multicolumn{6}{c}{6: $\psi^2X\phi D^2$} \\
\hline
$\mathcal{O}_{GquHD^2}^{(1)}$ & $\bin{q^p}{u_r}{D^\mu D_\nu \tilde{H} G_{\mu\lambda}\sigma^{\nu\lambda}}$         & $\mathcal{O}_{BquHD^2}^{(3)}$ & $\bin{q^p}{u_r}{D^\nu\tilde{H} \lrd^\mu}B_{\mu\nu}$                       & $\mathcal{O}_{BqdHD^2}^{(2)}$ & $\bin{q^p}{d_r}{D^\mu D_\nu H\sigma^{\nu\lambda}}\tilde{B}_{\mu\lambda}$\\
$\mathcal{O}_{GquHD^2}^{(2)}$ & $\bin{q^p}{u_r}{D^\mu D_\nu \tilde{H} \tilde{G}_{\mu\lambda}\sigma^{\nu\lambda}}$ & $\mathcal{O}_{GqdHD^2}^{(1)}$ & $\bin{q^p}{d_r}{D^\mu D_\nu H G_{\mu\lambda}\sigma^{\nu\lambda}}$         & $\mathcal{O}_{BqdHD^2}^{(3)}$ & $\bin{q^p}{d_r}{D^\nu H \lrd^\mu}B_{\mu\nu}$ \\
$\mathcal{O}_{GquHD^2}^{(3)}$ & $\bin{q^p}{u_r}{D^\nu\tilde{H} G_{\mu\nu}\lrd^\mu}$                               & $\mathcal{O}_{GqdHD^2}^{(2)}$ & $\bin{q^p}{d_r}{D^\mu D_\nu H \tilde{G}_{\mu\lambda}\sigma^{\nu\lambda}}$ & $\mathcal{O}_{WleHD^2}^{(1)}$ & $\bin{l^p}{e_r}{W_{\mu\lambda}D^\mu D_\nu H\sigma^{\nu\lambda}}$ \\
$\mathcal{O}_{WquHD^2}^{(1)}$ & $\bin{q^p}{u_r}{W_{\mu\lambda}D^\mu D_\nu \tilde{H}\sigma^{\nu\lambda}}$          & $\mathcal{O}_{GqdHD^2}^{(3)}$ & $\bin{q^p}{d_r}{D^\nu H G_{\mu\nu}\lrd^\mu}$                              & $\mathcal{O}_{WleHD^2}^{(2)}$ & $\bin{l^p}{e_r}{\tilde{W}_{\mu\lambda}D^\mu D_\nu H\sigma^{\nu\lambda}}$ \\
$\mathcal{O}_{WquHD^2}^{(2)}$ & $\bin{q^p}{u_r}{\tilde{W}_{\mu\lambda}D^\mu D_\nu \tilde{H}\sigma^{\nu\lambda}}$  & $\mathcal{O}_{WqdHD^2}^{(1)}$ & $\bin{q^p}{d_r}{W_{\mu\lambda}D^\mu D_\nu H\sigma^{\nu\lambda}}$          & $\mathcal{O}_{WleHD^2}^{(3)}$ & $\bin{l^p}{e_r}{W_{\mu\nu}D^\nu H\lrd^\mu}$\\
$\mathcal{O}_{WquHD^2}^{(3)}$ & $\bin{q^p}{u_r}{W_{\mu\nu}D^\nu\tilde{H}\lrd^\mu}$                                & $\mathcal{O}_{WqdHD^2}^{(2)}$ & $\bin{q^p}{d_r}{\tilde{W}_{\mu\lambda}D^\mu D_\nu H\sigma^{\nu\lambda}}$  & $\mathcal{O}_{BleHD^2}^{(1)}$ & $\bin{l^p}{e_r}{D^\mu D_\nu H \sigma^{\nu\lambda}}B_{\mu\lambda}$ \\
\hline
\end{tabular}
\end{minipage}

\begin{minipage}[c][11cm][t]{22em} 
\begin{tabular}{c|c||c|c||c|c}
\multicolumn{6}{c}{7: $\psi^4\phi D$} \\
\hline
$\mathcal{O}_{q^3uHD}^{(1)}$ & $\bin{q^p}{q_r}{\gamma^\mu}\bin{q^s}{u_t}{\tilde{H}\lrd_\mu}$                                     & $\mathcal{O}_{qud^2HD}^{(3)}$ & $\bin{q^p}{u_r}{D_\mu\tilde{H}\sigma^{\mu\nu}}\bin{d^s}{d_t}{\gamma_\nu}$                           & $\mathcal{O}_{qdl^2HD}^{(2)}$ & $\bin{q^{ip}}{u_r}{}\bin{l^s}{l_{it}}{D_\mu H \gamma^\mu}$ \\
$\mathcal{O}_{q^3uHD}^{(2)}$ & $\bin{q^{ip}}{q_r}{\gamma^\mu H^\dagger}\epsilon_{ij}\bin{q^{js}}{u_t}{\lrd_\mu}$                 & $\mathcal{O}_{qud^2HD}^{(4)}$ & $\bin{q^p}{u_r}{D_\mu\tilde{H}\lambda^A\sigma^{\mu\nu}}\bin{d^s}{d_t}{\lambda^A\gamma_\nu}$         & $\mathcal{O}_{qdl^2HD}^{(3)}$ & $\bin{q^p}{d_r}{D_\mu H\sigma^{\mu\nu}}\bin{l^s}{l_t}{\gamma_\nu}$ \\
$\mathcal{O}_{q^3uHD}^{(3)}$ & $\bin{q^p}{q_r}{\gamma^\mu}\bin{q^s}{u_t}{D_\mu\tilde{H}}$                                        & $\mathcal{O}_{qud^2HD}^{(5)}$ & $\bin{q^p}{u_r}{\tilde{H}\sigma^{\mu\nu}}\bin{d^s}{d_t}{\gamma_\nu\lrd_\mu}$                        & $\mathcal{O}_{qdl^2HD}^{(4)}$ & $\bin{q^{ip}}{d_r}{\sigma^{\mu\nu}}\bin{l^s}{l_{it}}{D_\mu H\gamma_\nu}$ \\
$\mathcal{O}_{q^3uHD}^{(4)}$ & $\bin{q^{ip}}{q_r}{\gamma^\mu D_\mu H^\dagger}\epsilon_{ij}\bin{q^{js}}{u_t}{}$                   & $\mathcal{O}_{qud^2HD}^{(6)}$ & $\bin{q^p}{u_r}{\tilde{H}\lambda^A\sigma^{\mu\nu}}\bin{d^s}{d_t}{\lambda^A\gamma_\nu\lrd_\mu}$      & $\mathcal{O}_{qdl^2HD}^{(5)}$ & $\bin{q^p}{d_r}{H\sigma^{\mu\nu}\lrd_\mu}\bin{l^s}{l_t}{\gamma_\nu}$ \\
$\mathcal{O}_{q^3uHD}^{(5)}$ & $\bin{q^p}{q_r}{\lambda^A\gamma^\mu}\bin{q^s}{u_t}{ D_\mu\tilde{H}\lambda^A}$                     & $\mathcal{O}_{qd^3HD}^{(1)}$ & $\bin{q^p}{d_r}{D_\mu H}\bin{d^s}{d_t}{\gamma^\mu}$                                                  & $\mathcal{O}_{qdl^2HD}^{(6)}$ & $\bin{q^{ip}}{d_r}{\sigma^{\mu\nu}\lrd_\mu}\bin{l^s}{l_{it}}{H\gamma_\nu}$ \\
\hline
\end{tabular}
\end{minipage}

\newpage

\begin{minipage}[c][4cm][t]{22em} 
\begin{tabular}{c|c||c|c||c|c}
\multicolumn{6}{c}{8: $\psi^2X\phi^2D$} \\
\hline
$\mathcal{O}_{Gq^2H^2D}^{(1)}$ & $\bin{q^p}{q_r}{G_{\mu\nu}\gamma^\nu}D^\mu(H^\dagger H)$                                 & $\mathcal{O}_{Gu^2H^2D}^{(1)}$ & $\bin{u^p}{u_r}{G_{\mu\nu}\gamma^\nu}D^\mu(H^\dagger H)$                & $\mathcal{O}_{Wl^2H^2D}^{(5)}$ & $\bin{l^p}{l_r}{W_{\mu\nu}\tau^I \gamma^\nu}D^\mu(H^\dagger \tau^I H)$ \\
$\mathcal{O}_{Gq^2H^2D}^{(2)}$ & $\bin{q^p}{q_r}{G_{\mu\nu}\gamma^\nu}(H^\dagger \lrd^\mu H)$                             & $\mathcal{O}_{Gu^2H^2D}^{(2)}$ & $\bin{u^p}{u_r}{G_{\mu\nu}\gamma^\nu}(H^\dagger \lrd^\mu H)$            & $\mathcal{O}_{Wl^2H^2D}^{(6)}$ & $\bin{l^p}{l_r}{W_{\mu\nu}\tau^I \gamma^\nu}(H^\dagger \tau^I \lrd^\mu H)$ \\
$\mathcal{O}_{Gq^2H^2D}^{(3)}$ & $\bin{q^p}{q_r}{\tilde{G}_{\mu\nu}\gamma^\nu}D^\mu(H^\dagger H)$                         & $\mathcal{O}_{Gu^2H^2D}^{(3)}$ & $\bin{u^p}{u_r}{\tilde{G}_{\mu\nu}\gamma^\nu}D^\mu(H^\dagger H)$        & $\mathcal{O}_{Wl^2H^2D}^{(7)}$ & $\bin{l^p}{l_r}{\tilde{W}_{\mu\nu}\tau^I \gamma^\nu}D^\mu(H^\dagger \tau^I H)$ \\
$\mathcal{O}_{Gq^2H^2D}^{(4)}$ & $\bin{q^p}{q_r}{\tilde{G}_{\mu\nu}\gamma^\nu}(H^\dagger \lrd^\mu H)$                     & $\mathcal{O}_{Gu^2H^2D}^{(4)}$ & $\bin{u^p}{u_r}{\tilde{G}_{\mu\nu}\gamma^\nu}(H^\dagger \lrd^\mu H)$    & $\mathcal{O}_{Wl^2H^2D}^{(8)}$ & $\bin{l^p}{l_r}{\tilde{W}_{\mu\nu}\tau^I \gamma^\nu}(H^\dagger \tau^I \lrd^\mu H)$ \\
$\mathcal{O}_{Gq^2H^2D}^{(5)}$ & $\bin{q^p}{q_r}{G_{\mu\nu}\tau^I\gamma^\nu}D^\mu(H^\dagger \tau^I H)$                    & $\mathcal{O}_{Wu^2H^2D}^{(1)}$ & $\bin{u^p}{u_r}{\gamma^\nu}D^\mu(H^\dagger W_{\mu\nu} H)$               & $\mathcal{O}_{Wl^2H^2D}^{(9)}$ & $\bin{l^p}{l_r}{\tau^I W_{\mu\nu} \gamma^\nu}D^\mu(H^\dagger \tau^I H)$ \\
\hline
\end{tabular}
\end{minipage}

\newpage

\begin{minipage}[c][9.5cm][t]{22em} 
\begin{tabular}{c|c||c|c||c|c}
\multicolumn{6}{c}{9: $\psi^2X^2\phi$} \\
\hline
$\mathcal{O}_{G^2quH}^{(1)}$ & $\bin{q^p}{u_r}{\tilde{H}\{G_{\mu\nu},G^{\mu\nu}\}}$                       & $\mathcal{O}_{BWquH}^{(1)}$ & $\bin{q^p}{u_r}{W_{\mu\nu}\tilde{H}}B^{\mu\nu}$                         & $\mathcal{O}_{BGqdH}^{(1)}$ & $\bin{q^p}{d_r}{H G_{\mu\nu}}B^{\mu\nu}$ \\
$\mathcal{O}_{G^2quH}^{(2)}$ & $\bin{q^p}{u_r}{\tilde{H}}\lra{G_{\mu\nu}G^{\mu\nu}}$                      & $\mathcal{O}_{BWquH}^{(2)}$ & $\bin{q^p}{u_r}{\tilde{W}_{\mu\nu}\tilde{H}}B^{\mu\nu}$                 & $\mathcal{O}_{BGqdH}^{(2)}$ & $\bin{q^p}{d_r}{H \tilde{G}_{\mu\nu}}B^{\mu\nu}$ \\
$\mathcal{O}_{G^2quH}^{(3)}$ & $\bin{q^p}{u_r}{\tilde{H}\{\tilde{G}_{\mu\nu},G^{\mu\nu}\}}$               & $\mathcal{O}_{BWquH}^{(3)}$ & $\bin{q^p}{u_r}{W^\mu{}_\lambda\tilde{H}\sigma^{\nu\lambda}}B_{\mu\nu}$ & $\mathcal{O}_{BGqdH}^{(3)}$ & $\bin{q^p}{d_r}{H G^\mu{}_\lambda\sigma^{\nu\lambda}}B_{\mu\nu}$\\
$\mathcal{O}_{G^2quH}^{(4)}$ & $\bin{q^p}{u_r}{\tilde{H}}\lra{\tilde{G}_{\mu\nu}G^{\mu\nu}}$              & $\mathcal{O}_{G^2qdH}^{(1)}$ & $\bin{q^p}{d_r}{H\{G_{\mu\nu},G^{\mu\nu}\}}$                           & $\mathcal{O}_{BWqdH}^{(1)}$ & $\bin{q^p}{d_r}{W_{\mu\nu}H}B^{\mu\nu}$ \\
$\mathcal{O}_{G^2quH}^{(5)}$ & $\bin{q^p}{u_r}{\tilde{H}[G_{\mu\nu},G^\mu{}_\lambda]\sigma^{\nu\lambda}}$ & $\mathcal{O}_{G^2qdH}^{(2)}$ & $\bin{q^p}{d_r}{H}\lra{G_{\mu\nu}G^{\mu\nu}}$                          & $\mathcal{O}_{BWqdH}^{(2)}$ & $\bin{q^p}{d_r}{\tilde{W}_{\mu\nu}H}B^{\mu\nu}$ \\
$\mathcal{O}_{W^2quH}^{(1)}$ & $\bin{q^p}{u_r}{\tilde{H}}\lra{W_{\mu\nu}W^{\mu\nu}}$                      & $\mathcal{O}_{G^2qdH}^{(3)}$ & $\bin{q^p}{d_r}{H\{\tilde{G}_{\mu\nu},G^{\mu\nu}\}}$                   & $\mathcal{O}_{BWqdH}^{(3)}$ & $\bin{q^p}{d_r}{W^\mu{}_\lambda H\sigma^{\nu\lambda}}B_{\mu\nu}$ \\
\hline
    \end{tabular}
\end{minipage}

\begin{minipage}[b][4cm][t]{22em} 
\begin{tabular}{c|c||c|c||c|c}
\multicolumn{6}{c}{10: $\psi^2X^2D$} \\
\hline
$\mathcal{O}_{G^2q^2D}^{(1)}$ & $\bin{q^p}{q_r}{[G^\mu{}_\nu, G^\nu{}_\lambda]\gamma^\lambda\lrd_\mu}$         & $\mathcal{O}_{G^2d^2D}^{(3)}$ & $\bin{d^p}{d_r}{[G^\mu{}_\nu, \tilde{G}^\nu{}_\lambda]\gamma^\lambda\lrd_\mu}$  & $\mathcal{O}_{GBq^2D}^{(2)}$ & $\bin{q^p}{q_r}{\tilde{G}^\mu{}_\lambda \gamma_\nu\lrd_\mu}B^{\nu\lambda}$ \\
$\mathcal{O}_{G^2q^2D}^{(2)}$ & $\bin{q^p}{q_r}{[\tilde{G}^\mu{}_\nu, G^\nu{}_\lambda]\gamma^\lambda\lrd_\mu}$ & $\mathcal{O}_{G^2d^2D}^{(4)}$ & $\bin{d^p}{d_r}{\{G^\mu{}_\nu, G^\nu{}_\lambda\}\gamma^\lambda\lrd_\mu}$        & $\mathcal{O}_{GBq^2D}^{(3)}$ & $\bin{q^p}{q_r}{G^\nu{}_\lambda \gamma_\nu\lrd_\mu}B^{\mu\lambda}$ \\
$\mathcal{O}_{G^2q^2D}^{(3)}$ & $\bin{q^p}{q_r}{[G^\mu{}_\nu, \tilde{G}^\nu{}_\lambda]\gamma^\lambda\lrd_\mu}$ & $\mathcal{O}_{G^2d^2D}^{(5)}$ & $\bin{d^p}{d_r}{\gamma^\lambda\lrd_\mu}\lra{G^\mu{}_\nu G^\nu{}_\lambda}$       & $\mathcal{O}_{GBq^2D}^{(4)}$ & $\bin{q^p}{q_r}{\tilde{G}^\nu{}_\lambda \gamma_\nu\lrd_\mu}B^{\mu\lambda}$ \\
$\mathcal{O}_{G^2q^2D}^{(4)}$ & $\bin{q^p}{q_r}{\{G^\mu{}_\nu, G^\nu{}_\lambda\}\gamma^\lambda\lrd_\mu}$       & $\mathcal{O}_{W^2d^2D}$ & $\bin{d^p}{d_r}{\gamma^\lambda\lrd_\mu}\lra{W^\mu{}_\nu W^\nu{}_\lambda}$             & $\mathcal{O}_{WBq^2D}^{(1)}$ & $\bin{q^p}{q_r}{W^\mu{}_\lambda \gamma_\nu\lrd_\mu}B^{\nu\lambda}$ \\
$\mathcal{O}_{G^2q^2D}^{(5)}$ & $\bin{q^p}{q_r}{\gamma^\lambda\lrd_\mu}\lra{G^\mu{}_\nu G^\nu{}_\lambda}$      & $\mathcal{O}_{B^2d^2D}$ & $\bin{d^p}{d_r}{\gamma^\lambda\lrd_\mu}B^\mu{}_\nu B^\nu{}_\lambda$                   & $\mathcal{O}_{WBq^2D}^{(2)}$ & $\bin{q^p}{q_r}{\tilde{W}^\mu{}_\lambda \gamma_\nu\lrd_\mu}B^{\nu\lambda}$ \\
\hline
\end{tabular}
\end{minipage}

\newpage

\begin{minipage}[c][11cm][t]{22em} 
\begin{tabular}{c|c||c|c||c|c}
\multicolumn{6}{c}{11: $\psi^4D^2$} \\
\hline
$\mathcal{O}_{q^4D^2}^{(1)}$ & $\bin{q^p}{q_r}{\gamma^\mu}D^2\bin{q^s}{q_t}{\gamma_\mu}$                                   & $\mathcal{O}_{d^4D^2}^{(1)}$ & $\bin{d^p}{d_r}{\gamma^\mu}D^2\bin{d^s}{d_t}{\gamma_\mu}$                              &  $\mathcal{O}_{d^2l^2D^2}^{(2)}$ & $\bin{d^p}{d_r}{\gamma^\mu\lrd^\nu}\bin{l^s}{l_t}{\gamma_\nu\lrd_\mu}$ \\
$\mathcal{O}_{q^4D^2}^{(2)}$ & $\bin{q^p}{q_r}{\lambda^A\gamma^\mu}D^2\bin{q^s}{q_t}{\lambda^A\gamma_\mu}$                 & $\mathcal{O}_{d^4D^2}^{(2)}$ & $\bin{d^p}{d_r}{\gamma^\mu\lrd^\nu}\bin{d^s}{d_t}{\gamma_\nu\lrd_\mu}$                 &  $\mathcal{O}_{d^2e^2D^2}^{(1)}$ & $\bin{d^p}{d_r}{\gamma^\mu}D^2\bin{e^s}{e_t}{\gamma_\mu}$ \\
$\mathcal{O}_{q^4D^2}^{(3)}$ & $\bin{q^p}{q_r}{\gamma^\mu\lrd^\nu}\bin{q^s}{q_t}{\gamma_\nu\lrd_\mu}$                      & $\mathcal{O}_{q^2udD^2}^{(1)}$ & $\bin{q^{ip}}{d_r}{}\epsilon_{ij}D^2\bin{q^{js}}{u_t}{}$                             &  $\mathcal{O}_{d^2e^2D^2}^{(2)}$ & $\bin{d^p}{d_r}{\gamma^\mu\lrd^\nu}\bin{e^s}{e_t}{\gamma_\nu\lrd_\mu}$ \\
$\mathcal{O}_{q^4D^2}^{(4)}$ & $\bin{q^p}{q_r}{\lambda^A\gamma^\mu\lrd^\nu}\bin{q^s}{q_t}{\lambda^A\gamma_\nu\lrd_\mu}$    & $\mathcal{O}_{q^2udD^2}^{(2)}$ & $\bin{q^{ip}}{d_r}{\lambda^A}\epsilon_{ij}D^2\bin{q^{js}}{u_t}{\lambda^A}$           &  $\mathcal{O}_{quleD^2}^{(1)}$ & $\bin{q^{ip}}{u_r}{}\epsilon_{ij}D^2\bin{l^{js}}{e_t}{}$ \\
$\mathcal{O}_{q^2u^2D^2}^{(1)}$ & $\bin{q^p}{q_r}{\gamma^\mu}D^2\bin{u^s}{u_t}{\gamma_\mu}$                                & $\mathcal{O}_{q^2udD^2}^{(3)}$ & $D_\mu\bin{q^{ip}}{d_r}{\lrd_\nu}\epsilon_{ij}\bin{q^{js}}{u_t}{\sigma^{\mu\nu}}$    &  $\mathcal{O}_{quleD^2}^{(2)}$ & $\bin{q^{ip}}{u_r}{\sigma^{\mu\nu}}\epsilon_{ij}D^2\bin{l^{js}}{e_t}{\sigma_{\mu\nu}}$ \\
\hline
\end{tabular}
\end{minipage}

\begin{minipage}[c][4cm][t]{22em} 
\begin{tabular}{c|c||c|c||c|c}
\multicolumn{6}{c}{12: $\psi^4\phi^2$} \\
\hline
$\mathcal{O}_{q^4H^2}^{(1)}$ & $\bin{q^p}{q_r}{\gamma^\mu}\bin{q^s}{q_t}{H\gamma_\mu H^\dagger}$                                     & $\mathcal{O}^{'(2)}_{q^2d^2H^2}$ & $\bin{q^p}{d_r}{H\lambda^A}\bin{q^s}{d_t}{H\lambda^A}$                             & $\mathcal{O}_{d^2l^2H^2}^{(2)}$ & $\bin{d^p}{d_r}{\gamma^\mu}\bin{l^s}{l_t}{\gamma_\mu}(H^\dagger H)$ \\
$\mathcal{O}_{q^4H^2}^{(2)}$ & $\bin{q^p}{q_r}{\lambda^A\gamma^\mu}\bin{q^s}{q_t}{H\lambda^A\gamma_\mu H^\dagger}$                   & $\mathcal{O}_{q^2udH^2}^{(1)}$ & $\bin{q^p}{u_r}{\tilde{H}}\bin{q^s}{d_t}{H}$                                         & $\mathcal{O}_{d^2e^2H^2}$ & $\bin{d^p}{d_r}{\gamma^\mu}\bin{e^s}{e_t}{\gamma_\mu}(H^\dagger H)$ \\
$\mathcal{O}_{q^4H^2}^{(3)}$ & $\bin{q^p}{q_r}{\gamma^\mu}\bin{q^s}{q_t}{\gamma_\mu}(H^\dagger H)$                                   & $\mathcal{O}_{q^2udH^2}^{(2)}$ & $\bin{q^p}{u_r}{\lambda^A\tilde{H}}\bin{q^s}{d_t}{\lambda^A H}$                      & $\mathcal{O}_{quleH^2}^{(1)}$ & $\bin{q^{ip}}{u_r}{}\epsilon_{ij}\bin{l^{js}}{e_t}{}(H^\dagger H)$ \\
$\mathcal{O}_{q^2u^2H^2}^{(1)}$ & $\bin{q^p}{q_r}{\tau^I\gamma^\mu}\bin{u^s}{u_t}{H^\dagger\gamma_\mu H}$                            & $\mathcal{O}_{q^2udH^2}^{(3)}$ & $\bin{q^{ip}}{u_r}{}\epsilon_{ij}\bin{q^{js}}{d_t}{}(H^\dagger H)$                   & $\mathcal{O}_{quleH^2}^{(2)}$ & $\bin{q^{ip}}{u_r}{\sigma^{\mu\nu}}\epsilon_{ij}\bin{l^{js}}{e_t}{\sigma_{\mu\nu}}(H^\dagger H)$ \\
$\mathcal{O}_{q^2u^2H^2}^{(2)}$ & $\bin{q^p}{q_r}{\lambda^A\tau^I\gamma^\mu}\bin{u^s}{u_t}{H^\dagger\lambda^A\gamma_\mu H}$          & $\mathcal{O}_{q^2udH^2}^{(4)}$ & $\bin{q^{ip}}{u_r}{\lambda^A}\epsilon_{ij}\bin{q^{js}}{d_t}{\lambda^A}(H^\dagger H)$ & $\mathcal{O}_{quleH^2}^{(3)}$ & $\bin{q^p}{u_r}{\tilde{H}}\bin{l^s}{e_t}{H}$ \\
\hline
\end{tabular}
\end{minipage}

\newpage
\begin{minipage}[c][11cm][t]{22em} 
\begin{tabular}{c|c||c|c||c|c}
\multicolumn{6}{c}{13: $\psi^4X$} \\
\hline
$\mathcal{O}_{Gq^4}^{(1)}$ & $\bin{q^p}{q_r}{G_{\mu\nu}\gamma^\mu}\bin{q^s}{q_t}{\gamma^\nu}$                                 & $\mathcal{O}_{Wq^2d^2}^{(1)}$ & $\bin{q^p}{q_r}{W_{\mu\nu}\gamma^\mu}\bin{d^s}{d_t}{\gamma^\nu}$                                           & $\mathcal{O}_{Wq^2ud}^{(3)}$ & $\bin{q^{ip}}{u_r}{\sigma^{\mu\nu}}(W_{\mu\nu})_i^k\epsilon_{jk}\bin{q^{js}}{d_t}{}$ \\
$\mathcal{O}_{Gq^4}^{(2)}$ & $\bin{q^p}{q_r}{G_{\mu\nu}\tau^I\gamma^\mu}\bin{q^s}{q_t}{\tau^I\gamma^\nu}$                     & $\mathcal{O}_{Wq^2d^2}^{(2)}$ & $\bin{q^p}{q_r}{\lambda^AW_{\mu\nu}\gamma^\mu}\bin{d^s}{d_t}{\lambda^A\gamma^\nu}$                         & $\mathcal{O}_{Bq^2ud}^{(1)}$ & $\bin{q^{ip}}{u_r}{}\epsilon_{ij}\bin{q^{js}}{d_t}{\sigma^{\mu\nu}}B_{\mu\nu}$ \\
$\mathcal{O}_{Gq^4}^{(3)}$ & $\bin{q^p}{q_r}{\tilde{G}_{\mu\nu}\gamma^\mu}\bin{q^s}{q_t}{\gamma^\nu}$                         & $\mathcal{O}_{Wq^2d^2}^{(3)}$ & $\bin{q^p}{q_r}{\tilde{W}_{\mu\nu}\gamma^\mu}\bin{d^s}{d_t}{\gamma^\nu}$                                   & $\mathcal{O}_{Bq^2ud}^{(2)}$ & $\bin{q^{ip}}{u_r}{\lambda^A}\epsilon_{ij}\bin{q^{js}}{d_t}{\lambda^A\sigma^{\mu\nu}}B_{\mu\nu}$ \\
$\mathcal{O}_{Gq^4}^{(4)}$ & $\bin{q^p}{q_r}{\tilde{G}_{\mu\nu}\tau^I\gamma^\mu}\bin{q^s}{q_t}{\tau^I\gamma^\nu}$             & $\mathcal{O}_{Wq^2d^2}^{(4)}$ & $\bin{q^p}{q_r}{\lambda^A\tilde{W}_{\mu\nu}\gamma^\mu}\bin{d^s}{d_t}{\lambda^A\gamma^\nu}$                 & $\mathcal{O}_{Bq^2ud}^{(3)}$ & $\bin{q^{ip}}{u_r}{\sigma^{\mu\nu}}\epsilon_{ij}\bin{q^{js}}{d_t}{}B_{\mu\nu}$ \\
$\mathcal{O}_{Wq^4}^{(1)}$ & $\bin{q^p}{q_r}{W_{\mu\nu}\gamma^\mu}\bin{q^s}{q_t}{\gamma^\nu}$                                 & $\mathcal{O}_{Bq^2d^2}^{(1)}$ & $\bin{q^p}{q_r}{\gamma^\mu}\bin{d^s}{d_t}{\gamma^\nu}B_{\mu\nu}$                                           & $\mathcal{O}_{Gq^2l^2}^{(1)}$ & $\bin{q^p}{q_r}{G_{\mu\nu}\gamma^\nu}\bin{l^s}{l_t}{\gamma^\mu}$ \\
\hline
\end{tabular}
\end{minipage}

\newpage 

\begin{minipage}[c][11cm][t]{22em} 
\begin{tabular}{c|c||c|c}
\multicolumn{4}{c}{13: $\psi^4X$} \\
\hline
$\mathcal{O}_{Wd^2l^2}^{(1)}$ & $\bin{d^p}{d_r}{\gamma^\nu}\bin{l^s}{l_t}{W_{\mu\nu}\gamma^\mu}$              & $\mathcal{O}_{Gqule}^{(1)}$ & $\bin{q^{ip}}{u_r}{G_{\mu\nu}\sigma^{\mu\nu}}\epsilon_{ij}\bin{l^{js}}{e_t}{}$ \\
$\mathcal{O}_{Wd^2l^2}^{(2)}$ & $\bin{d^p}{d_r}{\gamma^\nu}\bin{l^s}{l_t}{\tilde{W}_{\mu\nu}\gamma^\mu}$      & $\mathcal{O}_{Gqule}^{(2)}$ & $\bin{q^{ip}}{u_r}{G_{\mu\nu}}\epsilon_{ij}\bin{l^{js}}{e_t}{\sigma^{\mu\nu}}$ \\
$\mathcal{O}_{Bd^2l^2}^{(1)}$ & $\bin{d^p}{d_r}{\gamma^\nu}\bin{l^s}{l_t}{\gamma^\mu}B_{\mu\nu}$              & $\mathcal{O}_{Gqule}^{(3)}$ & $\bin{q^{ip}}{u_r}{G^{\mu\nu}\sigma_\nu{}^\rho}\epsilon_{ij}\bin{l^{js}}{e_t}{\sigma_{\mu\rho}}$ \\
$\mathcal{O}_{Bd^2l^2}^{(2)}$ & $\bin{d^p}{d_r}{\gamma^\nu}\bin{l^s}{l_t}{\gamma^\mu}\tilde{B}_{\mu\nu}$      & $\mathcal{O}_{Wqule}^{(1)}$ & $\bin{q^{jp}}{u_r}{\sigma^{\mu\nu}}\epsilon_{jk}(W_{\mu\nu})_i^k\bin{l^{is}}{e_t}{}$ \\
$\mathcal{O}_{Gd^2e^2}^{(1)}$ & $\bin{d^p}{d_r}{G_{\mu\nu}\gamma^\mu}\bin{e^s}{e_t}{\gamma^\nu}$              & $\mathcal{O}_{Wqule}^{(2)}$ & $\bin{q^{jp}}{u_r}{}\epsilon_{jk}(W_{\mu\nu})_i^k\bin{l^{is}}{e_t}{\sigma^{\mu\nu}}$ \\
$\mathcal{O}_{Gd^2e^2}^{(2)}$ & $\bin{d^p}{d_r}{\tilde{G}_{\mu\nu}\gamma^\mu}\bin{e^s}{e_t}{\gamma^\nu}$      & $\mathcal{O}_{Wqule}^{(3)}$ & $\bin{q^{jp}}{u_r}{\sigma^\nu{}_\rho}\epsilon_{jk}(W_{\mu\nu})_i^k\bin{l^{is}}{e_t}{\sigma^{\mu\rho}}$ \\
$\mathcal{O}_{Bd^2e^2}^{(1)}$ & $\bin{d^p}{d_r}{\gamma^\mu}\bin{e^s}{e_t}{\gamma^\nu}B_{\mu\nu}$              & $\mathcal{O}_{Bqule}^{(1)}$ & $\bin{q^{ip}}{u_r}{\sigma^{\mu\nu}}\epsilon_{ij}\bin{l^{js}}{e_t}{}B_{\mu\nu}$ \\
\hline
\end{tabular}
\end{minipage}

\end{appendix}

\bibliography{ref}

\providecommand{\href}[2]{#2}\begingroup\raggedright\begin{thebibliography}{10}

\bibitem{Weinberg:1979sa}
S.~Weinberg, \emph{{Baryon and Lepton Nonconserving Processes}},
  \href{http://dx.doi.org/10.1103/PhysRevLett.43.1566}{\emph{Phys. Rev. Lett.}
  {\bf 43} (1979) 1566--1570}.

\bibitem{Buchmuller:1985jz}
W.~Buchmuller and D.~Wyler, \emph{{Effective Lagrangian Analysis of New
  Interactions and Flavor Conservation}},
  \href{http://dx.doi.org/10.1016/0550-3213(86)90262-2}{\emph{Nucl. Phys. B}
  {\bf 268} (1986) 621--653}.

\bibitem{Grzadkowski:2010es}
B.~Grzadkowski, M.~Iskrzynski, M.~Misiak and J.~Rosiek, \emph{{Dimension-Six
  Terms in the Standard Model Lagrangian}},
  \href{http://dx.doi.org/10.1007/JHEP10(2010)085}{\emph{JHEP} {\bf 10} (2010)
  085}, [\href{https://arxiv.org/abs/1008.4884}{{\tt 1008.4884}}].

\bibitem{Henning:2015alf}
B.~Henning, X.~Lu, T.~Melia and H.~Murayama, \emph{{2, 84, 30, 993, 560, 15456,
  11962, 261485, ...: Higher dimension operators in the SM EFT}},
  \href{http://dx.doi.org/10.1007/JHEP08(2017)016}{\emph{JHEP} {\bf 08} (2017)
  016}, [\href{https://arxiv.org/abs/1512.03433}{{\tt 1512.03433}}].

\bibitem{Lehman:2014jma}
L.~Lehman, \emph{{Extending the Standard Model Effective Field Theory with the
  Complete Set of Dimension-7 Operators}},
  \href{http://dx.doi.org/10.1103/PhysRevD.90.125023}{\emph{Phys. Rev. D} {\bf
  90} (2014) 125023}, [\href{https://arxiv.org/abs/1410.4193}{{\tt
  1410.4193}}].

\bibitem{Liao:2016hru}
Y.~Liao and X.-D. Ma, \emph{{Renormalization Group Evolution of Dimension-seven
  Baryon- and Lepton-number-violating Operators}},
  \href{http://dx.doi.org/10.1007/JHEP11(2016)043}{\emph{JHEP} {\bf 11} (2016)
  043}, [\href{https://arxiv.org/abs/1607.07309}{{\tt 1607.07309}}].

\bibitem{Li:2020gnx}
H.-L. Li, Z.~Ren, J.~Shu, M.-L. Xiao, J.-H. Yu and Y.-H. Zheng, \emph{{Complete
  set of dimension-eight operators in the standard model effective field
  theory}}, \href{http://dx.doi.org/10.1103/PhysRevD.104.015026}{\emph{Phys.
  Rev. D} {\bf 104} (2021) 015026},
  [\href{https://arxiv.org/abs/2005.00008}{{\tt 2005.00008}}].

\bibitem{Murphy:2020rsh}
C.~W. Murphy, \emph{{Dimension-8 operators in the Standard Model Effective
  Field Theory}}, \href{http://dx.doi.org/10.1007/JHEP10(2020)174}{\emph{JHEP}
  {\bf 10} (2020) 174}, [\href{https://arxiv.org/abs/2005.00059}{{\tt
  2005.00059}}].

\bibitem{Li:2020xlh}
H.-L. Li, Z.~Ren, M.-L. Xiao, J.-H. Yu and Y.-H. Zheng, \emph{{Complete set of
  dimension-nine operators in the standard model effective field theory}},
  \href{http://dx.doi.org/10.1103/PhysRevD.104.015025}{\emph{Phys. Rev. D} {\bf
  104} (2021) 015025}, [\href{https://arxiv.org/abs/2007.07899}{{\tt
  2007.07899}}].

\bibitem{Liao:2020jmn}
Y.~Liao and X.-D. Ma, \emph{{An explicit construction of the dimension-9
  operator basis in the standard model effective field theory}},
  \href{http://dx.doi.org/10.1007/JHEP11(2020)152}{\emph{JHEP} {\bf 11} (2020)
  152}, [\href{https://arxiv.org/abs/2007.08125}{{\tt 2007.08125}}].

\bibitem{Harlander:2023psl}
R.~V. Harlander, T.~Kempkens and M.~C. Schaaf, \emph{{Standard model effective
  field theory up to mass dimension 12}},
  \href{http://dx.doi.org/10.1103/PhysRevD.108.055020}{\emph{Phys. Rev. D} {\bf
  108} (2023) 055020}, [\href{https://arxiv.org/abs/2305.06832}{{\tt
  2305.06832}}].

\bibitem{Cabibbo:1963yz}
N.~Cabibbo, \emph{{Unitary Symmetry and Leptonic Decays}},
  \href{http://dx.doi.org/10.1103/PhysRevLett.10.531}{\emph{Phys. Rev. Lett.}
  {\bf 10} (1963) 531--533}.

\bibitem{Kobayashi:1973fv}
M.~Kobayashi and T.~Maskawa, \emph{{CP Violation in the Renormalizable Theory
  of Weak Interaction}},
  \href{http://dx.doi.org/10.1143/PTP.49.652}{\emph{Prog. Theor. Phys.} {\bf
  49} (1973) 652--657}.

\bibitem{Alonso:2013hga}
R.~Alonso, E.~E. Jenkins, A.~V. Manohar and M.~Trott, \emph{{Renormalization
  Group Evolution of the Standard Model Dimension Six Operators III: Gauge
  Coupling Dependence and Phenomenology}},
  \href{http://dx.doi.org/10.1007/JHEP04(2014)159}{\emph{JHEP} {\bf 04} (2014)
  159}, [\href{https://arxiv.org/abs/1312.2014}{{\tt 1312.2014}}].

\bibitem{Remmen:2019cyz}
G.~N. Remmen and N.~L. Rodd, \emph{{Consistency of the Standard Model Effective
  Field Theory}}, \href{http://dx.doi.org/10.1007/JHEP12(2019)032}{\emph{JHEP}
  {\bf 12} (2019) 032}, [\href{https://arxiv.org/abs/1908.09845}{{\tt
  1908.09845}}].

\bibitem{Durieux:2024zrg}
G.~Durieux, G.~N. Remmen, N.~L. Rodd, O.~J.~P. \'Eboli, M.~C. Gonzalez-Garcia,
  D.~Kondo et~al., \emph{{LHC EFT WG Note: Basis for Anomalous Quartic Gauge
  Couplings}},  \href{https://arxiv.org/abs/2411.02483}{{\tt 2411.02483}}.

\bibitem{Kondo:2022wcw}
D.~Kondo, H.~Murayama and R.~Okabe, \emph{{23, 381, 6242, 103268, 1743183,
  \textellipsis{} : Hilbert series for CP-violating operators in SMEFT}},
  \href{http://dx.doi.org/10.1007/JHEP03(2023)107}{\emph{JHEP} {\bf 03} (2023)
  107}, [\href{https://arxiv.org/abs/2212.02413}{{\tt 2212.02413}}].

\bibitem{Bonnefoy:2021tbt}
Q.~Bonnefoy, E.~Gendy, C.~Grojean and J.~T. Ruderman, \emph{{Beyond Jarlskog:
  699 invariants for CP violation in SMEFT}},
  \href{http://dx.doi.org/10.1007/JHEP08(2022)032}{\emph{JHEP} {\bf 08} (2022)
  032}, [\href{https://arxiv.org/abs/2112.03889}{{\tt 2112.03889}}].

\bibitem{Bonnefoy:2023bzx}
Q.~Bonnefoy, E.~Gendy, C.~Grojean and J.~T. Ruderman, \emph{{Opportunistic CP
  violation}}, \href{http://dx.doi.org/10.1007/JHEP06(2023)141}{\emph{JHEP}
  {\bf 06} (2023) 141}, [\href{https://arxiv.org/abs/2302.07288}{{\tt
  2302.07288}}].

\bibitem{Wang:2021wdq}
Y.~Wang, B.~Yu and S.~Zhou, \emph{{Flavor invariants and renormalization-group
  equations in the leptonic sector with massive Majorana neutrinos}},
  \href{http://dx.doi.org/10.1007/JHEP09(2021)053}{\emph{JHEP} {\bf 09} (2021)
  053}, [\href{https://arxiv.org/abs/2107.06274}{{\tt 2107.06274}}].

\bibitem{Yu:2021cco}
B.~Yu and S.~Zhou, \emph{{Hilbert series for leptonic flavor invariants in the
  minimal seesaw model}},
  \href{http://dx.doi.org/10.1007/JHEP10(2021)017}{\emph{JHEP} {\bf 10} (2021)
  017}, [\href{https://arxiv.org/abs/2107.11928}{{\tt 2107.11928}}].

\bibitem{Yu:2022ttm}
B.~Yu and S.~Zhou, \emph{{CP violation and flavor invariants in the seesaw
  effective field theory}},
  \href{http://dx.doi.org/10.1007/JHEP08(2022)017}{\emph{JHEP} {\bf 08} (2022)
  017}, [\href{https://arxiv.org/abs/2203.10121}{{\tt 2203.10121}}].

\bibitem{Grojean:2024qdm}
C.~Grojean, J.~Kley, D.~Leflot and C.-Y. Yao, \emph{{The flavor invariants of
  the \ensuremath{\nu}SM}},
  \href{http://dx.doi.org/10.1007/JHEP12(2024)069}{\emph{JHEP} {\bf 12} (2024)
  069}, [\href{https://arxiv.org/abs/2406.00094}{{\tt 2406.00094}}].

\bibitem{Darvishi:2023ckq}
N.~Darvishi, Y.~Wang and J.-H. Yu, \emph{{Automated ring-diagram framework for
  classifying CP invariants}},
  \href{http://dx.doi.org/10.1103/PhysRevD.108.115030}{\emph{Phys. Rev. D} {\bf
  108} (2023) 115030}, [\href{https://arxiv.org/abs/2311.15422}{{\tt
  2311.15422}}].

\bibitem{Darvishi:2024cwe}
N.~Darvishi, Y.~Wang and J.-H. Yu, \emph{{Identifying CP Basis Invariants in
  SMEFT}},  \href{https://arxiv.org/abs/2403.18732}{{\tt 2403.18732}}.

\bibitem{Glashow:1970gm}
S.~L. Glashow, J.~Iliopoulos and L.~Maiani, \emph{{Weak Interactions with
  Lepton-Hadron Symmetry}},
  \href{http://dx.doi.org/10.1103/PhysRevD.2.1285}{\emph{Phys. Rev. D} {\bf 2}
  (1970) 1285--1292}.

\bibitem{ParticleDataGroup:2024cfk}
{\scshape Particle Data Group} collaboration, S.~Navas et~al., \emph{{Review of
  particle physics}},
  \href{http://dx.doi.org/10.1103/PhysRevD.110.030001}{\emph{Phys. Rev. D} {\bf
  110} (2024) 030001}.

\bibitem{Isidori:2010kg}
G.~Isidori, Y.~Nir and G.~Perez, \emph{{Flavor Physics Constraints for Physics
  Beyond the Standard Model}},
  \href{http://dx.doi.org/10.1146/annurev.nucl.012809.104534}{\emph{Ann. Rev.
  Nucl. Part. Sci.} {\bf 60} (2010) 355},
  [\href{https://arxiv.org/abs/1002.0900}{{\tt 1002.0900}}].

\bibitem{Chivukula:1987py}
R.~S. Chivukula and H.~Georgi, \emph{{Composite Technicolor Standard Model}},
  \href{http://dx.doi.org/10.1016/0370-2693(87)90713-1}{\emph{Phys. Lett. B}
  {\bf 188} (1987) 99--104}.

\bibitem{DAmbrosio:2002vsn}
G.~D'Ambrosio, G.~F. Giudice, G.~Isidori and A.~Strumia, \emph{{Minimal flavor
  violation: An Effective field theory approach}},
  \href{http://dx.doi.org/10.1016/S0550-3213(02)00836-2}{\emph{Nucl. Phys. B}
  {\bf 645} (2002) 155--187}, [\href{https://arxiv.org/abs/hep-ph/0207036}{{\tt
  hep-ph/0207036}}].

\bibitem{Bonnefoy:2020yee}
Q.~Bonnefoy, E.~Gendy and C.~Grojean, \emph{{Positivity bounds on Minimal
  Flavor Violation}},
  \href{http://dx.doi.org/10.1007/JHEP04(2021)115}{\emph{JHEP} {\bf 04} (2021)
  115}, [\href{https://arxiv.org/abs/2011.12855}{{\tt 2011.12855}}].

\bibitem{Aoude:2020dwv}
R.~Aoude, T.~Hurth, S.~Renner and W.~Shepherd, \emph{{The impact of flavour
  data on global fits of the MFV SMEFT}},
  \href{http://dx.doi.org/10.1007/JHEP12(2020)113}{\emph{JHEP} {\bf 12} (2020)
  113}, [\href{https://arxiv.org/abs/2003.05432}{{\tt 2003.05432}}].

\bibitem{Bruggisser:2021duo}
S.~Bruggisser, R.~Sch\"afer, D.~van Dyk and S.~Westhoff, \emph{{The Flavor of
  UV Physics}}, \href{http://dx.doi.org/10.1007/JHEP05(2021)257}{\emph{JHEP}
  {\bf 05} (2021) 257}, [\href{https://arxiv.org/abs/2101.07273}{{\tt
  2101.07273}}].

\bibitem{Bruggisser:2022rhb}
S.~Bruggisser, D.~van Dyk and S.~Westhoff, \emph{{Resolving the flavor
  structure in the MFV-SMEFT}},
  \href{http://dx.doi.org/10.1007/JHEP02(2023)225}{\emph{JHEP} {\bf 02} (2023)
  225}, [\href{https://arxiv.org/abs/2212.02532}{{\tt 2212.02532}}].

\bibitem{Bartocci:2023nvp}
R.~Bartocci, A.~Biek\"otter and T.~Hurth, \emph{{A global analysis of the SMEFT
  under the minimal MFV assumption}},
  \href{http://dx.doi.org/10.1007/JHEP05(2024)074}{\emph{JHEP} {\bf 05} (2024)
  074}, [\href{https://arxiv.org/abs/2311.04963}{{\tt 2311.04963}}].

\bibitem{Bartocci:2024fgj}
R.~Bartocci, \emph{{Global analysis of the $U(3)^5$ symmetric SMEFT}},  in
  \emph{{58th Rencontres de Moriond on Electroweak Interactions and Unified
  Theories}}, 5, 2024.
\newblock \href{https://arxiv.org/abs/2405.10101}{{\tt 2405.10101}}.

\bibitem{Barbieri:2011ci}
R.~Barbieri, G.~Isidori, J.~Jones-Perez, P.~Lodone and D.~M. Straub,
  \emph{{$U(2)$ and Minimal Flavour Violation in Supersymmetry}},
  \href{http://dx.doi.org/10.1140/epjc/s10052-011-1725-z}{\emph{Eur. Phys. J.
  C} {\bf 71} (2011) 1725}, [\href{https://arxiv.org/abs/1105.2296}{{\tt
  1105.2296}}].

\bibitem{Barbieri:2012uh}
R.~Barbieri, D.~Buttazzo, F.~Sala and D.~M. Straub, \emph{{Flavour physics from
  an approximate $U(2)^3$ symmetry}},
  \href{http://dx.doi.org/10.1007/JHEP07(2012)181}{\emph{JHEP} {\bf 07} (2012)
  181}, [\href{https://arxiv.org/abs/1203.4218}{{\tt 1203.4218}}].

\bibitem{Blankenburg:2012nx}
G.~Blankenburg, G.~Isidori and J.~Jones-Perez, \emph{{Neutrino Masses and LFV
  from Minimal Breaking of $U(3)^5$ and $U(2)^5$ flavor Symmetries}},
  \href{http://dx.doi.org/10.1140/epjc/s10052-012-2126-7}{\emph{Eur. Phys. J.
  C} {\bf 72} (2012) 2126}, [\href{https://arxiv.org/abs/1204.0688}{{\tt
  1204.0688}}].

\bibitem{Greljo:2015mma}
A.~Greljo, G.~Isidori and D.~Marzocca, \emph{{On the breaking of Lepton Flavor
  Universality in B decays}},
  \href{http://dx.doi.org/10.1007/JHEP07(2015)142}{\emph{JHEP} {\bf 07} (2015)
  142}, [\href{https://arxiv.org/abs/1506.01705}{{\tt 1506.01705}}].

\bibitem{Barbieri:2015yvd}
R.~Barbieri, G.~Isidori, A.~Pattori and F.~Senia, \emph{{Anomalies in
  $B$-decays and $U(2)$ flavour symmetry}},
  \href{http://dx.doi.org/10.1140/epjc/s10052-016-3905-3}{\emph{Eur. Phys. J.
  C} {\bf 76} (2016) 67}, [\href{https://arxiv.org/abs/1512.01560}{{\tt
  1512.01560}}].

\bibitem{Buttazzo:2017ixm}
D.~Buttazzo, A.~Greljo, G.~Isidori and D.~Marzocca, \emph{{B-physics anomalies:
  a guide to combined explanations}},
  \href{http://dx.doi.org/10.1007/JHEP11(2017)044}{\emph{JHEP} {\bf 11} (2017)
  044}, [\href{https://arxiv.org/abs/1706.07808}{{\tt 1706.07808}}].

\bibitem{Faroughy:2020ina}
D.~A. Faroughy, G.~Isidori, F.~Wilsch and K.~Yamamoto, \emph{{Flavour
  symmetries in the SMEFT}},
  \href{http://dx.doi.org/10.1007/JHEP08(2020)166}{\emph{JHEP} {\bf 08} (2020)
  166}, [\href{https://arxiv.org/abs/2005.05366}{{\tt 2005.05366}}].

\bibitem{Sun:2022snw}
H.~Sun, M.-L. Xiao and J.-H. Yu, \emph{{Complete NNLO operator bases in Higgs
  effective field theory}},
  \href{http://dx.doi.org/10.1007/JHEP04(2023)086}{\emph{JHEP} {\bf 04} (2023)
  086}, [\href{https://arxiv.org/abs/2210.14939}{{\tt 2210.14939}}].

\bibitem{Sun:2022ssa}
H.~Sun, M.-L. Xiao and J.-H. Yu, \emph{{Complete NLO operators in the Higgs
  effective field theory}},
  \href{http://dx.doi.org/10.1007/JHEP05(2023)043}{\emph{JHEP} {\bf 05} (2023)
  043}, [\href{https://arxiv.org/abs/2206.07722}{{\tt 2206.07722}}].

\bibitem{Song:2025snz}
C.-Q. Song, H.~Sun and J.-H. Yu, \emph{{Systematic Spurion Matching between Low
  Energy EFT and Chiral Lagrangian}},
  \href{https://arxiv.org/abs/2501.09787}{{\tt 2501.09787}}.

\bibitem{Grimus:1995zi}
W.~Grimus and M.~N. Rebelo, \emph{{Automorphisms in gauge theories and the
  definition of CP and P}},
  \href{http://dx.doi.org/10.1016/S0370-1573(96)00030-0}{\emph{Phys. Rept.}
  {\bf 281} (1997) 239--308}, [\href{https://arxiv.org/abs/hep-ph/9506272}{{\tt
  hep-ph/9506272}}].

\bibitem{Buchbinder:2000cq}
I.~L. Buchbinder, D.~M. Gitman and A.~L. Shelepin, \emph{{Discrete symmetries
  as automorphisms of the proper Poincare group}},
  \href{http://dx.doi.org/10.1023/A:1015244830241}{\emph{Int. J. Theor. Phys.}
  {\bf 41} (2002) 753--790}, [\href{https://arxiv.org/abs/hep-th/0010035}{{\tt
  hep-th/0010035}}].

\bibitem{Chau:1984fp}
L.-L. Chau and W.-Y. Keung, \emph{{Comments on the Parametrization of the
  Kobayashi-Maskawa Matrix}},
  \href{http://dx.doi.org/10.1103/PhysRevLett.53.1802}{\emph{Phys. Rev. Lett.}
  {\bf 53} (1984) 1802}.

\bibitem{Gerard1983FermionMS}
J.~M. Gerard, \emph{Fermion mass spectrum insu(2)l×u(1)}, {\emph{Zeitschrift
  f{\"u}r Physik C Particles and Fields} {\bf 18} (1983) 145--154}.

\bibitem{PhysRevD.19.3369}
D.~Wyler, \emph{Discrete symmetries in the six-quark su(2)
  \ifmmode\times\else\texttimes\fi{} u(1) model},
  \href{http://dx.doi.org/10.1103/PhysRevD.19.3369}{\emph{Phys. Rev. D} {\bf
  19} (Jun, 1979) 3369--3379}.

\bibitem{PhysRevD.21.3417}
G.~C. Branco, H.~P. Nilles and V.~Rittenberg, \emph{Fermion masses and
  hierarchy of symmetry breaking},
  \href{http://dx.doi.org/10.1103/PhysRevD.21.3417}{\emph{Phys. Rev. D} {\bf
  21} (Jun, 1980) 3417--3422}.

\bibitem{Frampton:1994rk}
P.~H. Frampton and T.~W. Kephart, \emph{{Simple nonAbelian finite flavor groups
  and fermion masses}},
  \href{http://dx.doi.org/10.1142/S0217751X95002187}{\emph{Int. J. Mod. Phys.
  A} {\bf 10} (1995) 4689--4704},
  [\href{https://arxiv.org/abs/hep-ph/9409330}{{\tt hep-ph/9409330}}].

\bibitem{Henning:2017fpj}
B.~Henning, X.~Lu, T.~Melia and H.~Murayama, \emph{{Operator bases,
  $S$-matrices, and their partition functions}},
  \href{http://dx.doi.org/10.1007/JHEP10(2017)199}{\emph{JHEP} {\bf 10} (2017)
  199}, [\href{https://arxiv.org/abs/1706.08520}{{\tt 1706.08520}}].

\bibitem{Graf:2020yxt}
L.~Graf, B.~Henning, X.~Lu, T.~Melia and H.~Murayama, \emph{{2, 12, 117, 1959,
  45171, 1170086, \textellipsis{}: a Hilbert series for the QCD chiral
  Lagrangian}}, \href{http://dx.doi.org/10.1007/JHEP01(2021)142}{\emph{JHEP}
  {\bf 01} (2021) 142}, [\href{https://arxiv.org/abs/2009.01239}{{\tt
  2009.01239}}].

\bibitem{Sun:2022aag}
H.~Sun, Y.-N. Wang and J.-H. Yu, \emph{{Hilbert Series and Operator Counting on
  the Higgs Effective Field Theory}},
  \href{https://arxiv.org/abs/2211.11598}{{\tt 2211.11598}}.

\bibitem{Jarlskog:1985ht}
C.~Jarlskog, \emph{{Commutator of the Quark Mass Matrices in the Standard
  Electroweak Model and a Measure of Maximal CP Nonconservation}},
  \href{http://dx.doi.org/10.1103/PhysRevLett.55.1039}{\emph{Phys. Rev. Lett.}
  {\bf 55} (1985) 1039}.

\bibitem{Jarlskog:1985cw}
C.~Jarlskog, \emph{{A Basis Independent Formulation of the Connection Between
  Quark Mass Matrices, CP Violation and Experiment}},
  \href{http://dx.doi.org/10.1007/BF01565198}{\emph{Z. Phys. C} {\bf 29} (1985)
  491--497}.

\bibitem{Bernabeu:1986fc}
J.~Bernabeu, G.~C. Branco and M.~Gronau, \emph{{CP Restrictions on Quark Mass
  Matrices}}, \href{http://dx.doi.org/10.1016/0370-2693(86)90659-3}{\emph{Phys.
  Lett. B} {\bf 169} (1986) 243--247}.

\bibitem{Jenkins:2009dy}
E.~E. Jenkins and A.~V. Manohar, \emph{{Algebraic Structure of Lepton and Quark
  Flavor Invariants and CP Violation}},
  \href{http://dx.doi.org/10.1088/1126-6708/2009/10/094}{\emph{JHEP} {\bf 10}
  (2009) 094}, [\href{https://arxiv.org/abs/0907.4763}{{\tt 0907.4763}}].

\bibitem{Song:2024fae}
C.-Q. Song, H.~Sun and J.-H. Yu, \emph{{Complete CP-eigen bases of meson-baryon
  chiral lagrangian up to p$^{5}$-order}},
  \href{http://dx.doi.org/10.1007/JHEP09(2024)171}{\emph{JHEP} {\bf 09} (2024)
  171}, [\href{https://arxiv.org/abs/2404.15047}{{\tt 2404.15047}}].

\bibitem{Grinstein:2023njq}
B.~Grinstein, X.~Lu, L.~Merlo and P.~Qu\'\i{}lez, \emph{{Hilbert series for
  covariants and their applications to minimal flavor violation}},
  \href{http://dx.doi.org/10.1007/JHEP06(2024)154}{\emph{JHEP} {\bf 2024}
  (2024) 154}, [\href{https://arxiv.org/abs/2312.13349}{{\tt 2312.13349}}].

\bibitem{Kagan:2009bn}
A.~L. Kagan, G.~Perez, T.~Volansky and J.~Zupan, \emph{{General Minimal Flavor
  Violation}}, \href{http://dx.doi.org/10.1103/PhysRevD.80.076002}{\emph{Phys.
  Rev. D} {\bf 80} (2009) 076002}, [\href{https://arxiv.org/abs/0903.1794}{{\tt
  0903.1794}}].

\bibitem{Feldmann:2008ja}
T.~Feldmann and T.~Mannel, \emph{{Large Top Mass and Non-Linear Representation
  of Flavour Symmetry}},
  \href{http://dx.doi.org/10.1103/PhysRevLett.100.171601}{\emph{Phys. Rev.
  Lett.} {\bf 100} (2008) 171601}, [\href{https://arxiv.org/abs/0801.1802}{{\tt
  0801.1802}}].

\bibitem{Greljo:2022cah}
A.~Greljo, A.~Palavri\'c and A.~E. Thomsen, \emph{{Adding Flavor to the
  SMEFT}}, \href{http://dx.doi.org/10.1007/JHEP10(2022)005}{\emph{JHEP} {\bf
  10} (2022) 010}, [\href{https://arxiv.org/abs/2203.09561}{{\tt 2203.09561}}].

\bibitem{Kobayashi:2021pav}
T.~Kobayashi, H.~Otsuka, M.~Tanimoto and K.~Yamamoto, \emph{{Modular symmetry
  in the SMEFT}},
  \href{http://dx.doi.org/10.1103/PhysRevD.105.055022}{\emph{Phys. Rev. D} {\bf
  105} (2022) 055022}, [\href{https://arxiv.org/abs/2112.00493}{{\tt
  2112.00493}}].

\bibitem{Calo:2022jqv}
S.~Cal\`o, C.~Marinissen and R.~Rahn, \emph{{Discrete symmetries and efficient
  counting of operators}},
  \href{http://dx.doi.org/10.1007/JHEP05(2023)215}{\emph{JHEP} {\bf 05} (2023)
  215}, [\href{https://arxiv.org/abs/2212.04395}{{\tt 2212.04395}}].

\bibitem{Kobayashi:2023zzc}
T.~Kobayashi and M.~Tanimoto, \emph{{Modular flavor symmetric models}},  7,
  2023.
\newblock \href{https://arxiv.org/abs/2307.03384}{{\tt 2307.03384}}.

\bibitem{Loisa:2024xuk}
E.~Loisa and J.~Talbert, \emph{{Froggatt-Nielsen meets the SMEFT}},
  \href{http://dx.doi.org/10.1007/JHEP10(2024)017}{\emph{JHEP} {\bf 10} (2024)
  017}, [\href{https://arxiv.org/abs/2402.16940}{{\tt 2402.16940}}].

\bibitem{Palavric:2024gvu}
A.~Palavri\'c, \emph{{Discrete leptonic flavor symmetries: UV mediators and
  phenomenology}},
  \href{http://dx.doi.org/10.1103/PhysRevD.110.115025}{\emph{Phys. Rev. D} {\bf
  110} (2024) 115025}, [\href{https://arxiv.org/abs/2408.16044}{{\tt
  2408.16044}}].

\end{thebibliography}\endgroup
\end{document}